 \documentclass[twocolumn,trackchanges,floatfix]{aastex62}

\newcommand{\hii}{{\sc Hii}}
\newcommand{\hi}{{\sc Hi}}

\newcommand{\kms}{{$\rm km\ s^{-1}$}}
\graphicspath{{./}{figures/}}
\usepackage{float}
\usepackage{dirtytalk}

\accepted{to \apj}
\shorttitle{In-Situ OB Star Formation in SMC}
\shortauthors{Vargas-Salazar et al.}

\begin{document}

\title{A Search for In-Situ Field OB Star Formation in the Small Magellanic Cloud}

\correspondingauthor{Irene Vargas-Salazar}
\email{ivargasa@umich.edu}

\author[0000-0001-7046-6517]{Irene Vargas-Salazar}
\affil{University of Michigan, 1085 S. University, Ann Arbor, MI 48109, USA}

\author[0000-0002-5808-1320]{M. S. Oey}
\affiliation{University of Michigan, 1085 S. University, Ann Arbor, MI 48109, USA}

\author{Jesse R. Barnes}
\affiliation{University of Michigan, 1085 S. University, Ann Arbor, MI 48109, USA}
\affiliation{Present address:  Private}

\author{Xinyi Chen}
\affiliation{University of Michigan, 1085 S. University, Ann Arbor, MI 48109, USA}
\affiliation{Present address: Yale University, New Haven, CT 06520, USA}

\author{N. Castro}
\affiliation{Leibniz-Institut für Astrophysik An der Sternwarte, 16 D-14482, Potsdam, Germany}

\author{Kaitlin M. Kratter}
\affiliation{University of Arizona, Tucson, AZ 85721, USA}

\author{Timothy A. Faerber}
\affiliation{University of Michigan, 1085 S. University, Ann Arbor, MI 48109, USA}
\affiliation{Present address: Department of Physics and Astronomy, Uppsala University, Box 516, SE-751 20 Uppsala,
Sweden}

\begin{abstract}

Whether any OB stars form in isolation is a question central to theories of massive star formation. To address this, we search for tiny, sparse clusters around 210 field OB stars from the Runaways and Isolated O-Type Star Spectroscopic Survey of the SMC (RIOTS4), using friends-of-friends (FOF) and nearest neighbors (NN) algorithms. We also stack the target fields to evaluate the presence of an aggregate density enhancement. Using several statistical tests, we compare these observations with three random-field datasets, and we also compare the known runaways to non-runaways. 
We find that the local environments of non-runaways show higher aggregate central densities than for runaways, implying the presence of some ``tips-of-iceberg'' (TIB) clusters.  We find that the frequency of these tiny clusters is low, 
$\sim 4-5\%$ 
of our sample. This fraction is much lower than some previous estimates, but is consistent with field OB stars being almost entirely runaway and walkaway stars.
The lack of TIB clusters implies that such objects either evaporate on short timescales, or do not form, implying a higher cluster lower-mass limit and consistent with a relationship between maximum stellar mass ($m_{\rm max}$) and the mass of the cluster ($M_{\rm cl}$).  On the other hand, we also cannot rule out that some OB stars may form in highly isolated conditions.  Our results set strong constraints on the formation of massive stars in relative isolation.

\end{abstract}


\section{Introduction} \label{sec:intro}

\citet{Roberts1957} first examined the question of whether all massive stars form in clusters or whether a significant number might form in isolation as field stars. Based on the limited data of that era, he concluded that OB stars rarely, if ever, form in the field. Additionally, there is evidence to suggest that star  formation occurs in unbound associations of OB stars \citep{Ward2020, Griffiths2018}. However, while it is widely accepted that most massive stars form in clusters or associations \citep[e.g.,][]{Lada2003, Zinnecker2007,Elmegreen1985}, it is well known that a significant population of massive stars is also found in sparse, field environments.  The reported frequency of field OB stars varies, depending on how the ``field'' is defined, but it is on the order of 20 -- 30\% \citep[e.g.,][]{Oey2004, Gies1987}.

In spite of their significant numbers, the nature and origin of the field massive stars has been unclear. Previous investigations on the cluster mass function imply that many, if not most, field OB stars formed {\it in situ} \citep[e.g.,][]{Oey2004, lamb2010}. On the other hand, large populations of runaway OB stars are also known to exist, and may dominate the field population \citep[e.g.,][]{Oey2018, deWit2005, Renzo2019}.

\subsection{In-Situ Field OB Star Formation}\label{subsec:InSitu}

The degree to which massive stars can form in isolation provides an important discriminant between the two dominant theories of massive star formation: competitive accretion and core accretion. The competitive accretion model theorizes that cores accrete matter from a shared reservoir of gas \citep{Zinnecker1982}. The core with the highest mass accretes the most matter due to its size and location at the center of the sub-cluster \citep{Bonnell2001} while lower mass cores accrete the remaining gas. Thus, this model requires that low-mass stars must form in the presence of massive stars, and vice versa \citep{Bonnell2004}, yielding a spectrum of stellar masses \citep[e.g.,][]{Zinnecker1982,Bonnell2001}. This stipulation implies a relationship between the mass of the most massive star formed in the cluster $m_{\rm max}$, and the total mass of the cluster $M_{\rm cl}$, by $m_{\rm max}\propto M_{\rm cl}^{2/3}$  \citep{Bonnell2004}.

In contrast, the core accretion model allows for occasional formation of isolated massive stars \citep{Krumholz2009,Li2003}. The model is a scaled-up version of low-mass star formation. It suggests that cores do not compete to accrete gas and instead, the amount of gas they accrete depends on the masses of the cores themselves before collapse \citep{Shu1987}. Clouds maintain their mass because fragmentation is prevented by heating from the accretion process \citep{KrumholzMcKee2008}. 
Monolithic collapse could then finally happen for sufficiently dense clouds with high column densities (at least 1 g/cm$^2$), forming massive stars.
Thus, if OB stars are able to form {\it in situ} in the field, this would provide substantial evidence favoring the core accretion model, whereas competitive accretion requires all OB stars to form in clusters. 

\citet{Oey2004} found that OB clusters, and the \hii\ region luminosity function, \citep[e.g.,][]{Oey1998, Oey2003}, follow a power-law distribution $\sim N_*^{-2}$ for the number of OB stars $N_*$ per cluster.  This power law extends to the extreme value of $N_*=1$, implying that OB stars with no other nearby massive stars appear as field stars, simply by populating the low end of the cluster mass function. These individual, field OB stars may simply be the ``tip of the iceberg'' (TIB) on tiny, sparse clusters at the low-mass extreme that are difficult to detect. 
\citet{lamb2010} provide evidence for the existence of such sparse clusters, or “minimal O star groups”, associated with field OB stars in the Small Magellanic Cloud (SMC). With observational data from the {\sl Hubble Space Telescope} on 8 SMC field OB stars, they find that 3 out of the 8 are in sparse clusters with $\leq10$ companion stars, each having masses of $1-4 M_{\odot}$.  

Additionally, the existence of these sparse clusters is consistent with stochastic nature of the cluster mass function and the stellar initial mass function (IMF). Monte Carlo simulations show that tiny clusters with OB stars can occur if clusters are built stochastically by randomly sampling stars from a universal IMF, which implies that the maximum stellar mass in a cluster is independent of cluster mass  \citep{lamb2010,ParkerGoodwin2007}. However, Monte Carlo simulations by \citet{WeidnerKroupa2006} tested various methods of populating clusters including a completely stochastic sampling. They found that clusters populated through random sampling do not fit observations of young clusters as well as a cluster populated through sorted sampling in which stellar masses are sorted in ascending order and their sum is constrained to be the cluster mass. This would imply that clusters form in an organized fashion and is consistent with the relation $m_{\rm max}\propto M_{\rm cl}^{2/3}$.

An important test for star formation and cluster models is thus the existence of truly isolated, {\it in situ} OB star formation.  We note that this is true whether or not "isolated" OB stars are binaries, since most binaries form from a single star-forming core.
While it is currently almost impossible to determine whether any OB stars form in true isolation, some tantalizing observations exist.  In the SMC, \citet{Oey2013} present a sample of 14 field OB stars that are strong candidates for {\it in situ} formation. These objects are found in the center of circular \hii\ regions, showing no bow shocks implying supersonic motion,
and having radial velocities matching those of the local {{\sc Hi}} components. Five of these targets are extremely isolated.  Also in the SMC, observations by \citet{Selier2011} show that the compact \hii\ region N33 is consistent with this object being a case of isolated massive-star formation. 
In the 30 Doradus region of the LMC, \citet{Bressert2012} identified 15 O stars as candidates for isolated formation. These stars are not in binary systems and show no evidence of clustering.
Additionally, \citet{Oskinova2013} suggests that one of the most massive stars in the Milky Way, WR102ka, may have been born in isolation. It is not a runaway, since it shows a circumstellar nebula with no bow shock, and there is no evidence of an associated star cluster.
\citet{deWit2004} found that $4\% \pm 2$\% (4-11 out of 193) of the Galactic O-star population either cannot be traced to OB associations, or have non-runwaway space velocities.

\subsection{Runaway and Walkaway OB Stars}

OB stars that are ejected from clusters could be runaway stars ($>30$ \kms) or walkaways, slower stars that are unbound, but below the runaway threshold velocity.
These are known to comprise a significant fraction of the field OB population \citep[e.g.,][]{Blaauw1961}.
Over the course of their lifetimes, these stars move far beyond their original birthplaces, and so by definition, they are distinct from stars in clusters. There are two mechanisms for producing runaways and walkaways: dynamical ejection \citep{Poveda1967} and binary supernova ejection \citep{Blaauw1961}. The dynamical mechanism ejects stars primarily from unstable three- or four-body systems \citep{LeonardDuncan1990} and is prevalent in dense cluster cores. In the binary supernova scenario, the primary star of a binary system explodes as a supernova which ejects the secondary star by a slingshot release that may be combined with a supernova ``kick''. 

The frequency of runaway OB stars is not well established, but it is generally believed to be large.  Some estimates suggest that it is on the order of 50\% of the field OB star population \citep[e.g.,][]{deWit2005, Gies1986}, while others suggest that almost all field OB stars are runaways \citep[e.g.,][]{Gies1987,Gvaramadze2012}. Recent work from our group using {\sl Gaia} DR2 proper motions 
is consistent with runaways strongly dominating the field OB population in the SMC \citep{Oey2018, DorigoJones2020}. 

\subsection{Remnants of Evaporated Clusters}

Another way to generate field OB stars is a hybrid between {\it in situ} star formation and dynamical effects.  The loss of stars from small clusters may result in some of these being much smaller than when they formed, and if an OB star is present, it would be observed as a TIB star, as described in Section~\ref{subsec:InSitu}.  In the most extreme case, the OB star could be completely abandoned by its cohorts, although studies are needed to determine the likelihood of this scenario.  Many clusters, especially at low mass, become unbound due to gas expulsion and feedback not long after the stars form, a phenomenon dubbed ``infant mortality'' or ``infant weight loss'' \citep[e.g.,][]{Lada2003, GoodwinBastian2006}.  However, \citet{Farias2018} suggest that gas expulsion may be more difficult than previously believed.  Alternatively, \citet{Ward2020} find that the formation of smaller, unbound associations with OB stars may be relatively commonplace, even in lower density environments, thus supporting scenarios where massive stars are not all formed in bound clusters. There is evidence of this in the Cyg OB2 association, which has a high frequency of wide binaries that would be disrupted through dynamical encounters in clusters \citep{Griffiths2018}, but are possible if they form in unbound associations.

\section{A Search for Field OB Star Formation} \label{sec:algorithms}

To understand the contribution, if any, of {\it in situ} OB star formation to the field massive star population, we search for small clusters associated with field OB stars in the SMC, to establish and quantify their existence. The SMC offers a complete sample of field OB stars in an external galaxy, and is located at a well-determined distance of 60~kpc \citep{Harries2003}. We employ two different cluster-finding algorithms, friends-of-friends (FOF) \citep{Battinelli1991} and nearest neighbors (NN) \citep{Schmeja2011}, and we also examine the stacked fields around the target OB stars for an aggregate density enhancement. 

We note that OB stars have a high multiplicity fraction. This implies that TIB stars in small clusters also may be binaries or multiples, which are difficult to discern. Studies have shown that field massive stars also may have significant binarity, from about half the binary frequency of those in clusters \citep{Stone1981, Gies1987} to frequencies on the order of those in clusters \citep{Mason2009, Lamb2016}.

To carry out our analysis, we use the Runaways and Isolated O-Type Star Spectroscopic Survey of the SMC  \citep[RIOTS4][]{Lamb2016}, which identifies a uniform, statistically complete sample of field massive stars in the SMC. RIOTS4 represents the field-star subset of OB star candidates that were photometrically identified by \citet{Oey2004} from the \citet{Massey2002} survey of the SMC, which covers the star-forming expanse of the galaxy. Field stars were differentiated from cluster stars by identifying stars that are at least 28~pc away from any other OB candidates in the analysis of \citet{Oey2004}. The RIOTS4 field stars are all spectroscopically confirmed OB stars \citep{Lamb2016}, and they represent $\sim$ 28\% of the total SMC OB population \citep{Oey2004}. \citet{Lamb2016} find a binary frequency of $\gtrsim 60$\% in this sample.

To search for small, sparse clusters associated with the field OB stars, we require deep stellar imaging of their fields.  The Optical Gravitational Lensing Experiment \citep[OGLE-III][]{Udalski2008} has accumulated \(I\)-band photometry on the SMC for many years. OGLE-III uses the 1.3-m Warsaw Telescope at Las Campanas. Each CCD image is $35 \times 35$ arcmin$^2$ with a scale of 0.26 arcsec/pixel. 

To carry out our cluster-finding algorithms, we require high-quality astrometry of all the stars near our target OB stars. Given the crowded fields in the SMC Bar region, we therefore performed PSF-fitting photometry and astrometry on $1000\times1000\ \rm px^2\ (76\times 76\ pc^2)$ subframes centered on the target stars.  We used the DAOphot software in IRAF\footnote{IRAF was distributed by the National Optical Astronomy Observatory, which was managed by the Association of Universities for Research in Astronomy (AURA) under a cooperative agreement with the National Science Foundation.} applied to the OGLE-III $I$-band images. 
The image PSFs are generally on the order of 3.0 pixels, or 0.23 pc FWHM with variation on the order of 10\%.
Within a 200-pixel (15-pc) radius of the target, each field was carefully vetted with both automatic and manual identification of the stellar objects, to optimize the sample completeness.  We then calibrated the photometry for the entire field using the published OGLE-III photometry \citep{Udalski2008}. Our photometric errors indicate that the data have excellent completeness for $I< 19.0$, and we applied this cutoff to our dataset.

Our final sample comprises 210 field OB stars. There are fewer stars in our sample than in the original RIOTS4 survey for several reasons. Firstly, the OGLE-III survey excludes the eastern-most region of the SMC Wing. Secondly, targets that are $< 200$ pixels from the edge of the OGLE-III CCD frames were discarded, since the cluster-finding algorithms rely on complete spatial distribution of stars near the target. In addition, there are targets for which we were unable to carry out the astrometry and photometry due to technical issues related to field placement within the frame, for example, the presence of an extremely bright, foreground star within the field, and spatial distortions near the detector edge.
Finally, 6 stars \citep[M2002-SMC 11802, 38893, 42654, 46022, 62981, and 67029;][]{Massey2002} in the RIOTS4 sample were inadvertently included even though they did not meet the criterion of being 28 pc from another OB candidate; these were also deleted from our sample.

\begin{figure}[h!]
\begin{center}
\includegraphics[scale=0.5,angle=0]{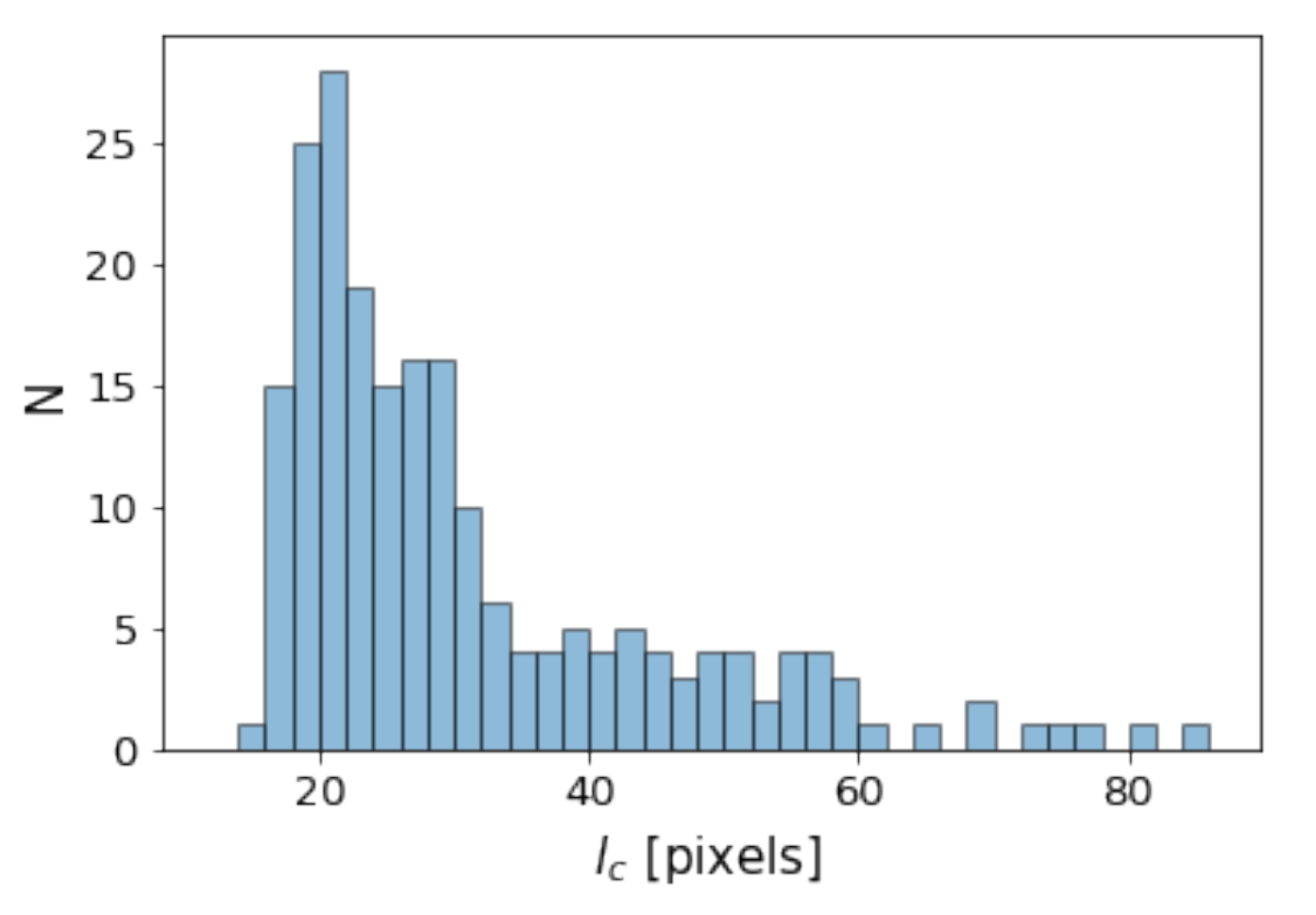}
\caption{The distribution of clustering length $l_c$ for stars having $I<19.0$ in our target fields. The lower stellar density in the Wing generates a long tail in the distribution to values much greater than those for bound stars in clusters. Therefore, a maximum value of $l_c =39$ px is adopted for fields with $l_c>50 $ px. 
\label{fig:1}}
\end{center}
\end{figure}

\subsection{Friends-of-Friends} \label{subsec:FOF}

The friends-of-friends cluster-finding method uses similar methodology as a minimum spanning tree algorithm.  It identifies associated stars, or ``friends'', as those that are located within a given clustering length (\(l_c\)) of another member.

We use the FOF algorithm in this manner
to search for faint companion stars in the OGLE-III images with $I < 19.0$ ($M_I = 0.11$, corresponding roughly to an A1 star), which might correspond to small clusters of which our target OB stars are the TIBs. 
The FOF clustering length, $l_c$, is the value characteristic of the background stellar density. Since our target OB stars are in locations of varying background density, we define $l_c$  specific to each field. To do this, we calculate the number of clusters found as a function of different clustering lengths
in a given field. The value that yields the maximum number of clusters is adopted as $l_c$ of the field \citep{Battinelli1991}. Our distribution of our final $l_c$ for our target fields is shown in Figure \ref{fig:1}.

The Wing of the SMC has a much lower stellar density than the Bar, generating a long tail in the $l_c$ distribution (Figure~\ref{fig:1}), to values much greater than what are likely to include bound stars in tiny clusters. We therefore adopt a maximum fixed \(l_c\) of 39 pixels (3.0 pc), which is 1-$\sigma$ above the median value, for any $l_c$ above 50 pixels. This is on the order of the core radii for small clusters 
\citep{Lada2003}. 

\begin{figure}[ht!]
\begin{center}
\includegraphics[scale=0.45,angle=0]{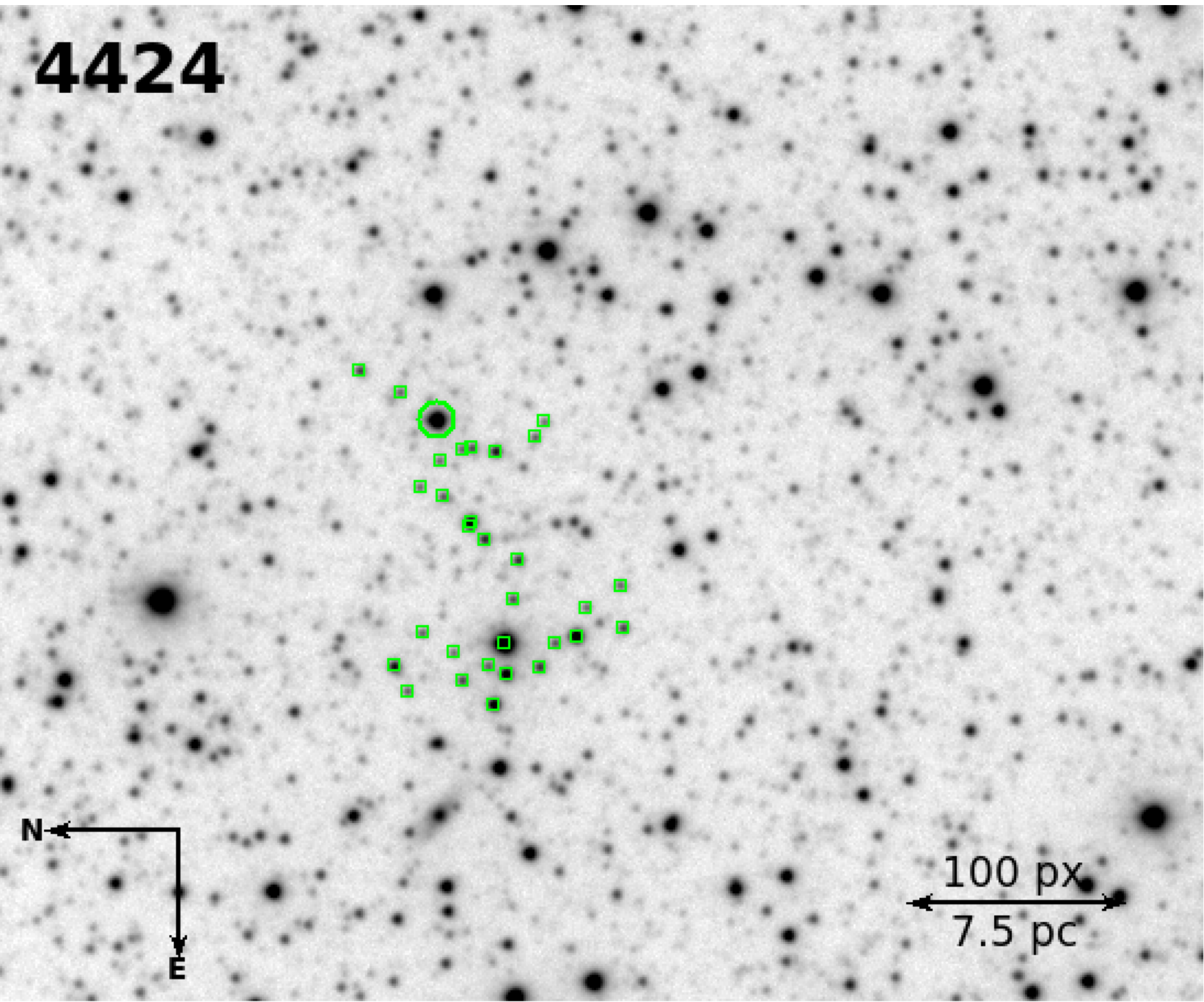}
\caption{An example of a filamentary string-like structure that occurs with FOF. Target 4424 is shown by the green circle.  \label{fig:4}}
\end{center}
\end{figure}

An inherent weakness in the FOF algorithm is that it has the tendency to associate stars in a filamentary structure that are physically unrelated.  
Although some clusters show real filamentary structure, the fact that this algorithm identifies such formations in artificial data demonstrates that it
could overestimate $N_*$ \citep{Schmeja2011}.
An example of this type of structure is shown in Figure \ref{fig:4}. 

We then apply the FOF algorithm, using the target OB stars as the origin for the algorithm. We evaluate the results using two criteria \citep{Schmeja2011,Campana2008}, one based on the number of stars (\(N_*\)) that are found to be associated; and another based on the $M$-value, which is a parameter that also takes into account the separation of the identified associated stars (see below). For both of these tests, a larger value indicates a higher probability that the associated stars correspond to a real cluster. 

However, some clusters do have filamentary structure, and we cannot reliably distinguish cases that are real from those that are not. Therefore, we generate three realizations of random-field datasets for each of our targets to serve as controls. The random fields have the same size ($1000\times1000\ \rm px^2\ $) and the same number of stars as the observed field for our targets, but with the stars randomly placed. 
This also allows us to evaluate the potential effects of random Poisson noise on the observed data. 

\begin{figure}[h!]
\begin{center}
\includegraphics[scale=0.5,angle=0]{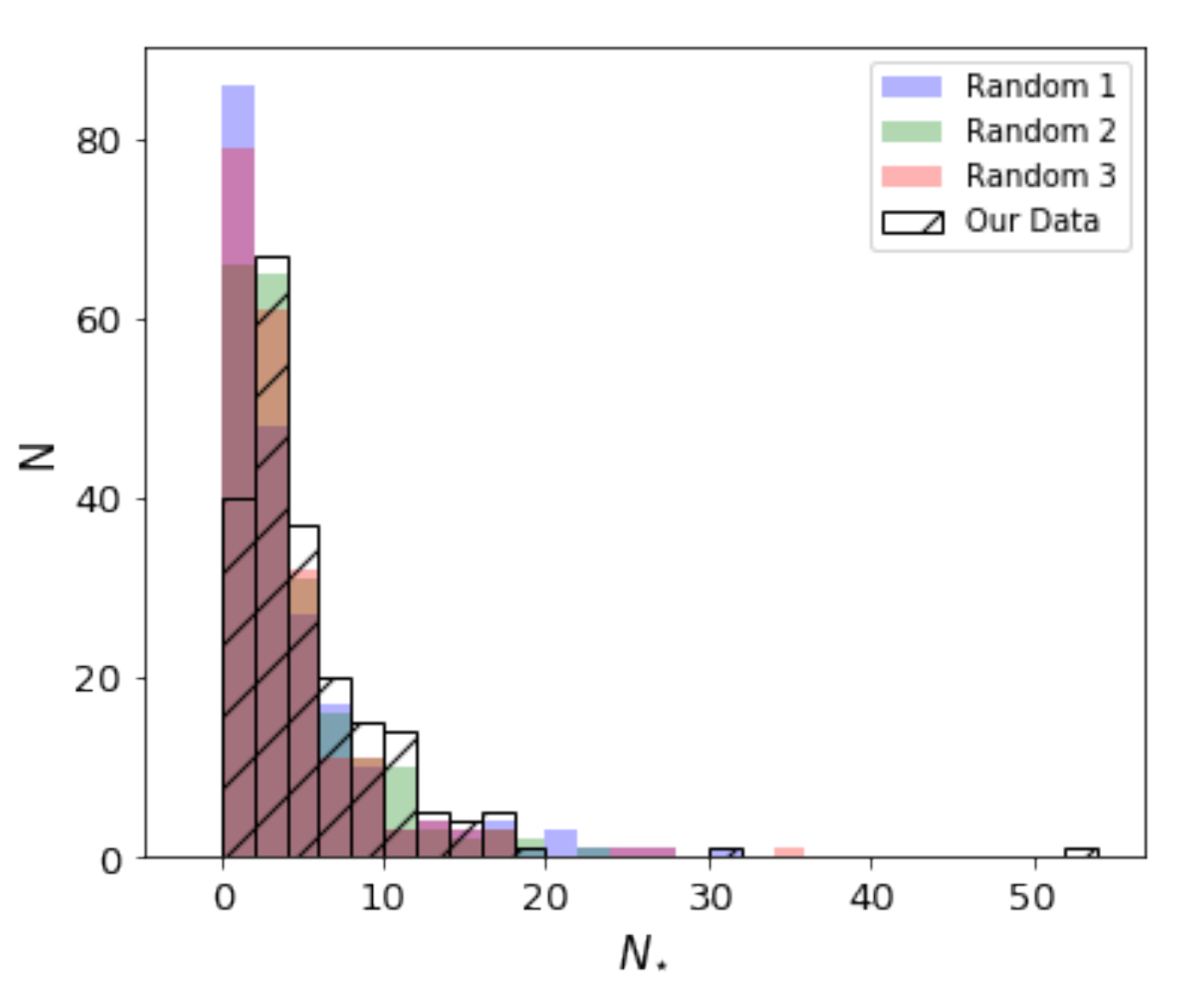}
\caption{The distribution of $N_*$, the number of stars associated by FOF for stars with $I<19.0$. Our observed dataset is plotted in black, and the random fields are colored. The Wilcoxon test indicates that our observed data and random fields are distinct, while our Rosenbaum test results reveal ambiguous results.
\label{fig:FOF1_Nstar}}
\end{center}
\end{figure}

Our \(N_*\) distribution, as well as those from the random fields, is shown in Figure \ref{fig:FOF1_Nstar}. We see that the observed distribution differs from the random fields at small $N_*$ values, but they are otherwise remarkably similar.

We use the Wilcoxon signed rank test for matched pairs to evaluate whether the observed and randomized distributions are statistically indistinguishable. This test is appropriate for two non-parametric datasets that are not independent, and is calculated from the difference between the pairs of data points for each field, the observed vs random field. The results are shown in Table \ref{table:statTests}. 
We adopt the conventional critical threshold of $p< 0.05$ for rejecting the null hypothesis that the two distributions originate from the same parent distribution.  Comparing the  $N_*$ distributions, we obtain  $p$-values that strongly indicate that the observed dataset differs from the randomized ones, indicating the potential presence of some TIB clusters.

We might expect that
the tails of the distributions should be sensitive to the presence of TIB clusters, which would have higher positive $N_*$ values, and would skew both the median and the tail to higher values. The Rosenbaum test \citep{Rosenbaum1965} is optimized to evaluate the significance of differences in the tails of two distributions, by counting the number of data points between the highest values of the two samples and quantifying the significance of the difference.
Using this test, we do not obtain a significant difference in spread, although the higher value does belong to our observed dataset in each case, when compared to the random data sets. The results are given in Table \ref{table:statTests} and show $p$-values higher than our threshold of 0.05, indicating that the tails of the observed data set vs the randomized ones do not differ statistically.

\startlongtable
\begin{deluxetable*}{lll|c}
\tablecaption{Statistical Test Results \label{table:statTests}}
\tablehead{
\colhead{Algorithm} & \colhead{Statistical Test} & \colhead{Dataset} & \colhead{$p$-values \tablenotemark{a}}
}
\startdata
FOF  $N_*$&  Wilcoxon& Full Data vs Random 1 & \textbf{0.001} \\ 
 & &   Full Data vs Random 2 & \textbf{0.006}   \\
    & & Full Data vs Random 3& \textbf{2.2e-5} \\
  & & Runaways vs Random 1    & \textbf{0.010} \\
     & & Runaways vs Random 2  & \textbf{0.034} \\
   & & Runaways vs Random 3   & \textbf{0.020}  \\
   & & Non-Runaways  vs Random 1   &  \textbf{0.003} \\
      & & Non-Runaways vs Random 2   & \textbf{0.044}  \\
   & & Non-Runaways vs Random 3    & \textbf{0.002} \\
    &Anderson-Darling &  Runaways vs non-Runaways   & 0.71 \\
    &Kolmogorov-Smirnov &  Runaways vs Non-Runaways   &  0.96\\
      & Rosenbaum&  Full Data vs Random 1  & 0.25  \\
       & &   Full Data vs Random 2 &  0.25  \\
    & & Full Data vs Random 3& 0.5 \\
            & &  Runaways vs Random 1  &  0.5 \\
                 & & Runaways vs Random 2  & 0.5 \\
   & & Runaways vs Random 3   & 0.5 \\
    & &  Non-Runaways vs Random 1  &  0.5  \\
          & & Non-Runaways vs Random 2   & 0.5 \\
   & & Non-Runaways vs Random 3    & 0.5 \\
 FOF  M-Test &Wilcoxon& Full Data vs Random 1 & \textbf{0.001}  \\
  & &   Full Data vs Random 2 & \textbf{0.010}   \\
    & & Full Data vs Random 3& \textbf{0.001} \\
   & & Runaways vs Random 1    &  \textbf{0.006} \\
        & & Runaways vs Random 2  & \textbf{0.015} \\
   & & Runaways vs Random 3   & \textbf{0.003} \\
   & & Non-Runaways vs Random 1   & 0.058 \\
         & & Non-Runaways vs Random 2   & 0.14 \\
   & & Non-Runaways vs Random 3    & \textbf{0.001} \\
    &Anderson-Darling &  Runaways vs non-Runaways   &  0.44\\
    &Kolmogorov-Smirnov &  Runaways vs Non-Runaways   & 0.51 \\
      & Rosenbaum&  Full Data vs Random 1   & 0.5  \\
       & &   Full Data vs Random 2 &  0.25  \\
    & & Full Data vs Random 3& 0.5 \\
      & &  Runaways vs Random 1 &  0.5  \\
              & & Runaways vs Random 2  & 0.5 \\
   & & Runaways vs Random 3   & 0.5 \\
    & &  Non-Runaways vs Random 1  & 0.5 \\
          & & Non-Runaways vs Random 2   & 0.5 \\
   & & Non-Runaways vs Random 3    & -0.5 \\
 \\
 NN Average &Wilcoxon& Full Data vs Random 1 & \textbf{1.4e-5} \\
 & &   Full Data vs Random 2 &  \textbf{2.3e-5}  \\
    & & Full Data vs Random 3&  \textbf{0.001}\\
    & & Runaways vs Random 1    & \textbf{0.028} \\
   & & Runaways vs Random 2  & \textbf{0.016} \\
   & & Runaways vs Random 3   & \textbf{0.020}  \\
   & & Non-Runaways vs Random 1    &  \textbf{5.8e-5}   \\
   & & Non-Runaways vs Random 2   & \textbf{2.6e-4} \\
   & & Non-Runaways vs Random 3    & \textbf{0.013} \\
 &Anderson-Darling &  Runaways vs Non-Runaways   & 0.067\\
   &Kolmogorov-Smirnov &  Runaways vs Non-Runaways   &  \textbf{0.022}\\
  & Rosenbaum&   Full Data  vs Random 1 & 0.25 \\
    & & Full Data  vs Random 2 & \textbf{0.016}   \\
      & & Full Data  vs Random 3 & 0.5   \\
      & &  Runaways vs Random 1   & -0.25  \\
    & &  Runaways vs Random 2  & 0.5   \\
    & &  Runaways vs Random 3  & -0.5  \\
   & &  Non-Runaways vs Random 1   & \textbf{0.031} \\
    & &  Non-Runaways vs Random 2   & \textbf{0.0078}  \\
    & &  Non-Runaways vs Random 3  & 0.5   \\
  NN Median &Wilcoxon& Full Data vs Random 1 &  \textbf{1e-5} \\ 
 & & Full Data vs Random 2    & \textbf{4.7e-5}\\
 & & Full Data Random 3   &\textbf{0.001} \\
    & & Runaways vs Random 1  & \textbf{0.020} \\
  & &  Runaways vs Random 2   & \textbf{0.019}  \\
   & & Runaways vs Random 3   & \textbf{0.023}  \\
   & & Non-Runaways vs Random 1   & \textbf{7.5e-5}  \\
   & & Non-Runaways vs Random 2    & \textbf{0.001} \\
   & & Non-Runaways vs Random 3   & \textbf{0.018}   \\
   &Anderson-Darling &  Runaways vs Non-Runaways   & 0.090\\
   &Kolmogorov-Smirnov &  Runaways vs Non-Runaways   &  \textbf{0.035}\\
  & Rosenbaum&  Full Data  vs Random 1 & 0.25 \\
    & & Full Data  vs Random 2 & \textbf{0.016}   \\
      & & Full Data  vs Random 3 & 0.5   \\
     & &  Runaways vs Random 1   & -0.25  \\
    & &  Runaways vs Random 2  & 0.5   \\
    & &  Runaways vs Random 3  & -0.5  \\
    & &  Non-Runaways vs Random 1   & \textbf{0.031} \\
    & &  Non-Runaways vs Random 2   & \textbf{0.0078}  \\
    & &  Non-Runaways vs Random 3  & 0.5   \\
\enddata
\tablenotetext{a}{A higher $p$-value indicates a higher probability for the null hypothesis that the two distributions originate from the same parent distribution.
The negative $p$-values indicate comparisons for which the cluster-finding test favors the random dataset. The bold $p$-values indicate statistically significant differences ($p < 0.05)$. 
}
\end{deluxetable*}

\begin{figure}[h!]
\begin{center}
\includegraphics[scale=0.5,angle=0]{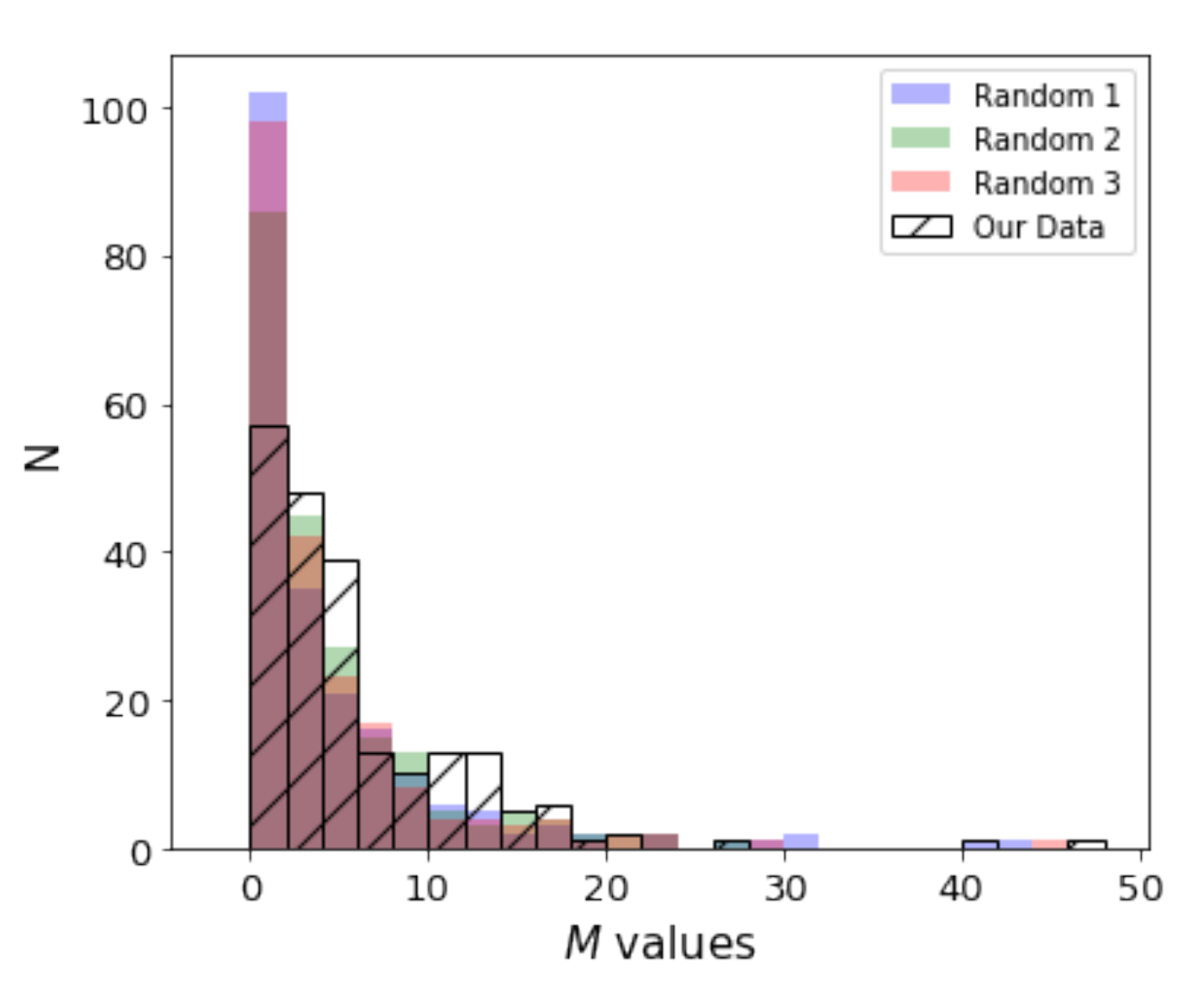}
\caption{The distributions of $M$-values for FOF applied to stars with $I<19.0$. Our observed dataset is plotted in black while the random fields are colored. 
These results are similar to our $N_*$ results.
\label{fig:3}}
\end{center}
\end{figure}

The second criterion we use to search for cluster candidates uses the  \(M\)-values \citep{Schmeja2011, Campana2008}, where:
\begin{equation} \label{eq:1}
M = \frac{l_{c,\rm SMC}}{l_{c,\rm field}} N_* \quad .
\end{equation}
This takes the ratio of the average clustering length for all the SMC fields ($l_{c,\rm SMC}$)
and the clustering length of a given field ($l_{c,\rm field}$) and multiplies it by the number of stars associated by FOF for that particular field. Therefore, this test not only uses $N_*$, but also takes into account the separation between these stars. A higher $M$-value corresponds to a larger number of stars that are closely spaced together. The distribution of $M$ values is shown in Figure \ref{fig:3} for our observed dataset and the random fields. 

We again find that the $M$-value distributions for our observed data and the random fields have different statistical test results. 
The Wilcoxon test $p$-values indicate that the observed and random datasets are statistically distinct while the Rosenbaum test results do not show a significant difference in spread.
It is possible that the Wilcoxon signal results from presence of some tiny clusters that are too small to affect the Rosenbaum tests.  On the other hand, this outcome could also be due to the background stars having positions that are not purely random. This is further discussed in Section \ref{sec:subpops}.

\subsection{Nearest Neighbors} \label{subsec:NN}

NN is an algorithm that measures the stellar density (\(\Sigma_j\)) associated with a given target star. This is calculated by counting the number of stars enclosed within the radius to its $j$th nearest neighbor in 2D as shown in equation \ref{eq:2} \citep[e.g.,][]{Schmeja2011}: 
\begin{equation} \label{eq:2}
\Sigma_j = \frac{j - 1}{S_j} \quad ,
\end{equation}
where \(j\) is the $j$th nearest neighbor and \(S_j\) is the area defined by the radius to the $j$th nearest neighbor.

\begin{figure}[h!]
\begin{center}
\includegraphics[scale=0.5,angle=0]{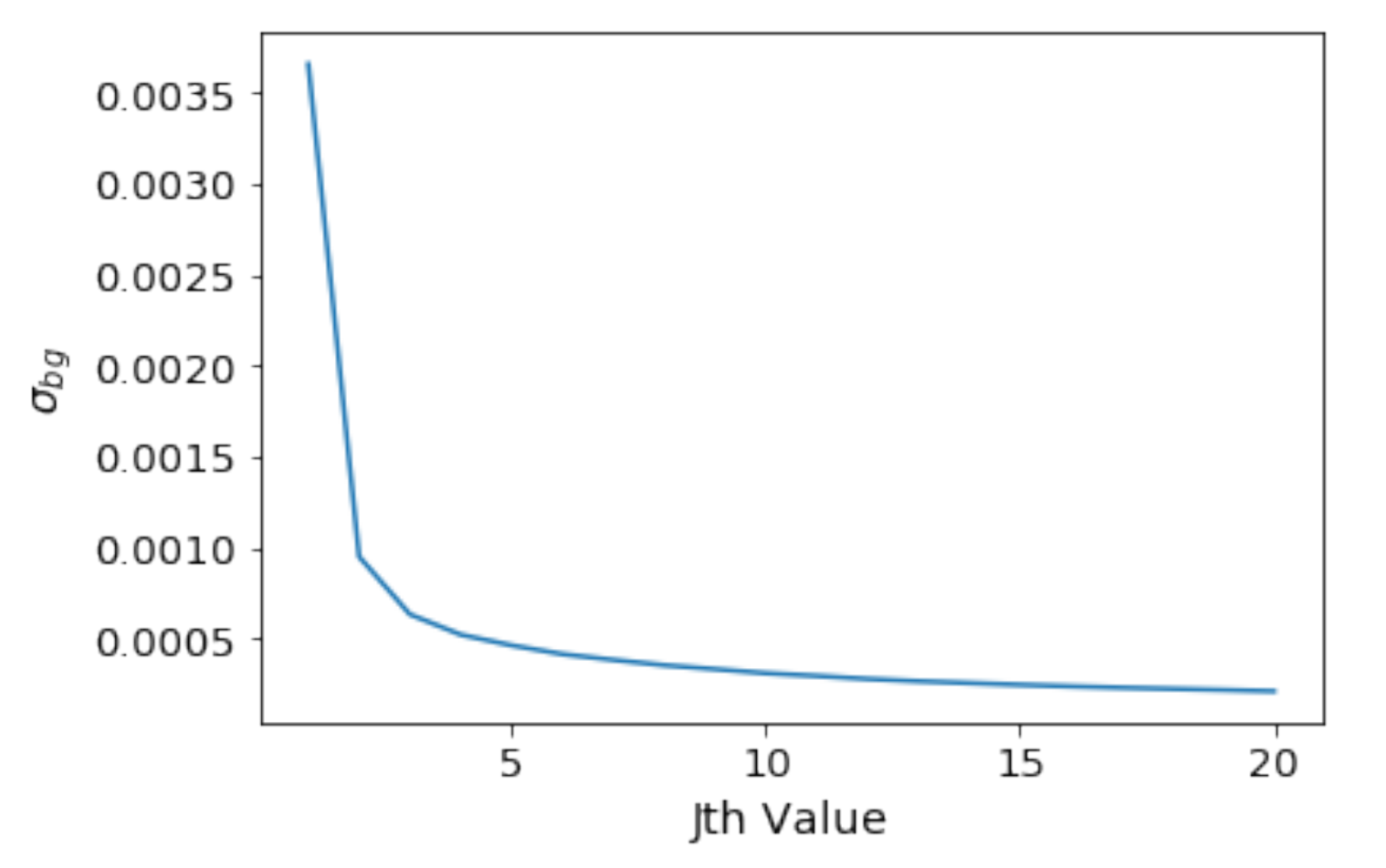}
\caption{ 
Average $\sigma_{bg}$ fluctuations as a function of $j$, showing that 
statistical fluctuations produce less noise at higher $j$ values.
\label{fig:9}}
\end{center}
\end{figure}

We compare the resulting stellar density to the background density. Since the background density varies greatly across the SMC, we perform background density calculations for each target individually. The average background density (\(\Sigma_{bg}\)) is calculated from the total number of stars in the field $N_t$, the total number of stars $N_j$ within \(S_j\), and the area outside \(S_j\), so as to not include the area within a potential cluster, as shown in equation \ref{eqn:avgbckd}:
\begin{equation} \label{eqn:avgbckd}
\Sigma_{bg} = \frac{N_{t} - N_j}{S_t- S_j} \quad ,
\end{equation}
where \(S_t\) is the total area of the field ($1000\times1000\ \rm px^2\ $). 
The Poisson error of $\Sigma_{bg}$ is therefore,
\begin{equation} \label{eqn:sigmabkgd}
\sigma_{bg} = \Sigma_{bg} \times N_{bg,j}^{-1/2} \quad ,
\end{equation}
where $N_{bg,j} = \Sigma_{bg}\times S_j$ is the number of background stars expected in area $S_j$. We caution that some of our fields may occasionally include external clusters or overdensities in the background, which would overestimate the background.

The NN algorithm works best for small clusters when $\Sigma_j$ is averaged over a few values of 
$j$ in the range $3 < j < 20$ \citep{Schmeja2011, CasertanoHut1985};
the lowest $j$ values are sensitive to statistical fluctuations, while at high $j$ values, the signal from small clusters becomes too diluted. We would need to select $j$ values as low as 3 to probe within the cluster radius. However, after reviewing results for various ranges from $j = 3$ to $j = 12$, we find that statistical fluctuations 
are sufficiently damped around $j > 8$ (Figure \ref{fig:9}). 

\begin{figure}[h!]
\begin{center}
\includegraphics[scale=0.5,angle=0]{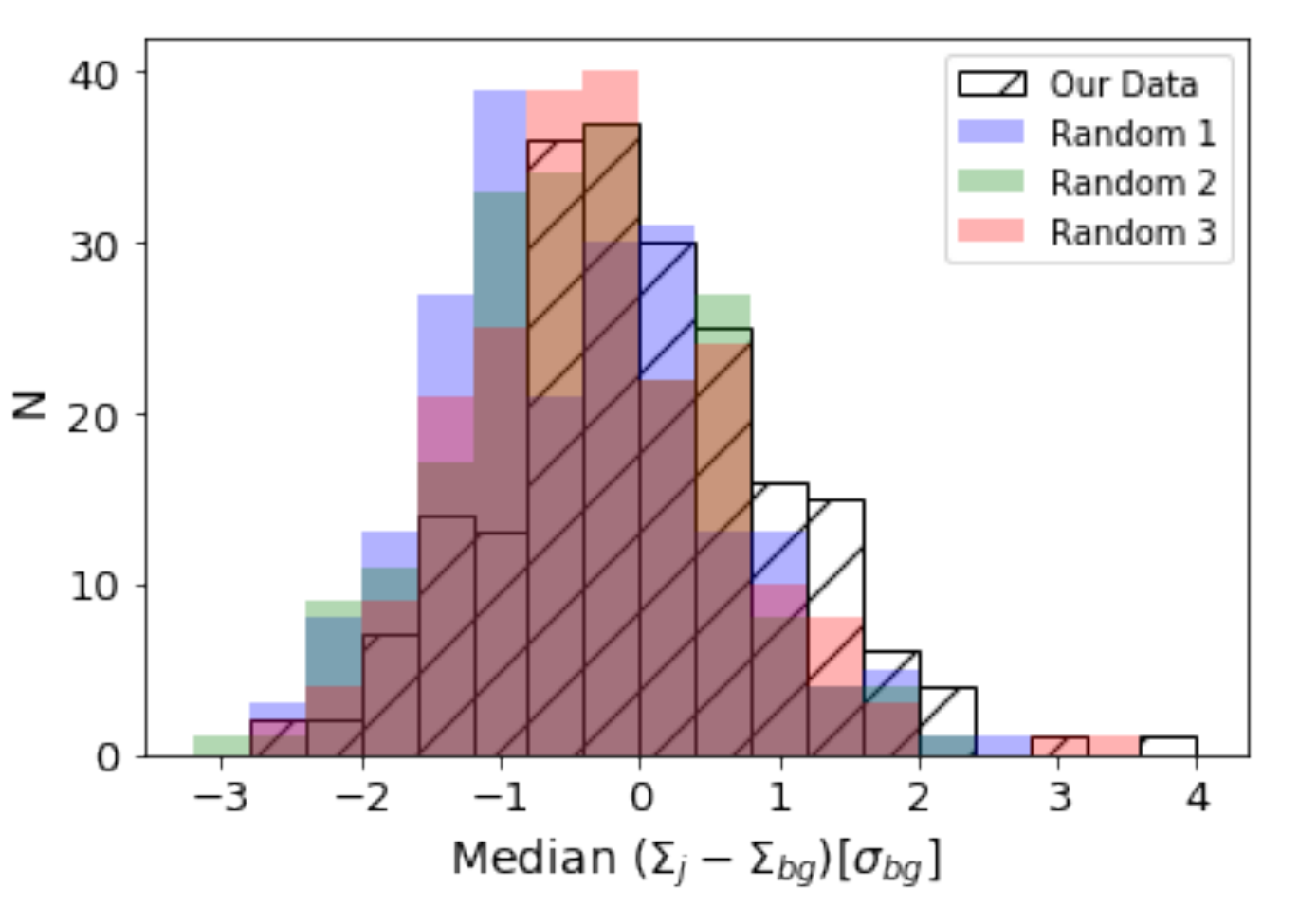}
\includegraphics[scale=0.5,angle=0]{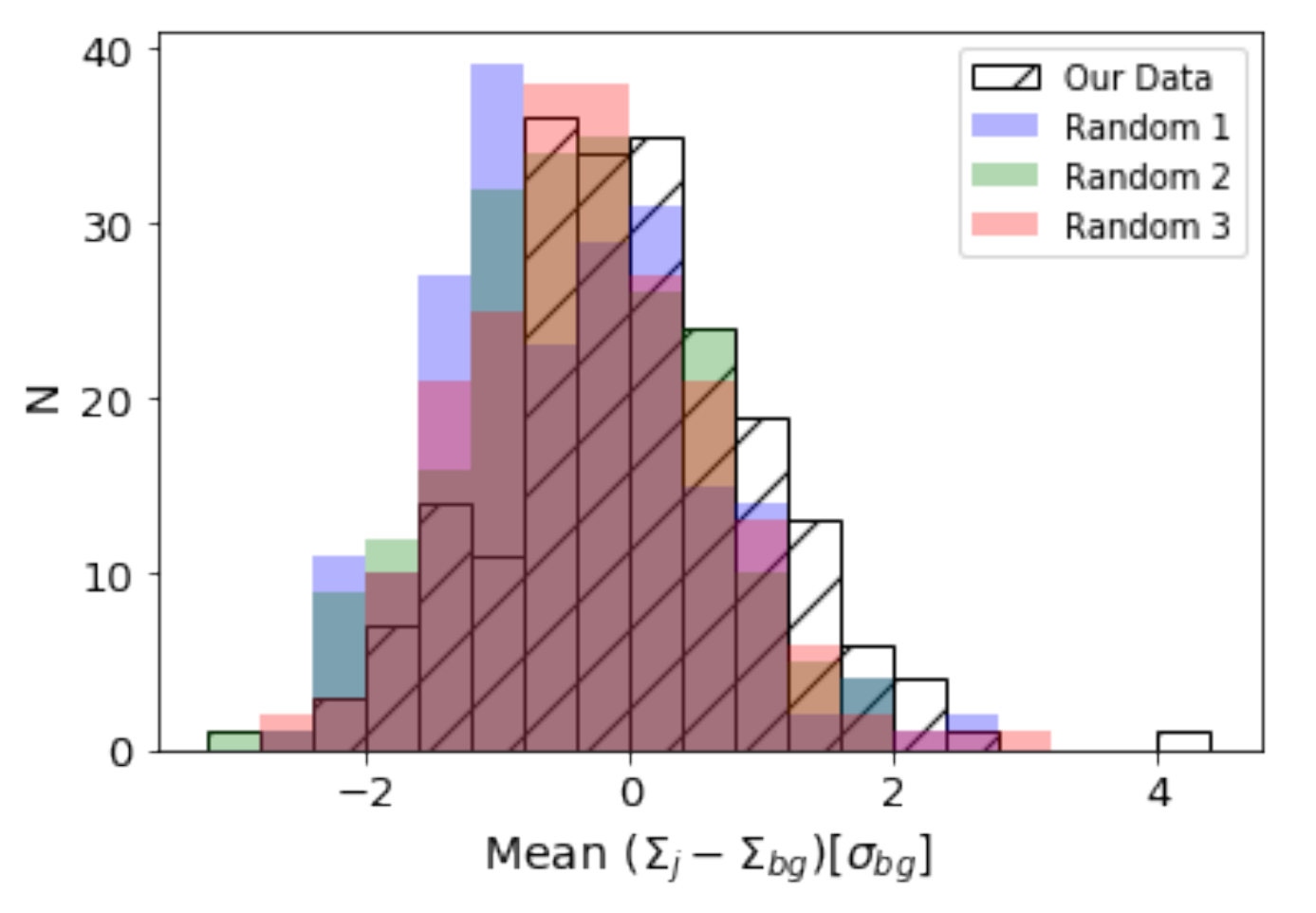}
\caption{Our NN results for the average (bottom) and median (top) overdensities for $j= 8$ to 12 as well as the three random datasets. Our observed data are shown in black while the random-field datasets are colored as shown. 
The Wilcoxon test indicates that our observed data and random fields are distinct, while two out of 3 of our Rosenbaum test results indicate the contrary. \label{fig:NNResults}}
\end{center}
\end{figure}

We therefore use the range $j=8 - 12$ as the basis for our cluster-finding analysis.
We calculate the difference between $\Sigma_j$ and the background density $\Sigma_{bg}$ in units of $\sigma_{bg}$ for $j=8-12$.  We then obtain the average and median differences across these $j$ values for each target field (Figure \ref{fig:NNResults}). Systems with higher $\Sigma_j$ above the background $\Sigma_{bg}$ are more likely to be physical clusters. 

We again compare our results with the random field data. Since NN is also calculated from a specific target star, we choose the star nearest to the center in the random datasets as the origin of this algorithm. These results are plotted together with our NN results in Figure \ref{fig:NNResults}.

The density distributions peak at slightly negative values because the density measurement is centered on a star, rather than a random, star-less point; it is caused by the fact that positions centered on stars are necessarily farther from the nearest star than random positions between them, which causes the stellar densities to be underestimated at the lowest $j$ values \citep{CasertanoHut1985}.

The Wilcoxon and Rosenbaum tests comparing the observed and random-field data are given in Table \ref{table:statTests}.
The statistical tests yield ambiguous results.
While the Wilcoxon test shows a significant likelihood of TIB clusters being present, the Rosenbaum test for NN, like the previous results for FOF, find that our observed data are indistinguishable from 2 of the 3 random-field datasets.
\begin{figure*}[ht!]
\begin{center}
\gridline{
\includegraphics[scale=0.3,angle=0]{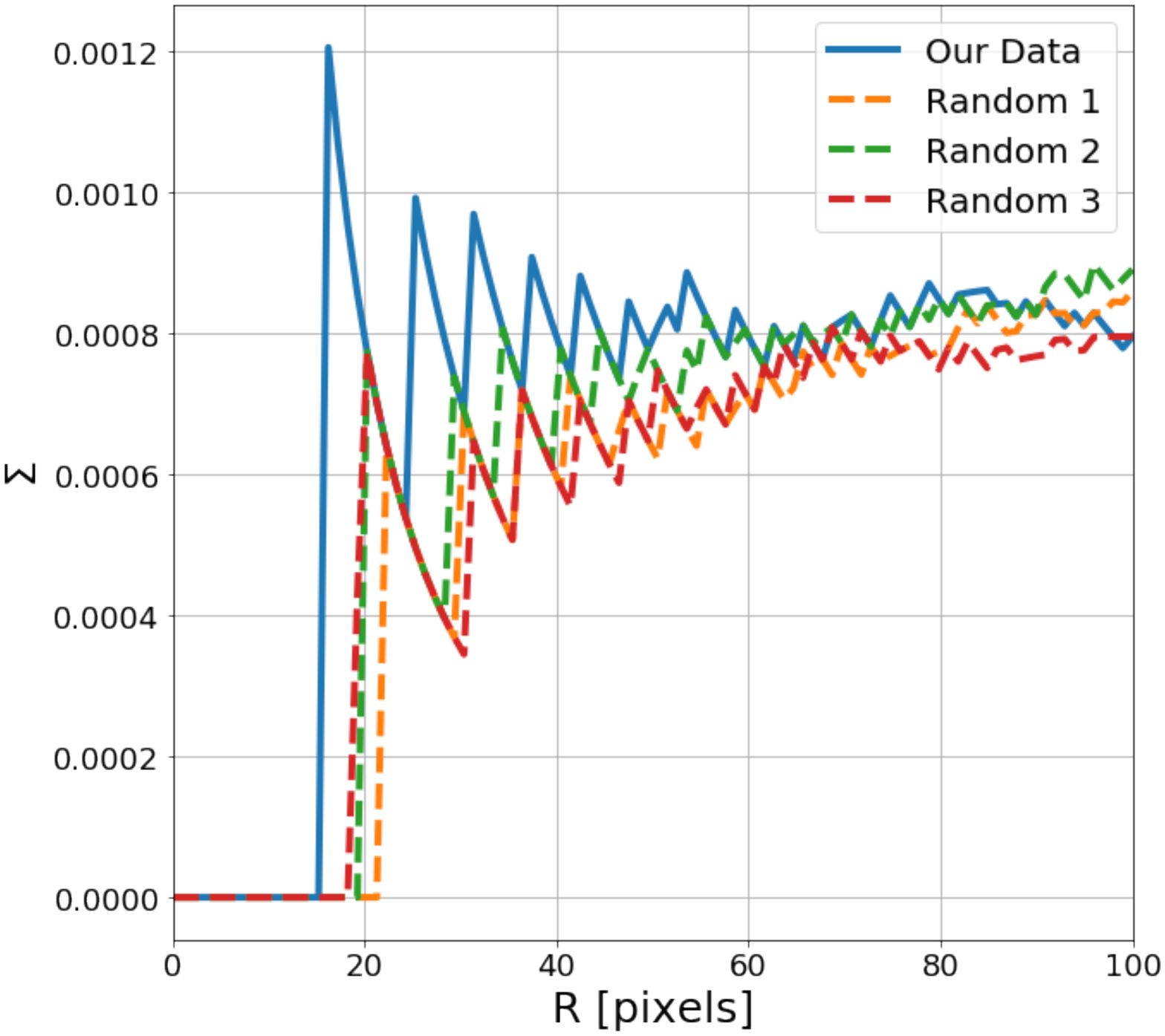}
\includegraphics[scale=0.3,angle=0]{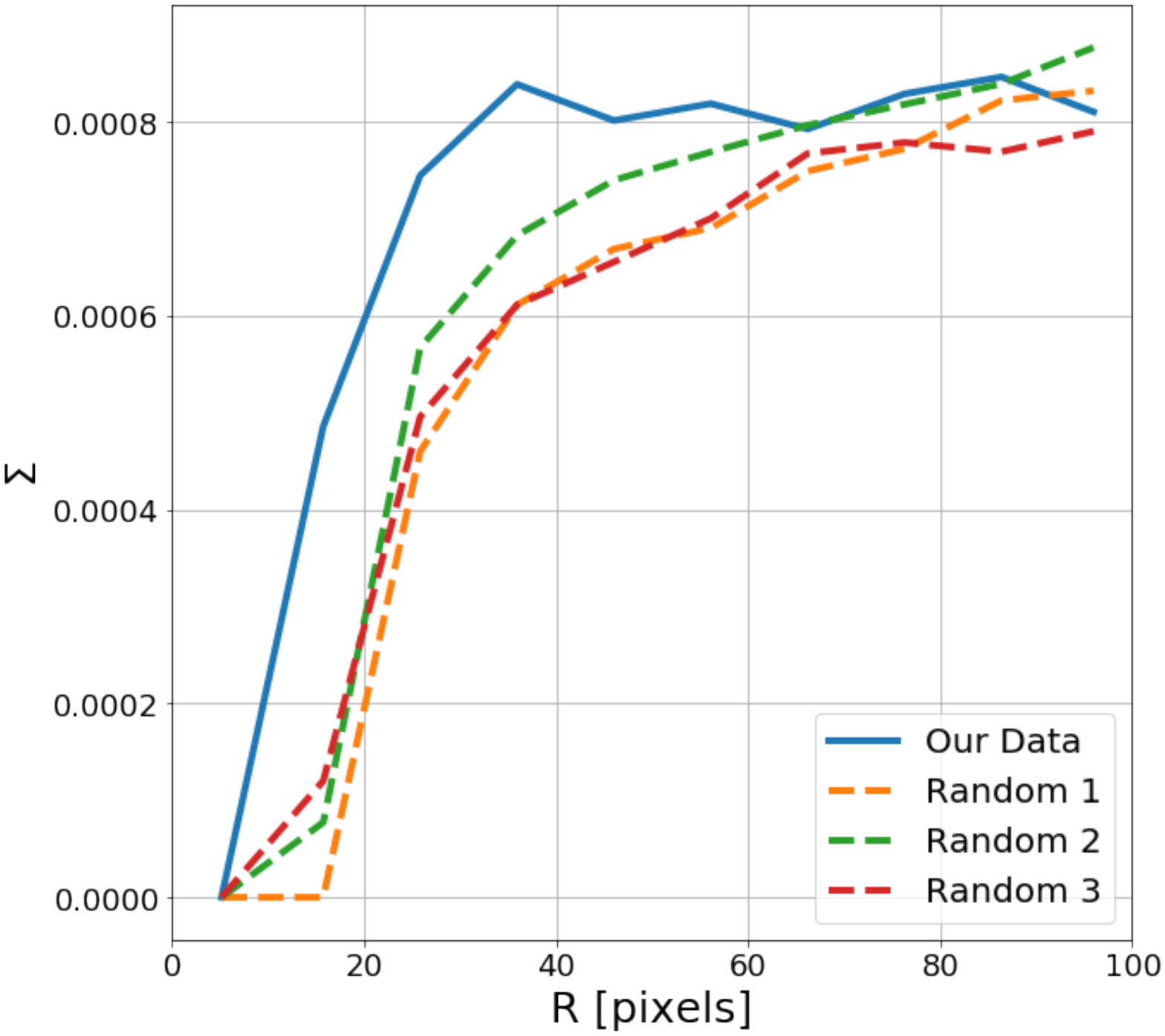}
}
\caption{Comparison of stacked target fields with stacked random datasets. This plot shows the stellar density as a function of radius from the target for our observed data (solid blue), along with those for our three random data sets (dashed lines). It is binned by 1 (left) and by 10 (right).
The observed data show a modest excess relative to the random fields at radii $< 60$ px suggesting the presence of a real aggregate enhancement.  The sawtooth pattern in the unsmoothed plot are due to undersampling of the data (see text). \label{fig:densitygrad_all}}
\end{center}
\end{figure*}
These contradicting statistical test results suggest that the observations are in a regime where TIB clusters are marginally detected, which is further discussed below in Section \ref{sec:subpops}.

\subsection{Stacked Fields} \label{subsec:stackedfields}

Given the non-detection, or at best, marginal detection, of any clusters by the above methods, we can improve our detection sensitivity for the aggregate sample by stacking the data for all the fields.
We measure the stellar density as a function of radius from each target star and
then take the median of all of our target fields at each radial step.
We do the same for the random fields.  For these, the densities are measured by centering on the star closest to the center of the field, as before. The radial density profiles
for the observed data and the random datasets are shown in Figure \ref{fig:densitygrad_all}. 
The sawtooth pattern in the unsmoothed plots result from oversampling the relatively small number of discrete stars relative to the higher resolution pixel grid:
the value of the stellar density associated with individual stars decreases with radius until additional stars are included within the target area.
The random-field data show a trend of increasing stellar density with radius that flattens out around 60 px. This is again the statistical effect caused by selecting a star, rather than a truly random position, as the origin for the counting algorithm.

The observed data do show an excess relative to the random fields at radii $< 60$ px, 
suggesting the presence of a real aggregate enhancement that may be due to the presence of some TIB clusters.  This will be discussed further below in Section~\ref{sec:implications}.

\section{Subpopulations} \label{sec:subpops}

\subsection{Runaways and Non-runaways} \label{subsec:runaways}

We know that a large fraction of the field OB stars are runaways, which would not be in TIB clusters. \citet{Oey2018} identified runaways with transverse velocities $> 30$ \kms\ from {\sl Gaia} proper motions.  Thus, we can evaluate the reliability of the cluster-finding algorithms by determining how many of the best cluster candidates are identified as runaways. For FOF, the best TIB candidates are those with the highest $N_*$ and $M$-values, and for NN, they are the target fields with the highest overdensities relative to the background. We use the residual transverse velocities $v_{\rm loc,\perp}$ that were measured by \citet{Oey2018} relative to the local velocity fields, adopting their runaway definition of $v_{\rm loc,\perp} > 30$ \kms.

Of the top 20 TIB candidates from the FOF analysis,
8 are runaways among the top $N_*$ candidates, and 7 among the top $M$-test candidates.  For the NN algorithm, 9 and 8 are among the top 20 candidates based on the median and average overdensities, respectively. These findings are summarized in Table \ref{table:runawayID}, where these runaways are identified.
Thus we see that on the order of half of even the top 20 TIB candidates for both FOF and NN are runaways.  This is reasonably consistent with the 2/3 fraction of runaways in the RIOTS4 sample \citep{DorigoJones2020}.
Although the Wilcoxon tests in Section~\ref{sec:algorithms} show a significant difference between the observed and random fields,
the number of runaways among the best TIB candidates underscores the role of random density fluctuations in generating signals suggestive of TIB clusters by our algorithms. 

The FOF algorithm shows significantly more runaways among the top 5 TIB candidates 
for both the $N_*$ and $M$ criteria than obtained by NN. Even 3 of the top 5 candidates identified by the $N_*$ criterion are runaways. On the other hand, for both NN criteria, none of the top 5 candidates include known runaways.  We caution that the {\sl Gaia} measurement errors are relatively large ($\sim 28$ \kms), and there is moreover uncertainty regarding the proper motion for any individual object; since the RIOTS4 runaway threshold is 30 \kms, the measurement errors leave open the possibility that a significant fraction of non-runaways are mis-identified as runaways, and could therefore be TIBs. 
However, note that this interpretation also depends on, and is consistent with, the difference between the FOF and NN results being due to the existence of a few real TIB clusters among the top candidates identified by NN.

\startlongtable
\begin{deluxetable*}{c|cccc}
\tablecaption{Runaways Among Top TIB Candidates  \label{table:runawayID} \tablenotemark{a}}
\tablehead{
\colhead{Target}\tablenotemark{b} & \colhead{FOF $N_*$} &
\colhead{FOF $M$-value} &\colhead{NN Average}& \colhead{NN Median}
}
\startdata
3815 & \nodata & $\circ$ & \nodata &$\circ$\\
13896 & \nodata & \nodata & $\circ$ &$\circ$\\ 
27600 & \nodata & $\circ$ & $\circ$ &$\circ$\\ 
27712 & \nodata & $\circ$ & \nodata&\nodata\\ 
27884 & $\bullet$ & $\bigstar$ &\nodata &\nodata\\ 
30492&\nodata & \nodata& $\circ$ & $\circ$ \\ 
43411& \nodata& \nodata& $\circ$ & $\circ$ \\ 
67893 & $\circ$ & \nodata & \nodata&\nodata   \\
70149 & $\bullet$ &\nodata& \nodata&\nodata   \\
71409 & $\bigstar$&$\circ$ & \nodata&\nodata \\
71652& $\bullet$& \nodata& $\circ$& $\circ$ \\ 
72208& $\bigstar$& $\circ$& \nodata& \nodata \\ 
73952& $\circ$& \nodata& \nodata& \nodata \\ 
74946& $\bigstar$& $\bigstar$& $\bullet$& $\bullet$ \\ 
75626& \nodata& \nodata& $\bullet$ & $\bullet$ \\ 
77734& \nodata& \nodata& $\bullet$ & $\bullet$
\enddata
\tablenotetext{a}{Open circles, filled circles, and stars correspond to objects identified among the top 20, 10, and 5 TIB candidates, respectively.}
\tablenotetext{b}{ID from \citet{Massey2002}}
\end{deluxetable*}

Separating our sample into runaway and non-runaway targets should strengthen the signal of any real TIBs among the latter.  Thus, we compare the results of our cluster-finding algorithms for these subsamples below in Figures \ref{fig:NNrunaways}, \ref{fig:FOFrunaways}  and \ref{fig:densitygradrunaway}.
We also apply the Wilcoxon and the Rosenbaum tests to compare the runaway and non-runaway targets to their respective random fields, as well as to each other. Our results are shown in Table \ref{table:statTests}.

The Wilcoxon and Rosenbaum test results for the runaways are essentially identical to those for the full sample.
For runaways, we would expect to not see any statistical differences from random fields. We therefore believe that the positive detections from the Wilcoxon test are not due to cluster detections, but instead result from other effects, like the possible non-random spatial distribution of field stars suggested earlier. This is consistent with the stacked field results for runaways, where at small radii they appear to have ambiguous, but slightly higher, densities.

\begin{figure*}[ht!]
\begin{center}
\gridline{
\includegraphics[scale=0.5,angle=0]{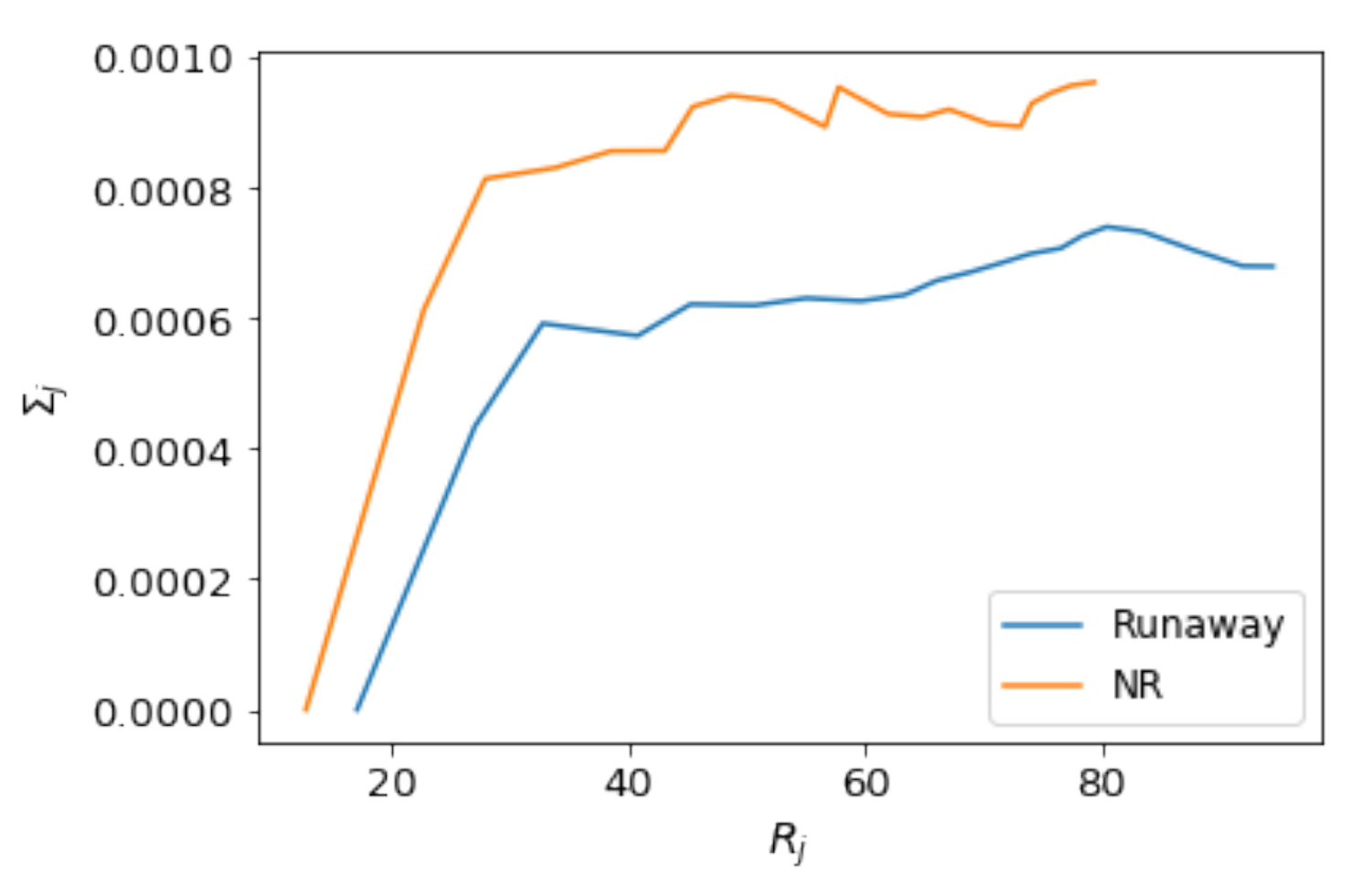}
\includegraphics[scale=0.5,angle=0]{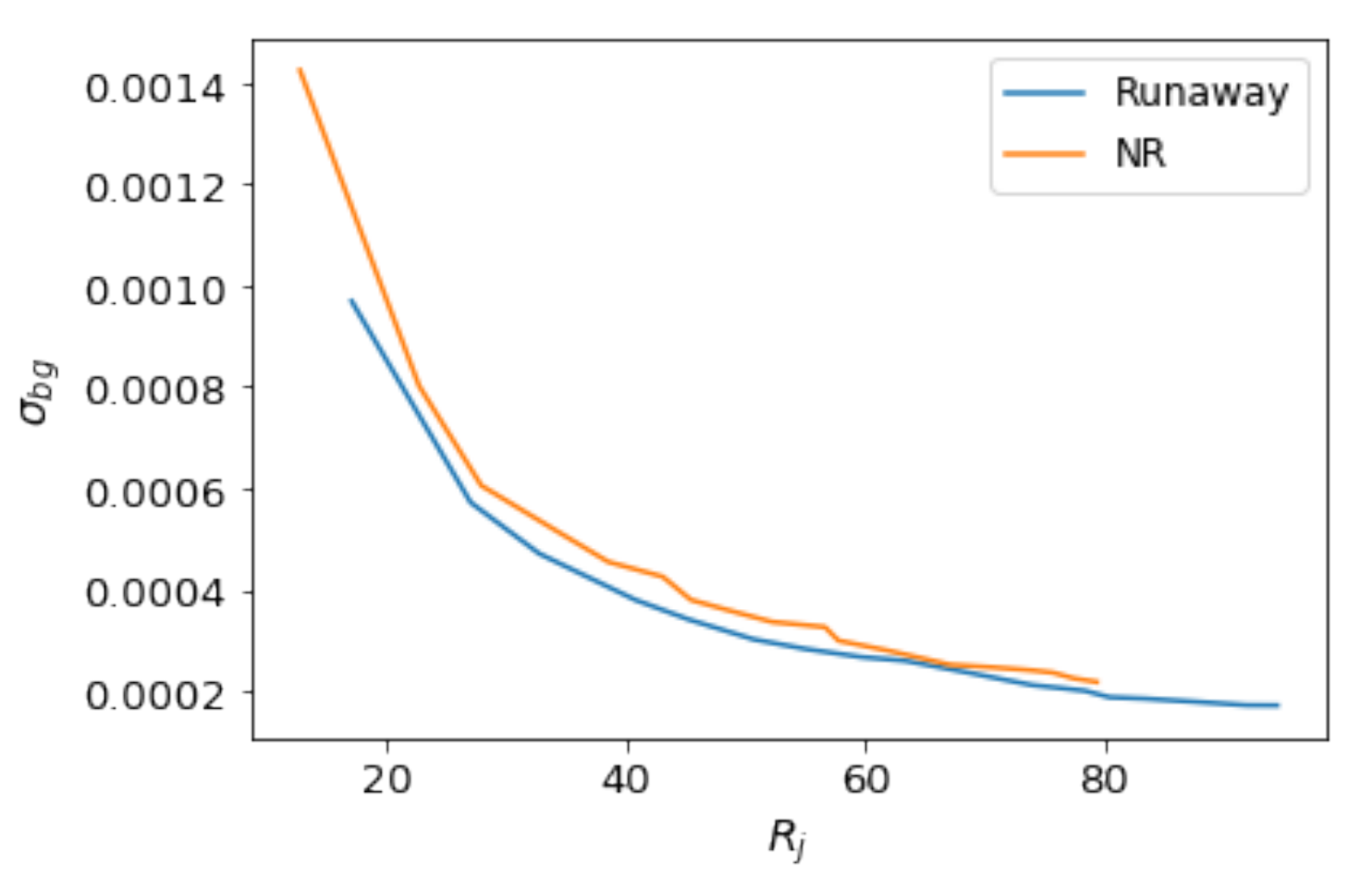}
}
\gridline{
\includegraphics[scale=0.5,angle=0]{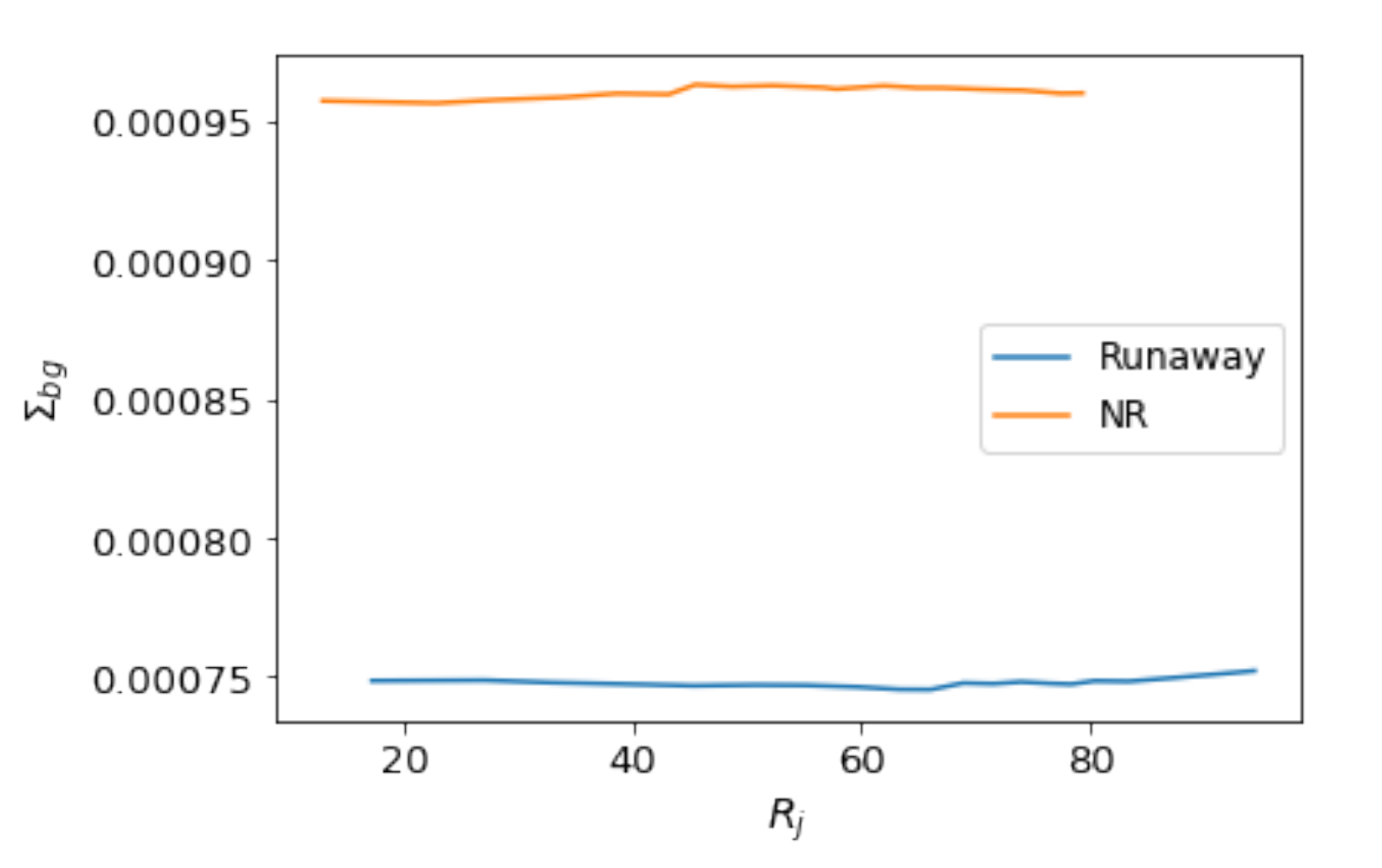}
\includegraphics[scale=0.5,angle=0]{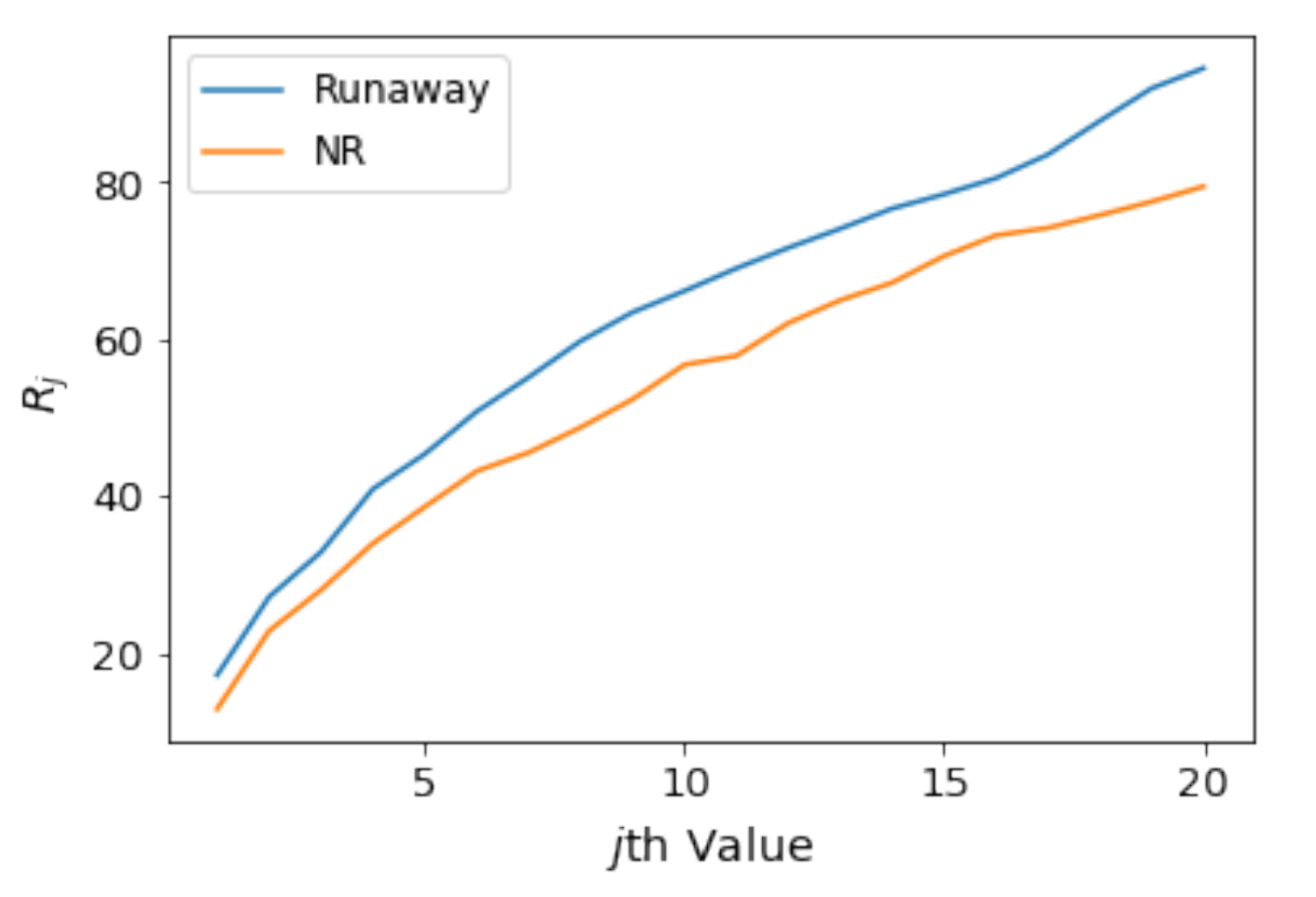}
}
\caption{Distribution of various NN values for our runaway and non-runaway (NR) data sets: the median $\Sigma_j$(upper left), the median $\sigma_{bg}$ (upper right) and median $\Sigma_{bg}$ 
(lower left) as a function of median $R_j$ values. The lower right plot is of the median $R_j$ values as a function of $j$. On average, the runaway data set shows higher values for $R_j$, but lower $\Sigma_j$ and $\Sigma_{bg}$ values, than the non-runaways, indicating that they are found in lower density environments. \label{fig:NNproperties}}
\end{center}
\end{figure*}

\begin{figure*}[ht!]
\begin{center}
\gridline{
\includegraphics[scale=0.5,angle=0]{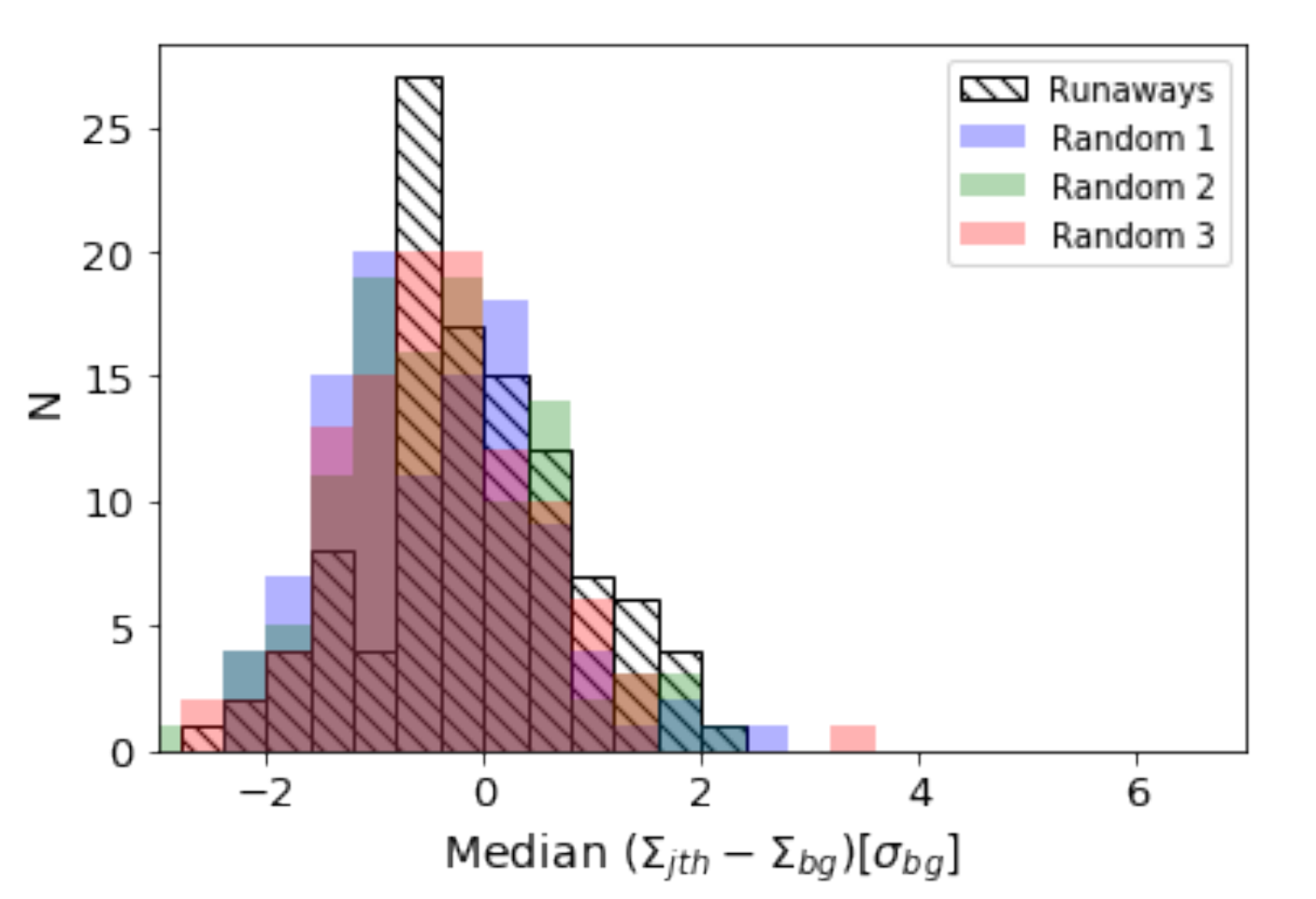}
\includegraphics[scale=0.5,angle=0]{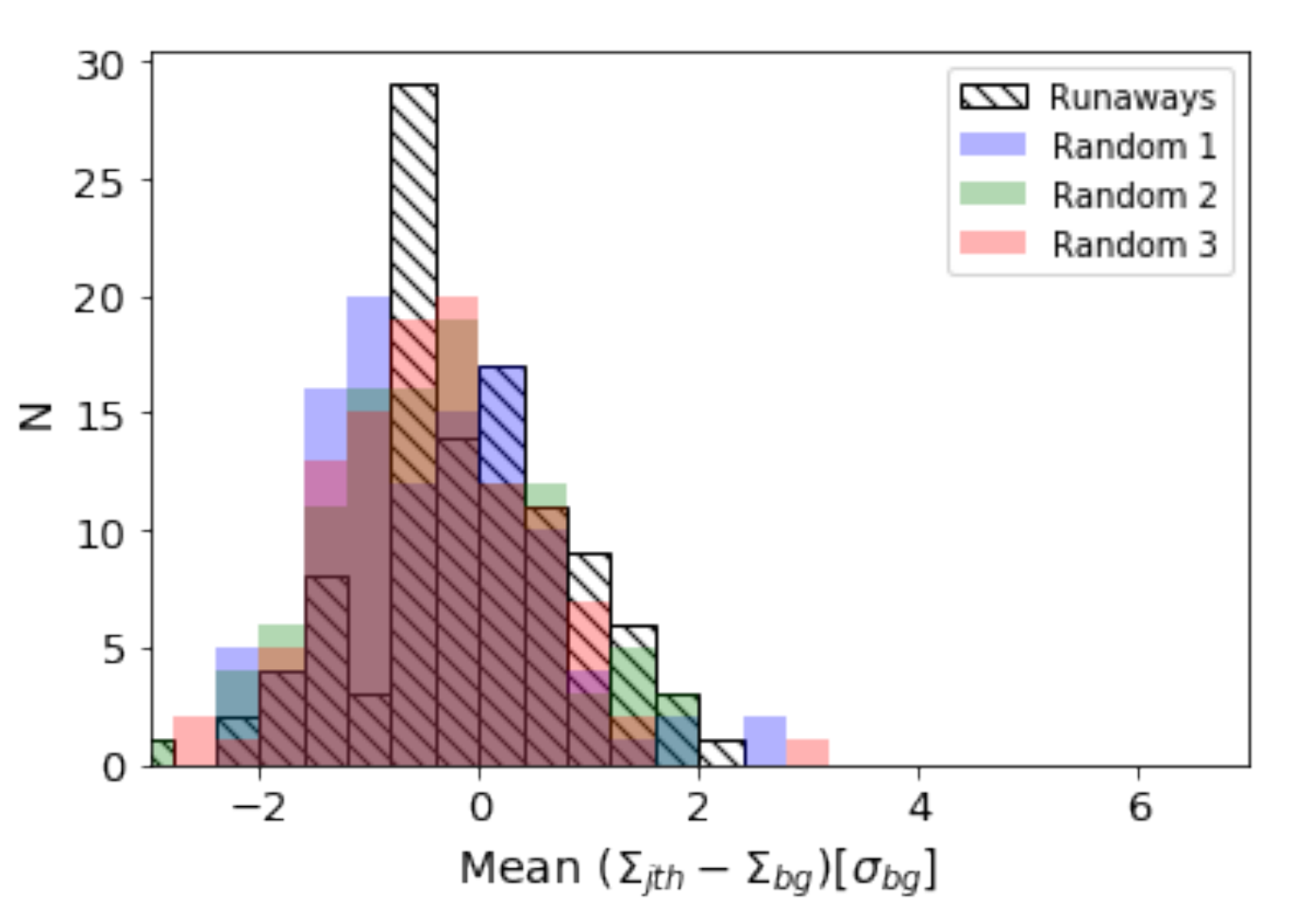}
}
\gridline{
\includegraphics[scale=0.5,angle=0]{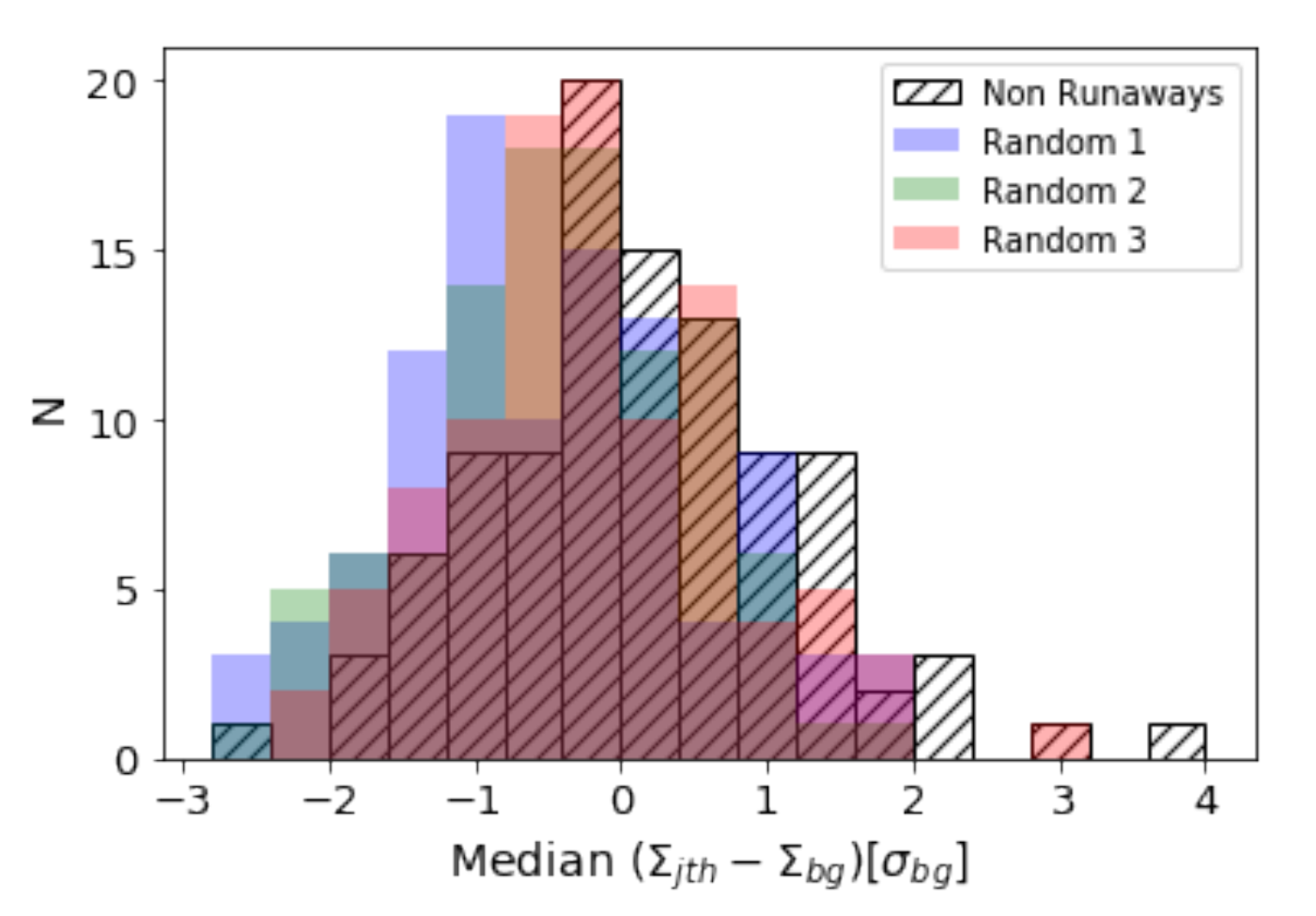}
\includegraphics[scale=0.5,angle=0]{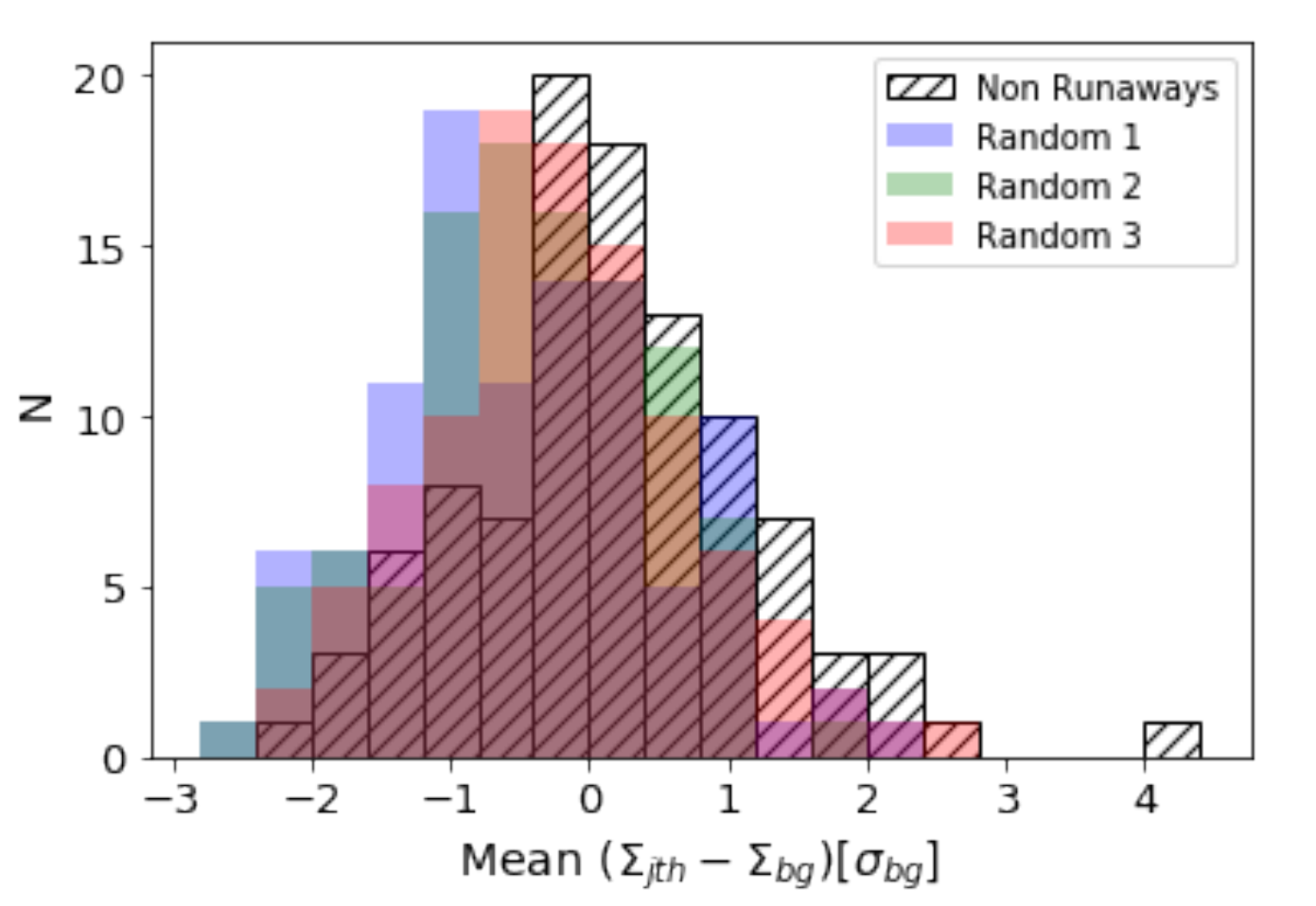}
}
\gridline{
\includegraphics[scale=0.5,angle=0]{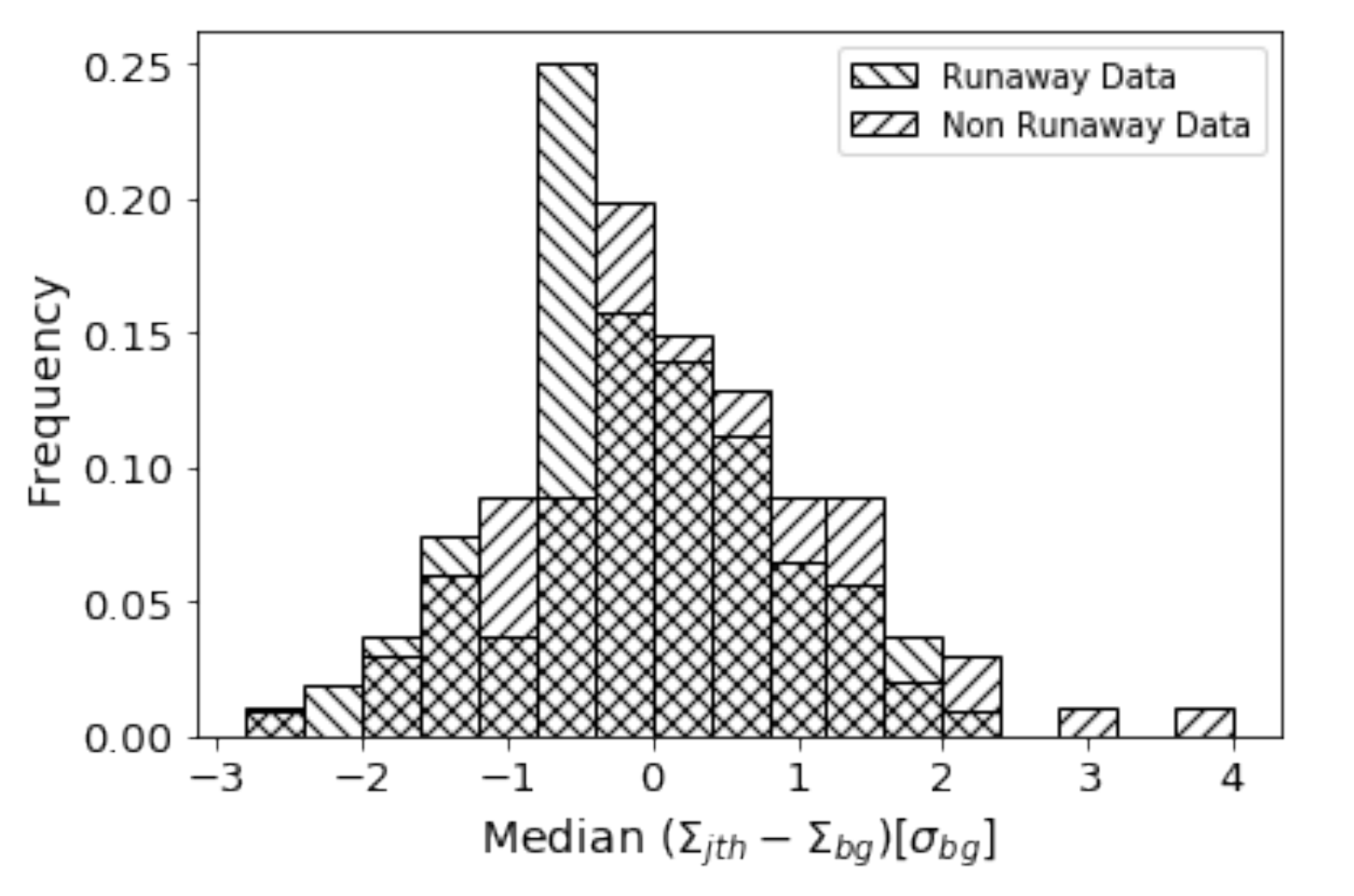}
\includegraphics[scale=0.5,angle=0]{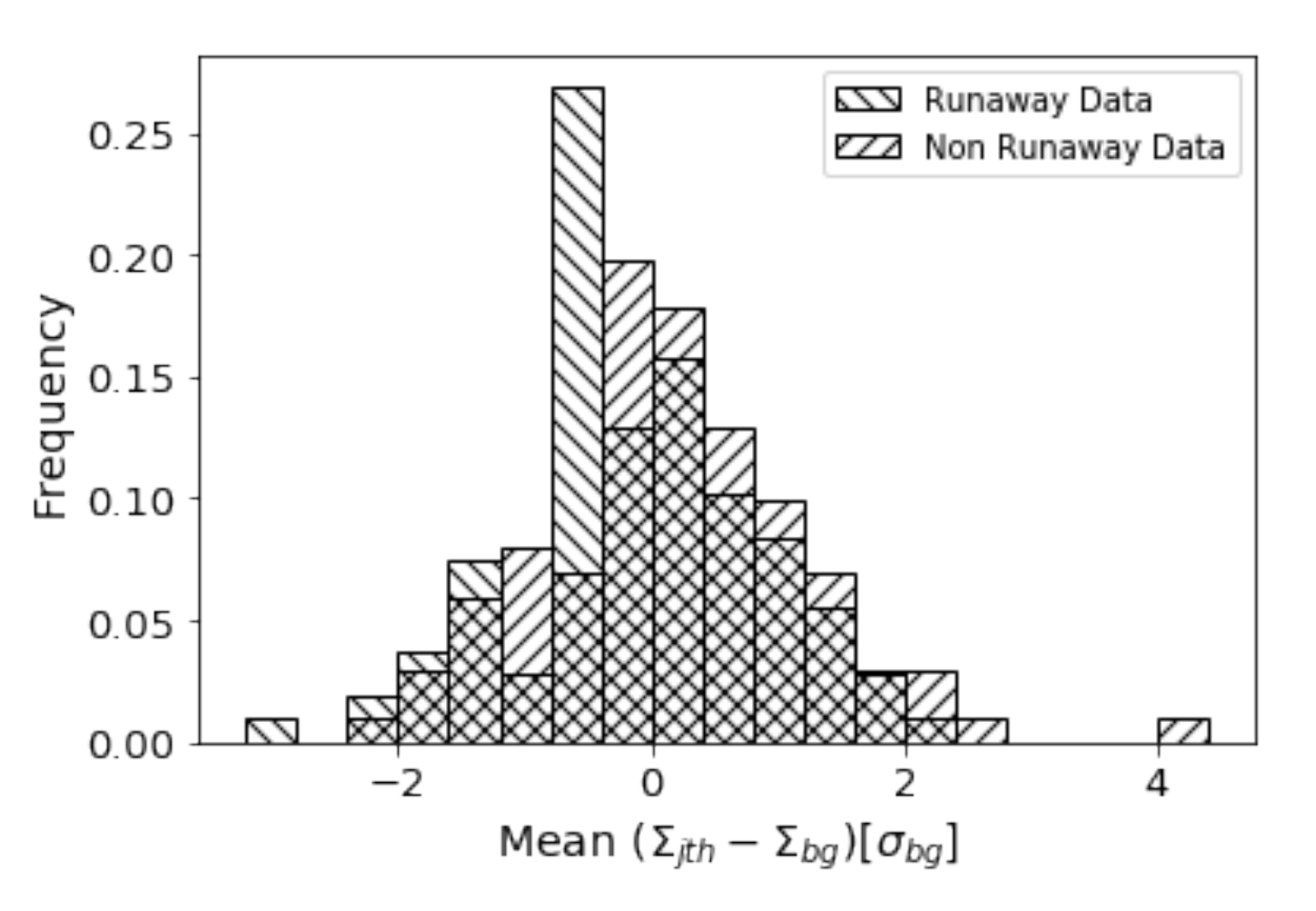}
}
\caption{NN density distributions comparing our observed data for runaway (top) and non-runaway targets (middle) with their respective fields in the random datasets. The observed data are shown in black while the three random datasets are colored as shown.  The left column shows the median overdensities for $j= 8$ to 12, and right column shows the averages. The bottom row compares the runaway (stars) and non-runaway (lines) distributions normalized with respect to their total number. 
The Rosenbaum and Wilcoxon tests give contradicting results on whether our runaway targets are distinct from a random distribution.  However, for non-runaway targets, these tests show strong statistical differences from a random distribution.
The non-runaway and runaway distributions are also statistically different from each other in their KS test.
\label{fig:NNrunaways}
}
\end{center}
\end{figure*}

\begin{figure*}[ht!]
\begin{center}
\gridline{
\includegraphics[scale=0.5,angle=0]{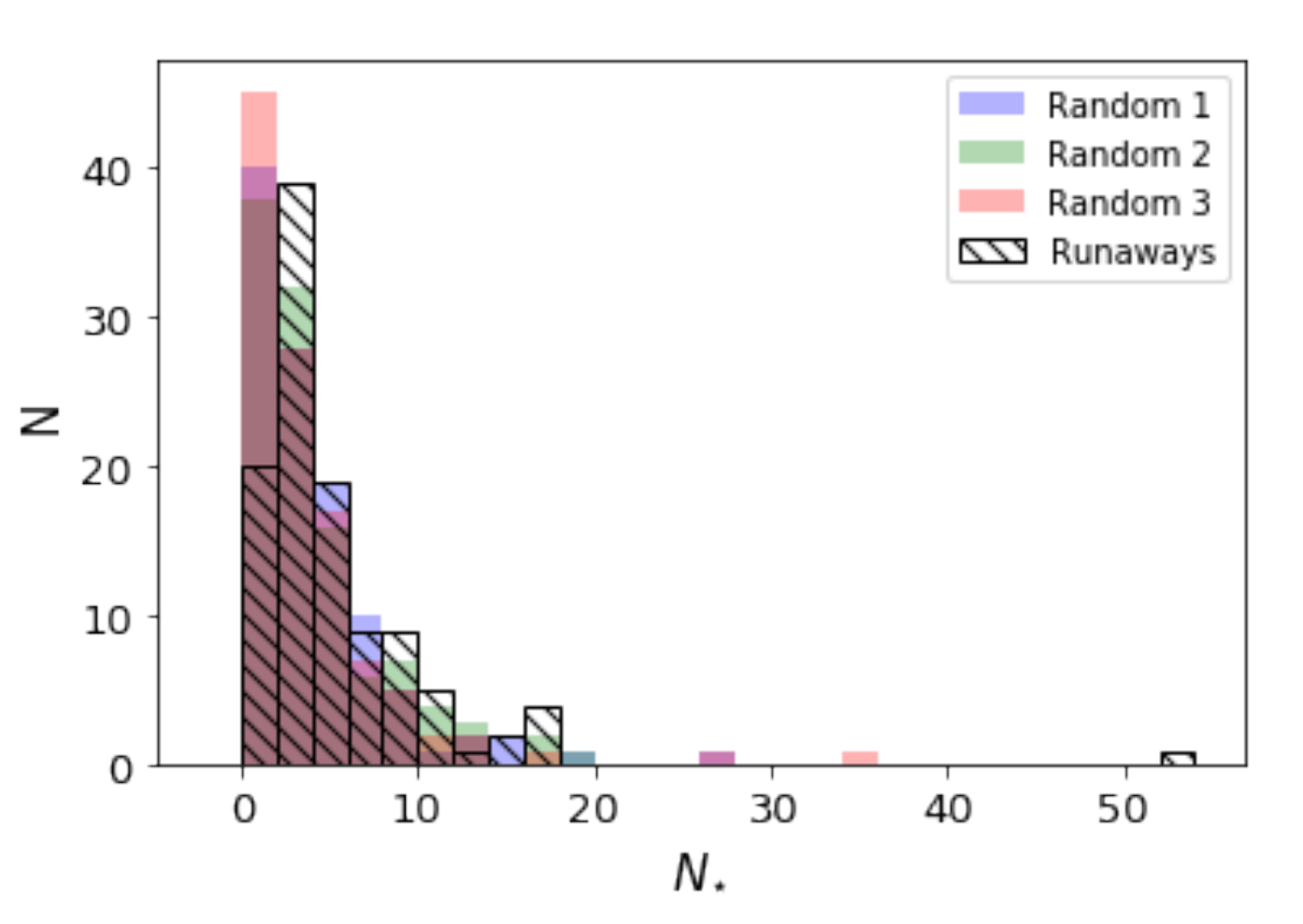}
\includegraphics[scale=0.5,angle=0]{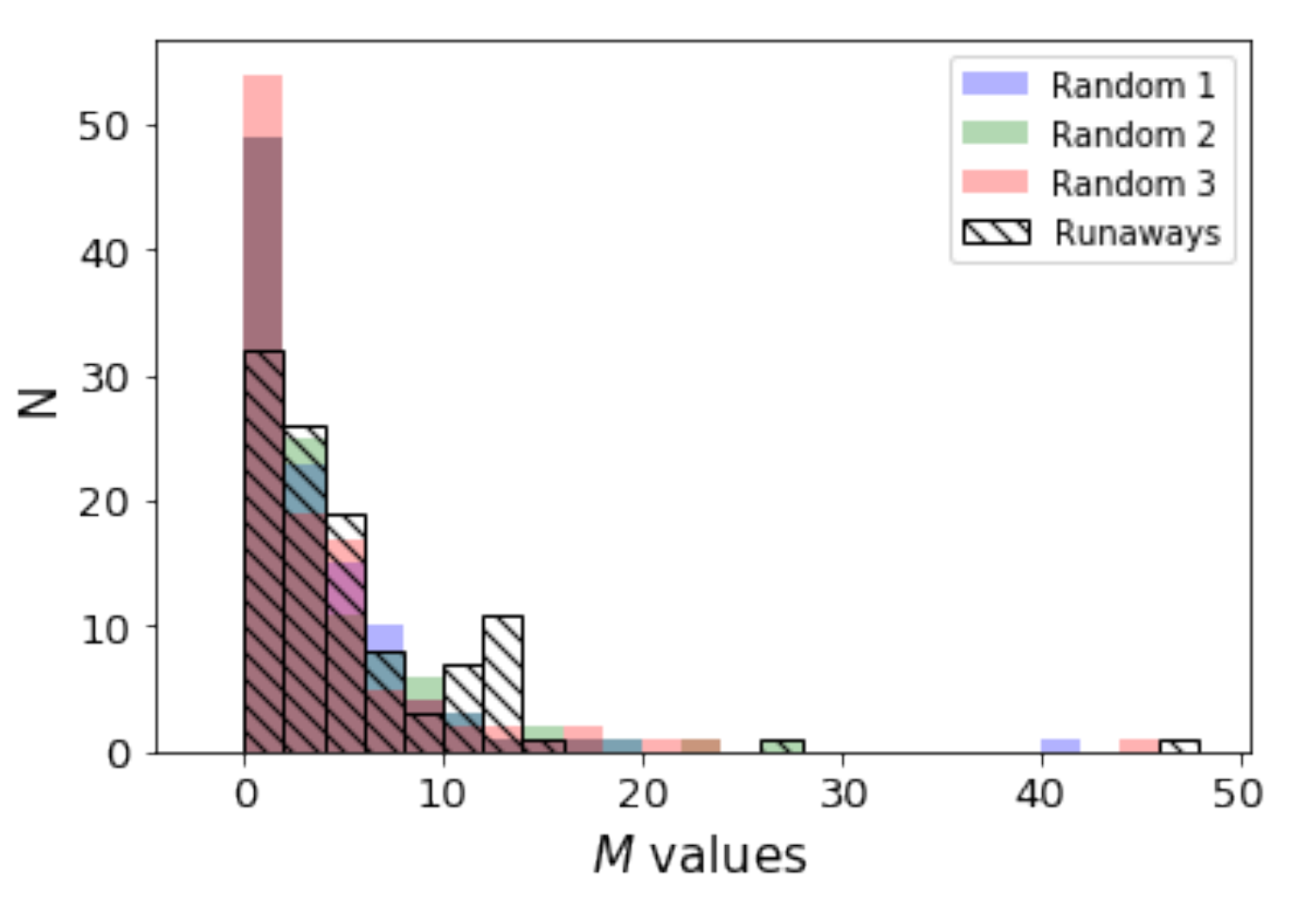}
}
\gridline{
\includegraphics[scale=0.5,angle=0]{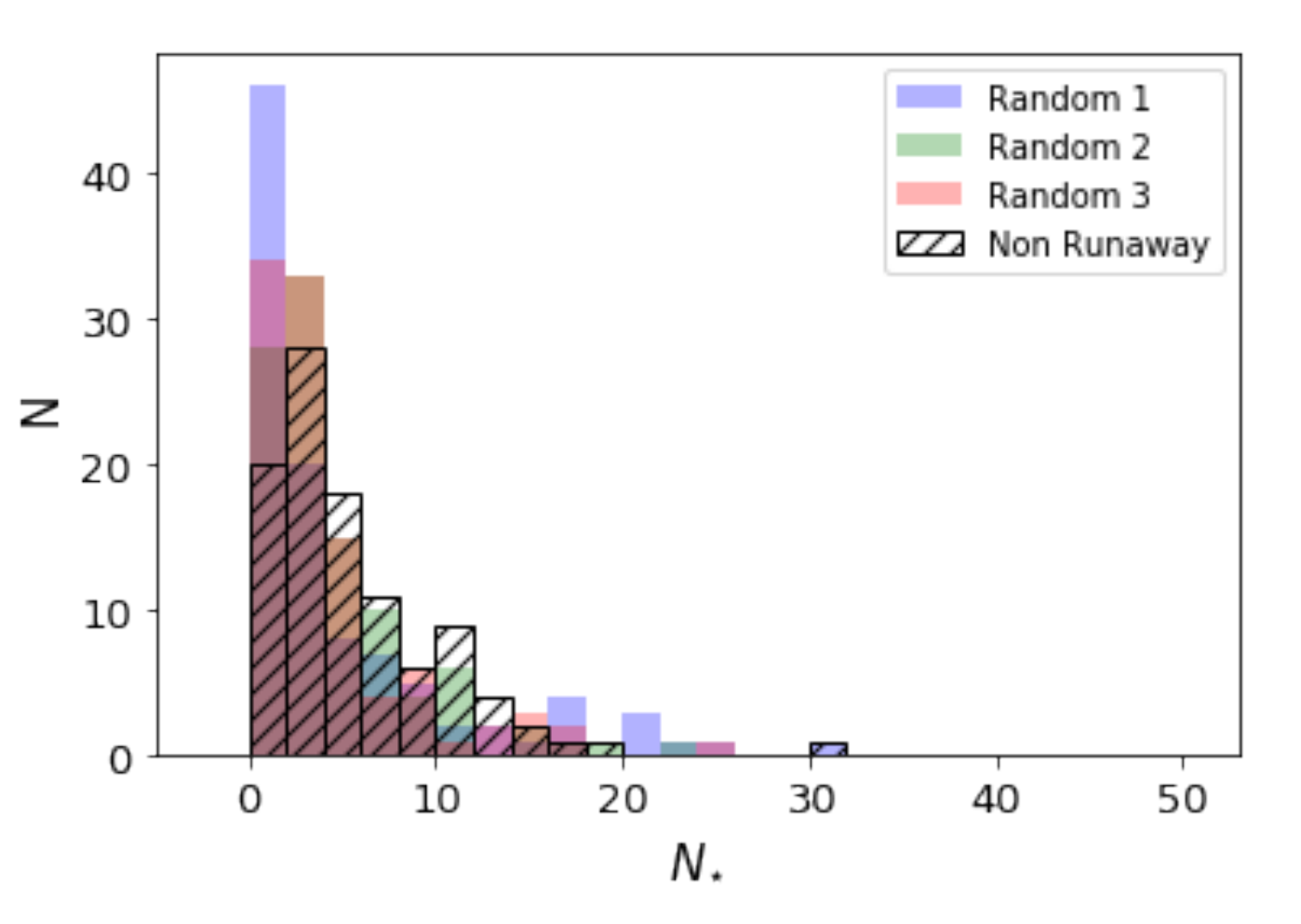}
\includegraphics[scale=0.5,angle=0]{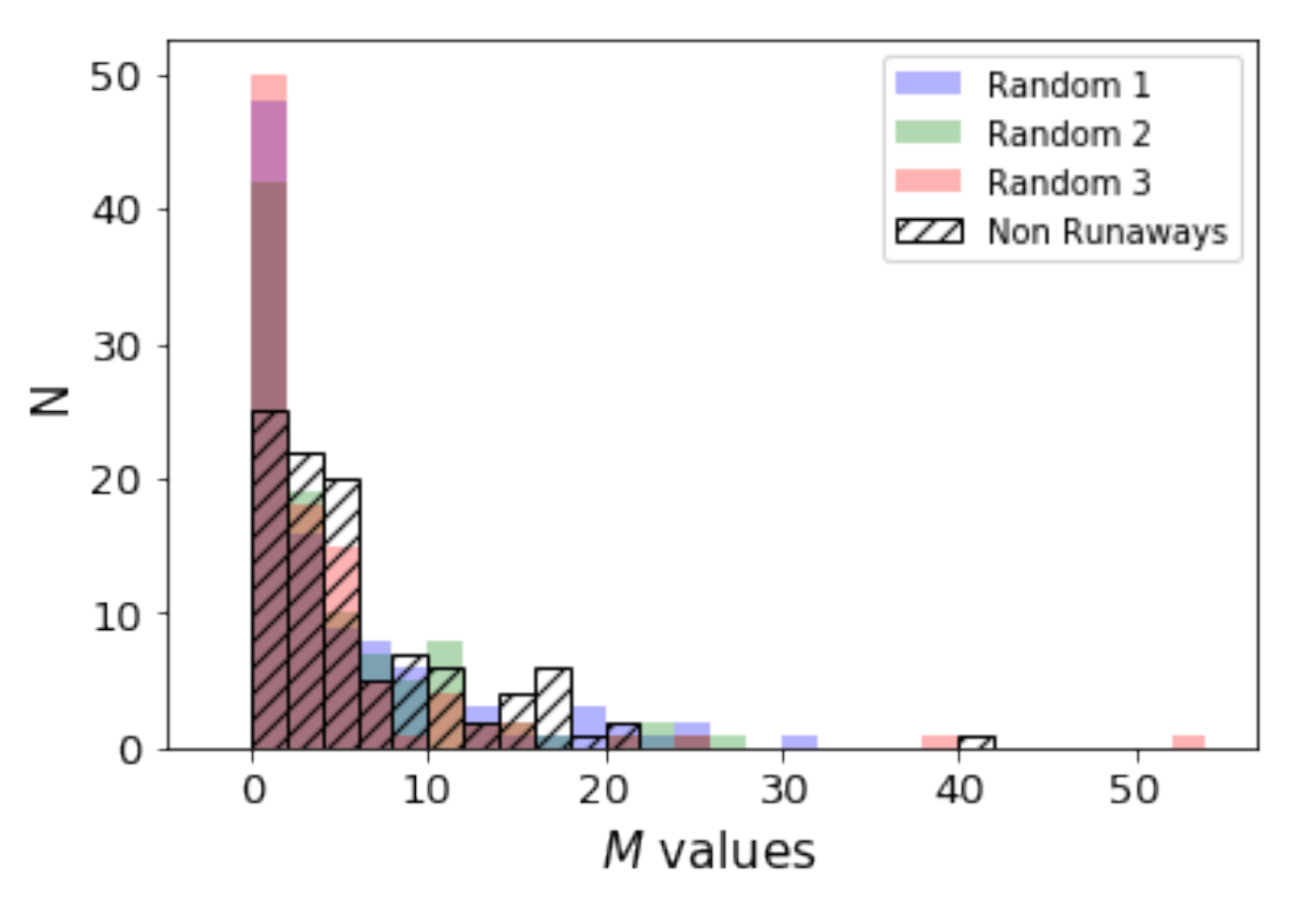}
}
\gridline{
\includegraphics[scale=0.5,angle=0]{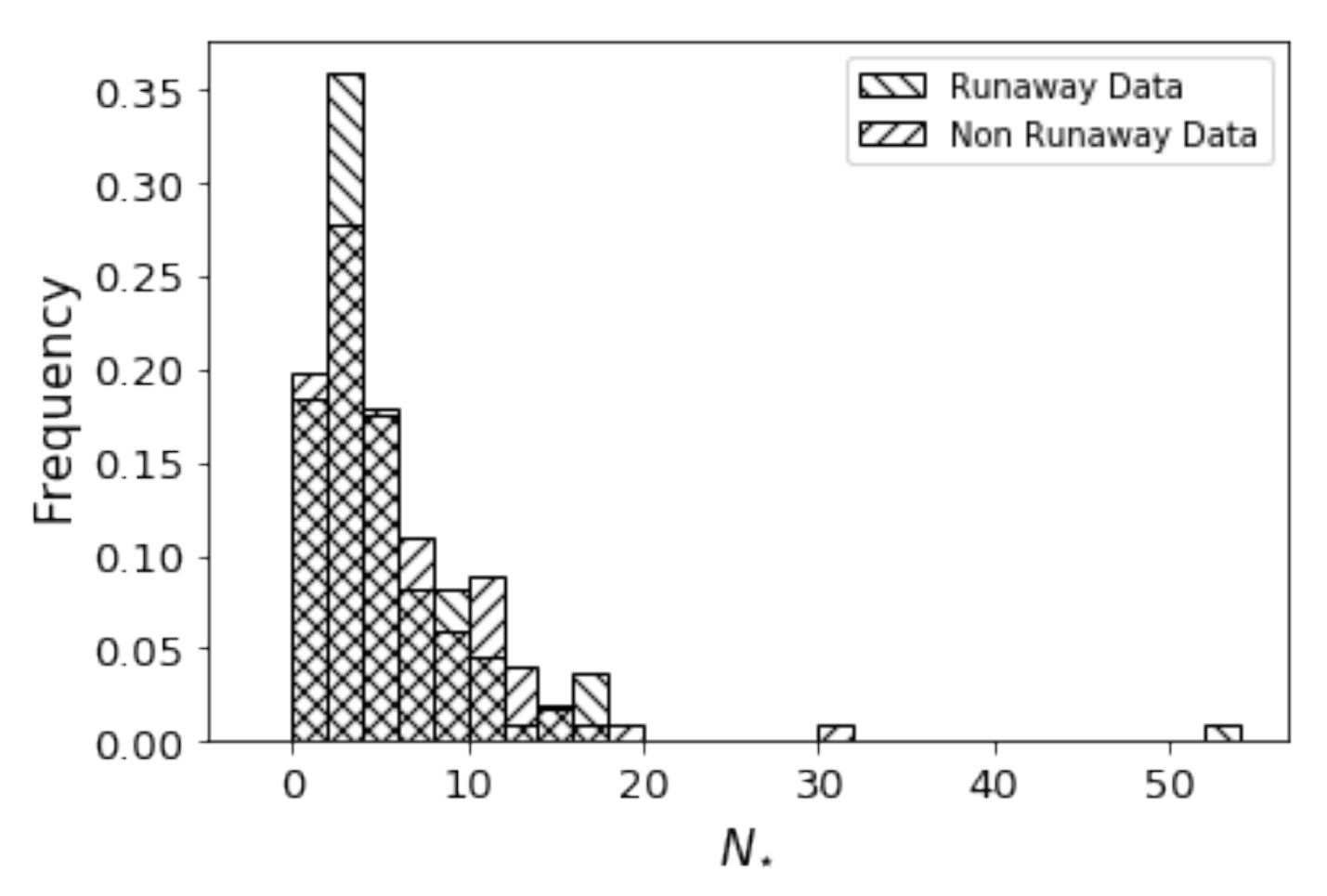}
\includegraphics[scale=0.5,angle=0]{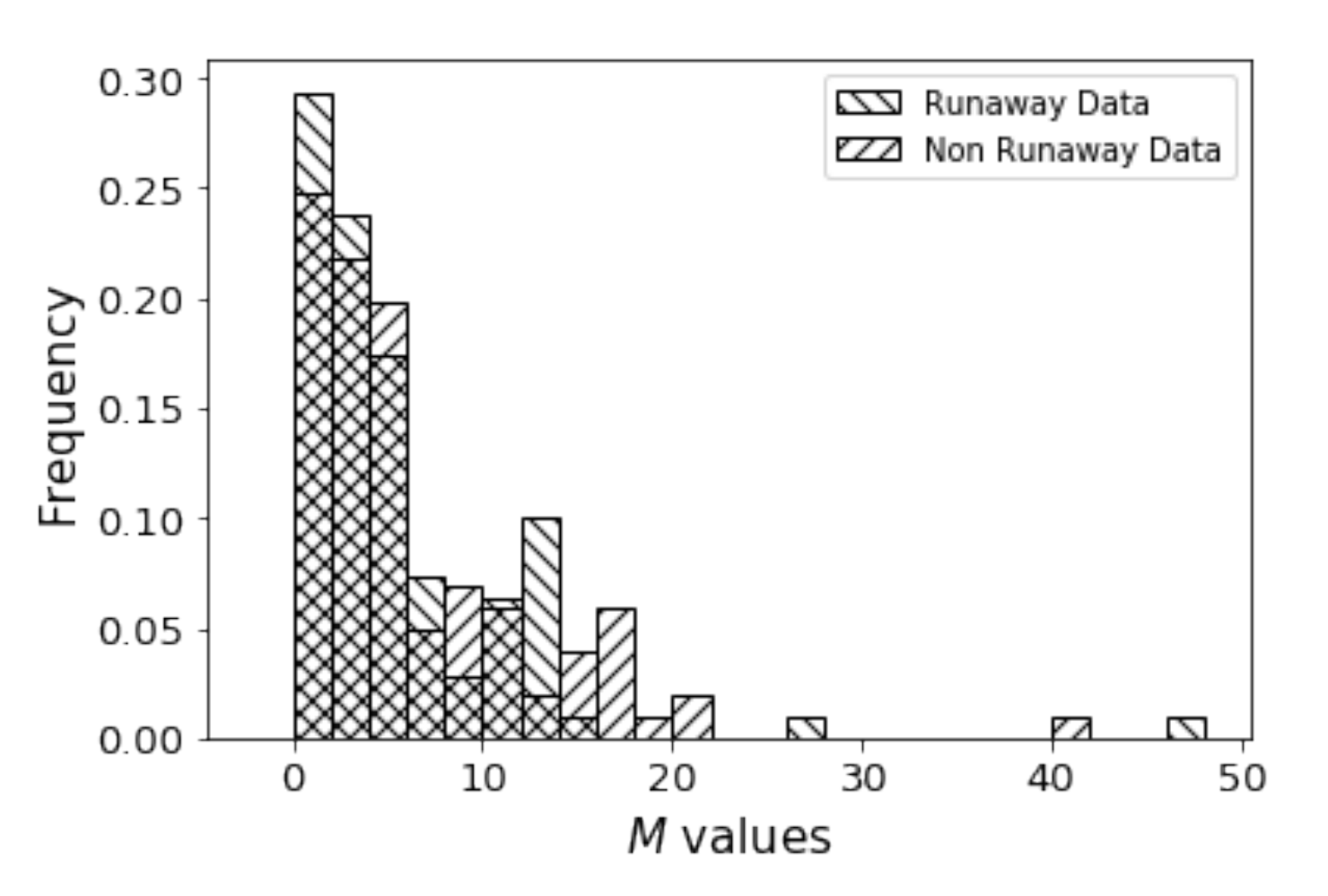}
}
\caption{
FOF results comparing our observed data for runaway (top) and non-runaway targets (middle) with their respective fields in the random dataset. The bottom row compares the runaway and non-runaway distributions normalized with respect to their total number. The left column shows the $N_*$ distribution, while the right column shows the $M$-value distribution. Our observed data are in black, while the random dataset is in blue. In the bottom row, non-runaway and runaway normalized distributions are shown with different hatching as shown. The Wilcoxon test identifies the runaway and non-runaway distributions as distinct from random distributions, but the Rosenbaum does not. The AD and KS tests are unable to distinguish the runaway and non-runaway distributions from each other.
\label{fig:FOFrunaways}
}
\end{center}
\end{figure*}

\begin{figure*}[ht!]
\begin{center}
\gridline{
\includegraphics[scale=0.3,angle=0]{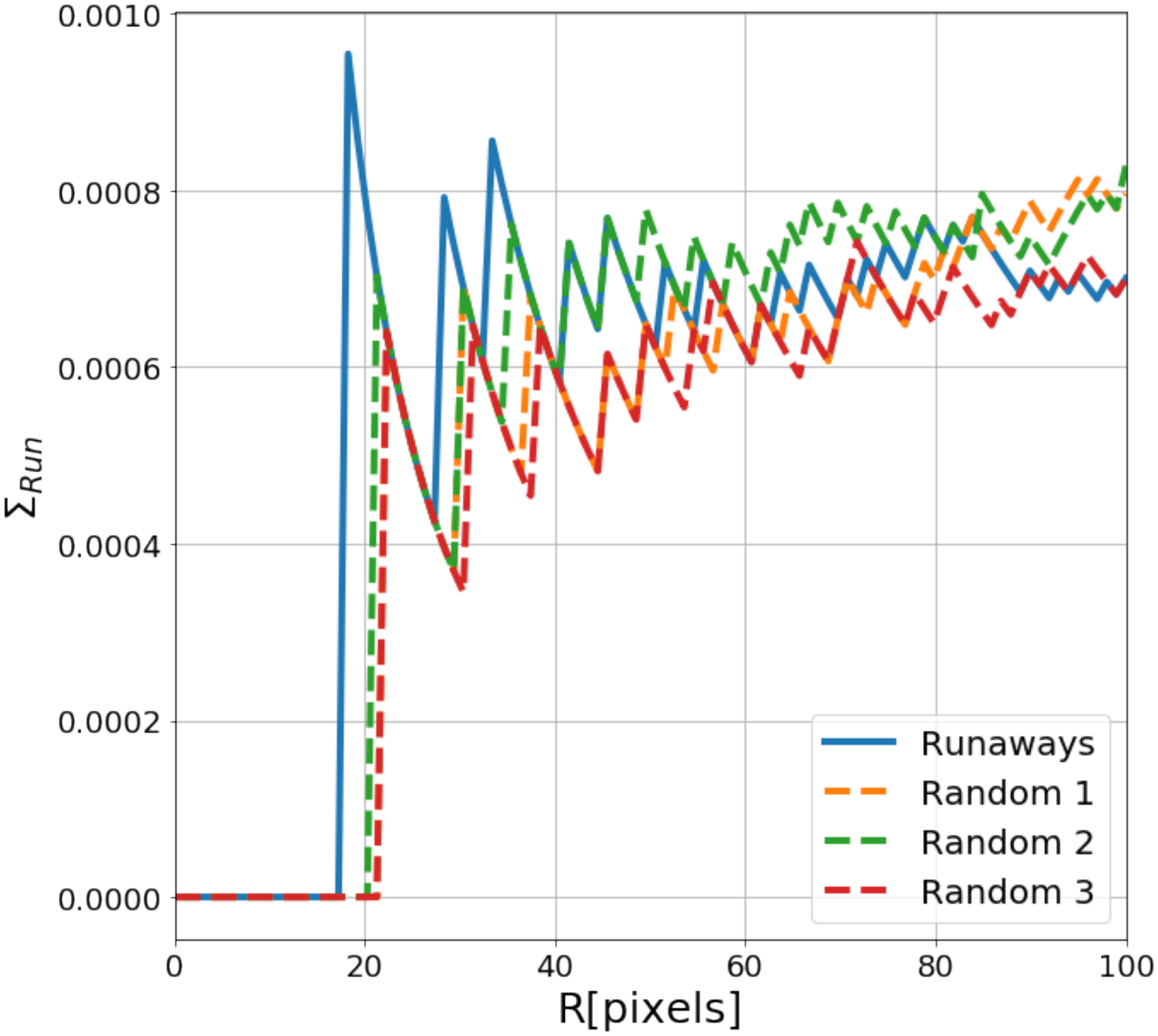}
\includegraphics[scale=0.3,angle=0]{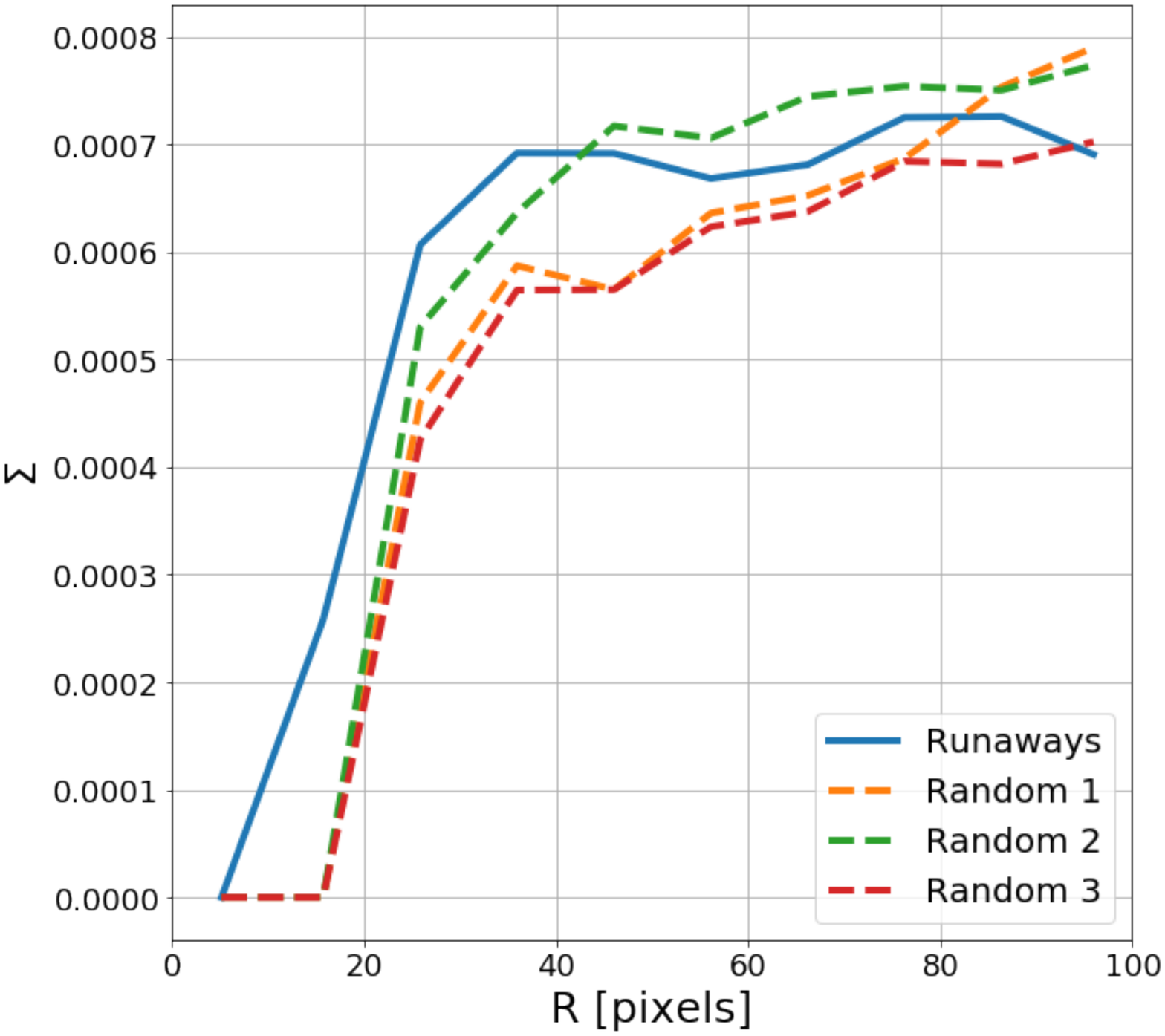}
}
\gridline{
\includegraphics[scale=0.3,angle=0]{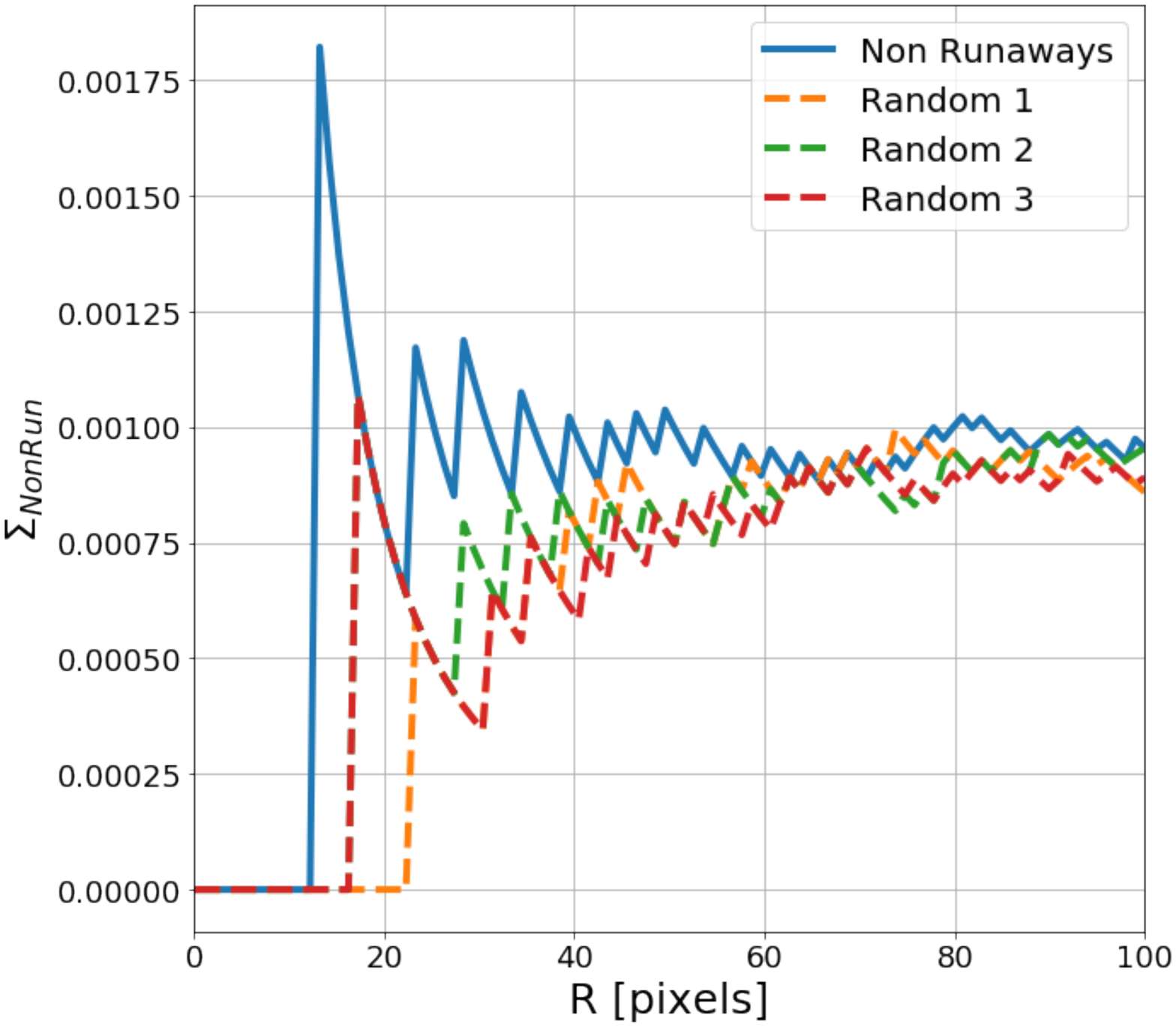}
\includegraphics[scale=0.3,angle=0]{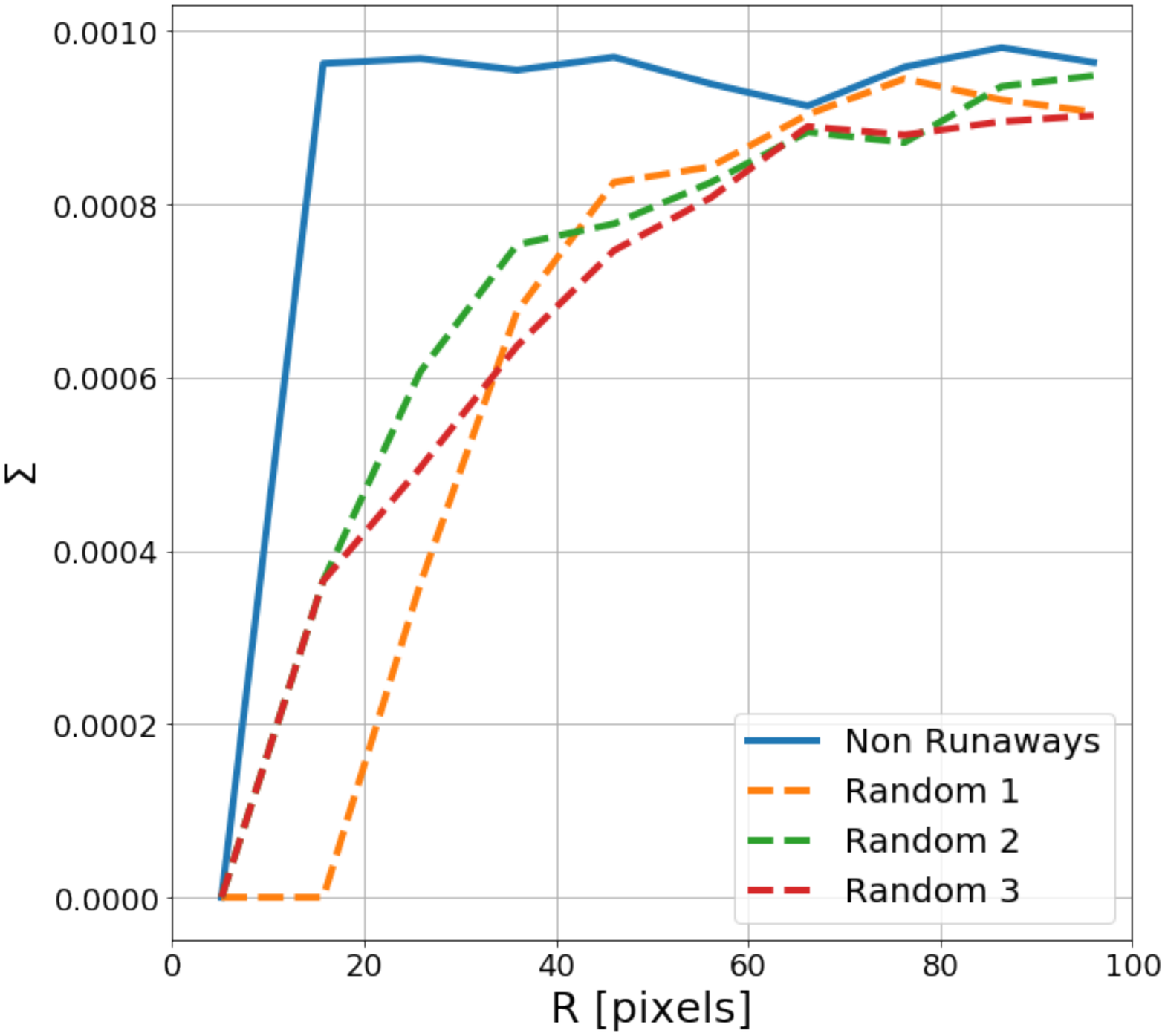}
}
\caption{Plots of observed overdensities to those for random fields, as a function of radius, for our runaway targets (top) and non-runaway targets (bottom). They are binned by 1 (left) and by 10 (right).
Our runaway targets appear to have ambiguously higher densities at very small radii that are not due to real density enhancements. However, non-runaway targets do show centrally concentrated densities. \label{fig:densitygradrunaway}}
\end{center}
\end{figure*}

Additionally, runaways show a significant fraction of targets found in lower density environments, as expected since they quickly move away from dense, cluster forming environments where they originated.
Figure~\ref{fig:NNproperties} shows $\Sigma_j$ and the corresponding background density, $\Sigma_{bg}$,
as a function of $R_j$, the mean radius of the $j$th nearest neighbor. The non-runaways on average have greater $\Sigma_j$ and $\Sigma_{bg}$ 
while runaways on average have higher $R_j$ values, indicating that runaways tend to be in target fields with lower stellar density. This is also reflected in Figure \ref{fig:NNrunaways} wherein the peaks of the non-runaway distributions are at larger values than those for the runaways.

For FOF, the non-runaways do not 
show statistically significant differences from the runaways. They 
exhibit the same behavior that is seen for both the full and the runaway datasets, showing significant Wilcoxon test results but not Rosenbaum test results.
For NN, however, the non-runaways do show statistically significant differences from the runaways. 
In the NN density distribution comparisons with the random fields, the Wilcoxon results give $p$-values that are statistically significant and lower by an order of magnitude than those for the runaways. 
However, the Rosenbaum results are more ambiguous because only two NN comparisons with random fields 
show a statistically significant difference.
This again demonstrates that the frequency of any TIB clusters in the observed dataset is on the order of the random noise in our fields.

When compared to each other, the runaway and non-runaway distributions also look noticeably different (Figures~  \ref{fig:NNrunaways} and \ref{fig:FOFrunaways}), confirming these trends.
Since the runaway and non-runaway datasets are independent of each other, we are able to use the Kolmogorov-Smirnov (KS) test and also the Anderson-Darling (AD) test when comparing these distributions. This version of the KS test gives more weight to the tails of the distributions, which in our case are more sensitive to the detection of TIBs. 
The resulting $p$-values of the AD and KS tests are also shown in Table \ref{table:statTests}. For FOF, we have similar statistical test results to NN between the runaways and non-runaways distributions of $N_*$ and $M$-values (Figure \ref{fig:FOFrunaways}) however, their AD and KS test are unable to distinguish the runaways and non-runaway distributions from each other, although we are able to see that non-runaways are skewed towards higher values.

Meanwhile, the NN data (Figure \ref{fig:NNrunaways}) do show evidence that the runaway and non-runaway density distributions are different, with $p$-values of 0.022 and 0.035 in the KS test; while the AD results are close to the critical range, between $p=0.05$ and 1.0. In Figure \ref{fig:NNrunaways}, runaways peak at lower values than the non-runaways, and their maximum overdensities are less than those of the random field datasets. On the other hand, non-runaways have both higher peak and maximum values that are distinct from 
the random-field data.

Our stacked fields also show differences between our runaway and non-runaway datasets (Figure \ref{fig:densitygradrunaway}). We find that the runaways show less variation from the random-field datasets, although they do appear to show, with some ambiguity, a significant density enhancement at the lowest radii that may be a product of the systematic errors similar to those found in NN.
In contrast, the non-runaways clearly show higher, and more centrally concentrated, densities than the non-runaways. 
In Figure \ref{fig:densitygradrunaway}, we can see that these higher densities are present even when smoothing our distribution of observed overdensities. These higher densities may be due to the presence of TIB clusters, and are likely the cause of the aggregate enhancement seen in Section \ref{subsec:stackedfields} before. This is further discussed below in Section \ref{sec:implications}.

In summary, any signal of TIBs in our sample should be strengthened in non-runaway datasets, while we expect runaway fields to behave more like random fields.  Despite its positive detections, the FOF algorithm is not sensitive to the differences between non-runaway and runaway fields.  Therefore, its results are not a reliable indicator of the presence of TIBS in our sample, and cannot be used to estimate the number of TIB clusters.  On the other hand, the results given by NN and the stacked fields do show significant differences between the runaways and non-runaways, which are consistent with the presence of a small, but real, number of TIB clusters.  This is supported by the statistical differences between non-runaway and runaway distributions given by their respective KS and AD test results.  The relative effectiveness of NN and FOF is consistent with the analysis by \citet{Schmeja2011}, who find that NN is a superior cluster-finding algorithm when compared to other methods, including the minimum spanning tree, on which FOF is based.

\subsection{In-Situ Field OB Stars} \label{subsec:insitufindings}

Conversely to runaways, we can also examine objects that have been identified as field OB stars that formed {\it in situ}.
\citet{Oey2013} identified 14 strong candidates in the RIOTS4 survey, based on their dense, symmetric \hii\ regions and radial velocities consistent with local \hi\ systemic velocities.  Of these, 
10 are in our dataset.  We might expect that few of these should be runaways, and we might expect some to be among the best TIB candidates from our cluster-finding algorithms.
\startlongtable
\begin{deluxetable*}{c|ccccccccc}
\tablecaption{Kinematic Data for Isolated {\it In-Situ} Candidates Identified as Runaways  \label{table:insitu}}
\tablehead{
\colhead{Target\tablenotemark{a}} & \colhead{RA Velocity \tablenotemark{b}} &
\colhead{err} &\colhead{DEC Velocity\tablenotemark{b} }& \colhead{err} & \colhead{$v_{loc,\perp}$\tablenotemark{c}}& \colhead{err} & \colhead{Radial Velocity \tablenotemark{d}}& \colhead{3D Velocity \tablenotemark{e}}&\colhead{err}\\
\colhead{ } & \colhead{km/s} & \colhead{km/s} & \colhead{km/s}& \colhead{km/s}& \colhead{km/s} & \colhead{km/s} & \colhead{km/s}& \colhead{km/s} & \colhead{km/s}
}
\startdata
35491 & 22 & 20 &25 &19& 34 & 27 &-23 & 40 & 29 \\ 
36514 & 90 &29 &-51&24  &103&38 &-8&104&39 \\
67334 & 119 &24& -66&20 &136 & 32&54 & 146&33  \\
70149 & 82&29 & -32&25 & 88 &39 &1 & 88&40\\
71409 & 28& 26& 19& 15& 33 &31 & \nodata & \nodata& \nodata \\ 
\enddata
\tablenotetext{a}{ID from \citet{Massey2002}}
\tablenotetext{b}{The RA and DEC velocities are relative to local systemic velocity and are calculated using the RA and DEC velocities from \citet{Oey2018}.}
\tablenotetext{c}{Residual transverse velocities from \citet{Oey2018}.  
}
\tablenotetext{d}{From \citet{Lamb2016}. The error in RV is on average 10 \kms.}
\tablenotetext{e}{Space velocity relative to local frame.}

\end{deluxetable*}
We find that only 2 of the {\it in situ} targets ([M2002] SMC-70149, 71409) are among the top 20 FOF TIB candidates, and another 2 of them ([M2002] SMC-69598, 75984) are among the top NN candidates. Instead, 5 of the {\it in situ} candidates, including two in the top 20 for FOF, are identified as runaways. Images of these runaway-star fields are shown in Figure \ref{fig:insitucand}, and their kinematic information \citep{Oey2018} \footnote{In Table~1 of \citet{Oey2018}, columns 11 and 13 correspond to systemic RA and Dec velocities, respectively, of the local fields for each target star.}
is shown in Table \ref{table:insitu}. Two of these have low runaway velocities (targets [M2002] SMC-35491 and 71409), and are also consistent with being non-runaways within the errors. However, others have velocities far above the runaway threshold of $v_{loc,\perp} > 30$ \kms 
(targets [M2002] SMC-36514, 67334 and 70149).

\begin{figure*}
\begin{center}
\gridline{
\includegraphics[scale=0.3,angle=0]{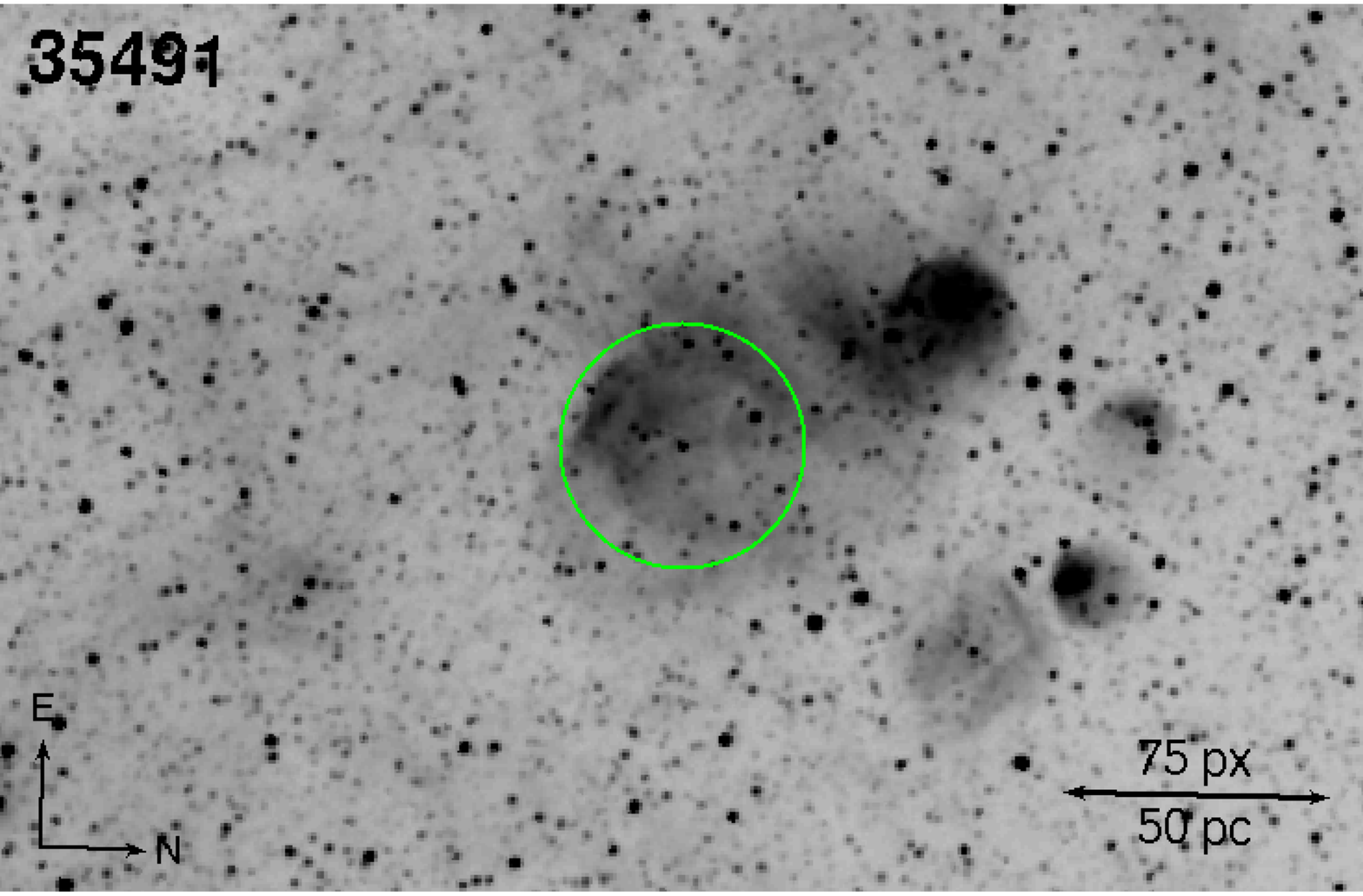}
\includegraphics[scale=0.3,angle=0]{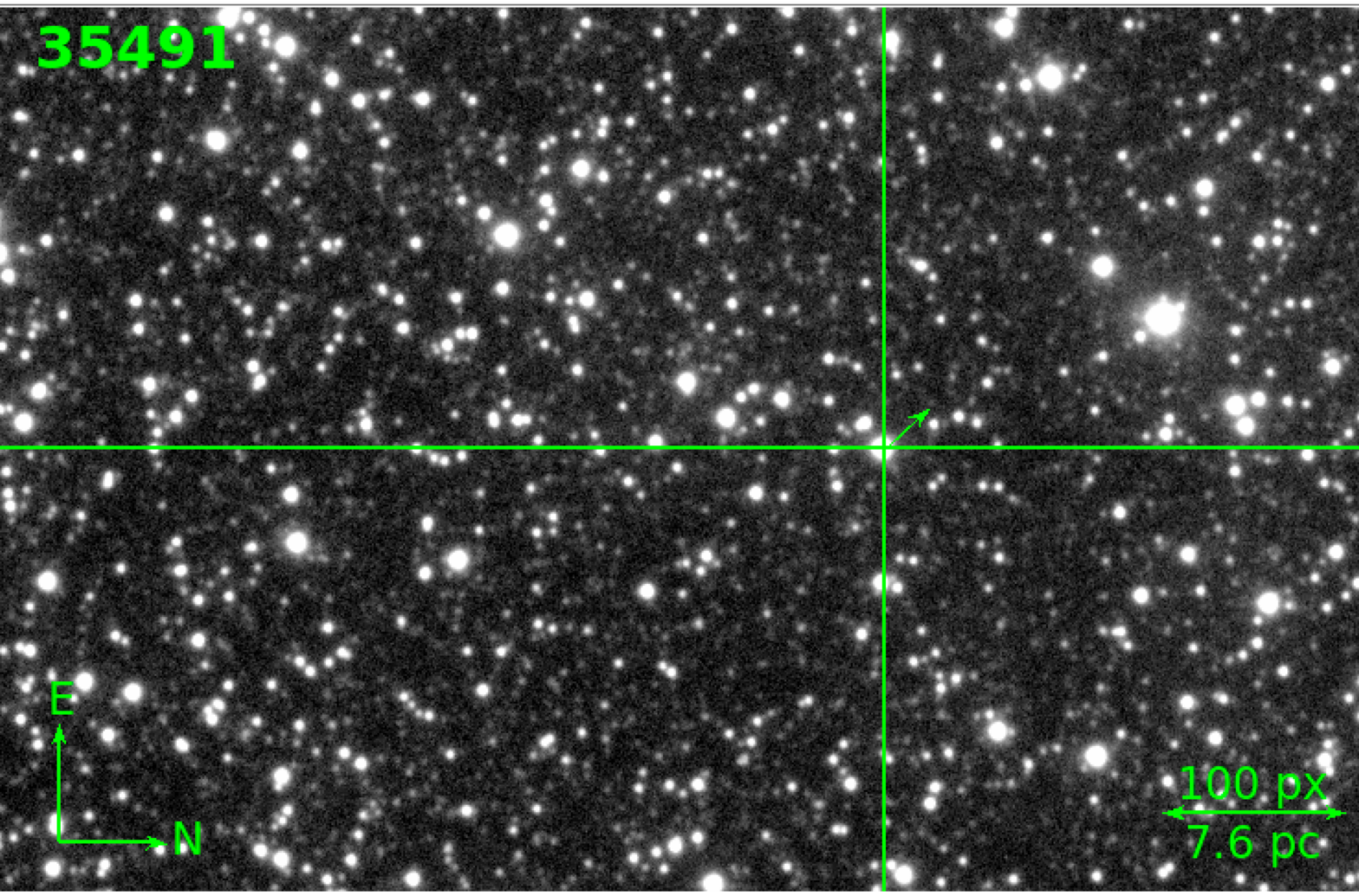}
}
\gridline{
\includegraphics[scale=0.3,angle=0]{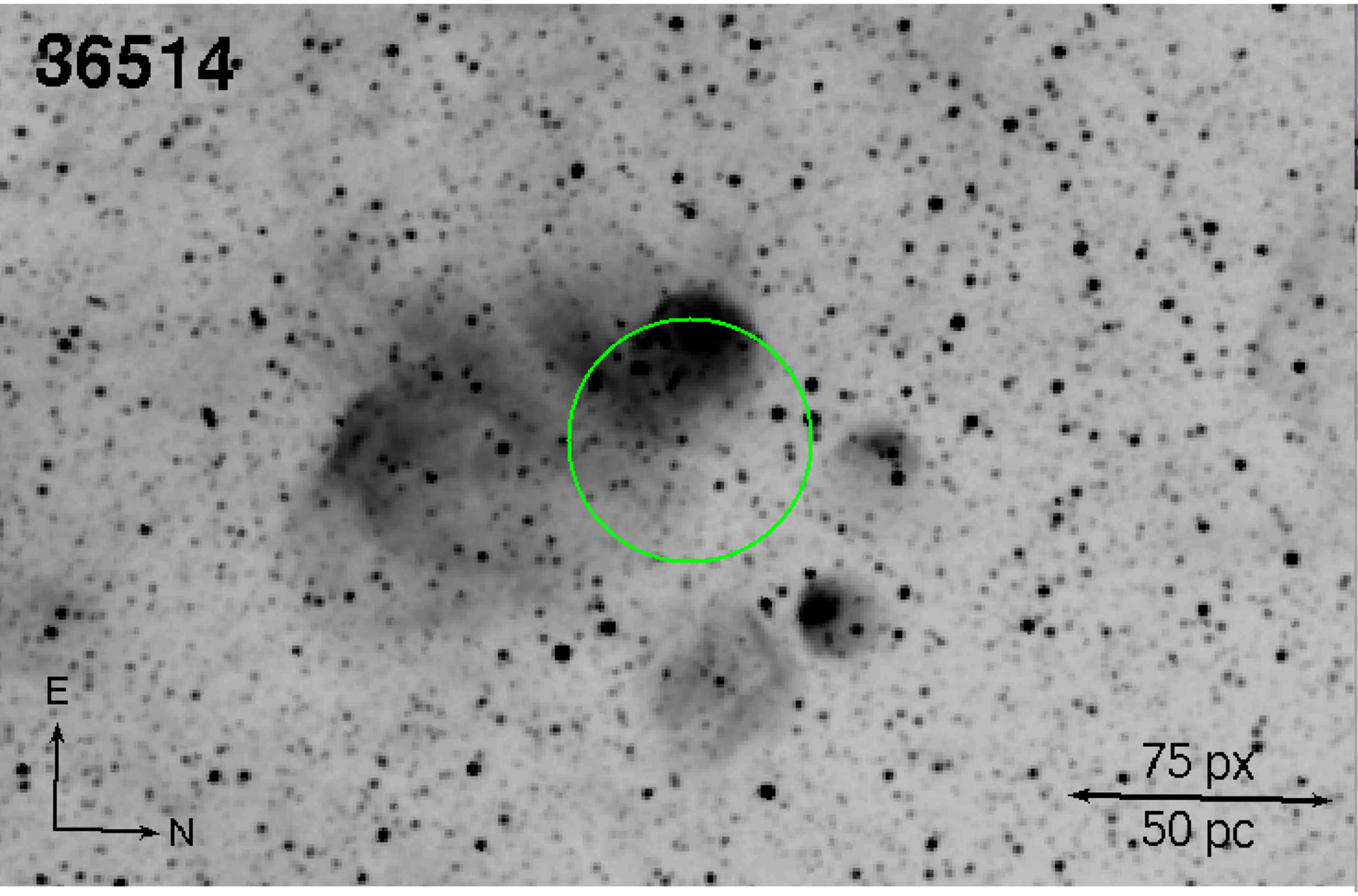}
\includegraphics[scale=0.3,angle=0]{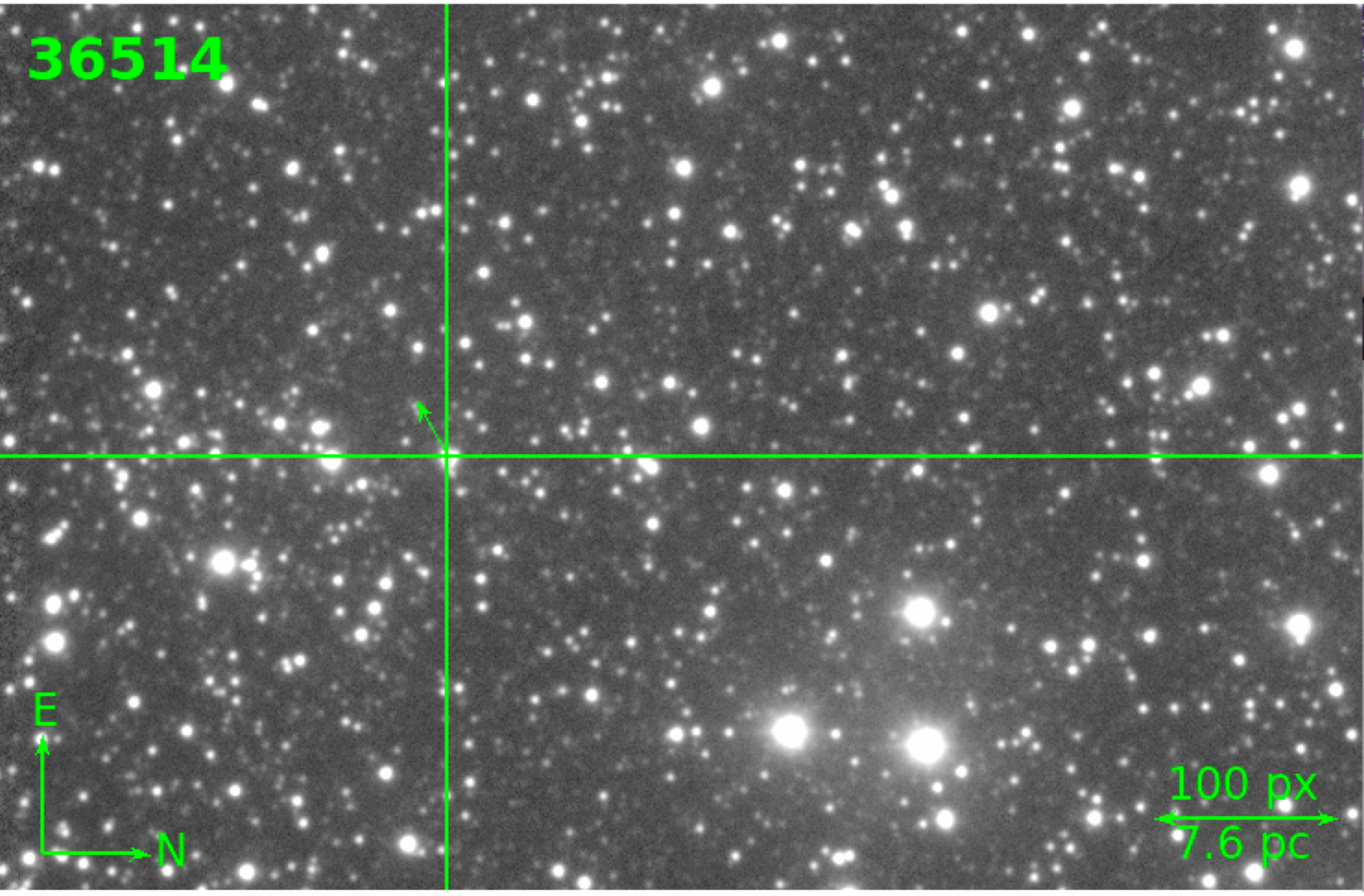}
}
\gridline{
\includegraphics[scale=0.3,angle=0]{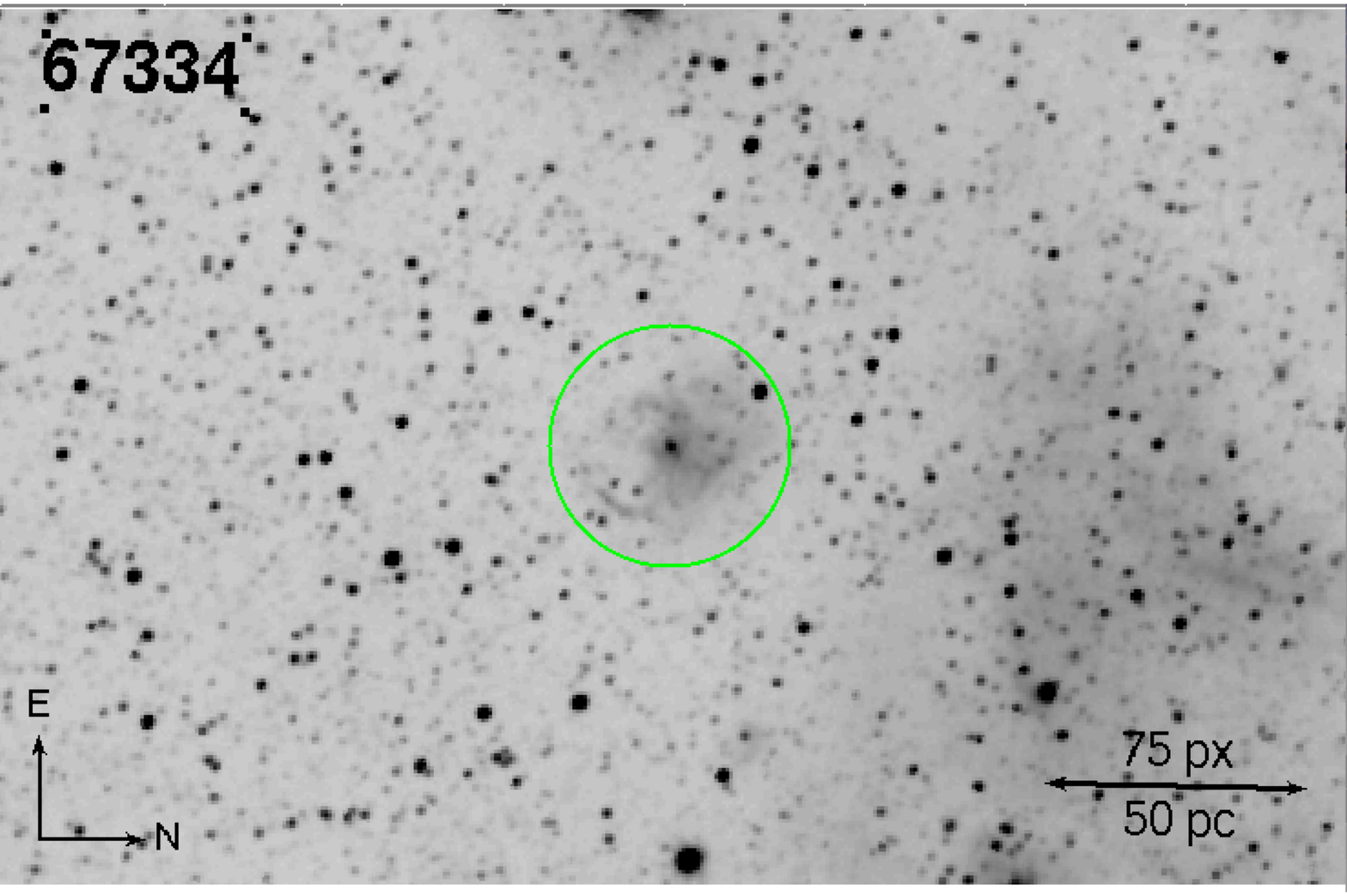}
\includegraphics[scale=0.3,angle=0]{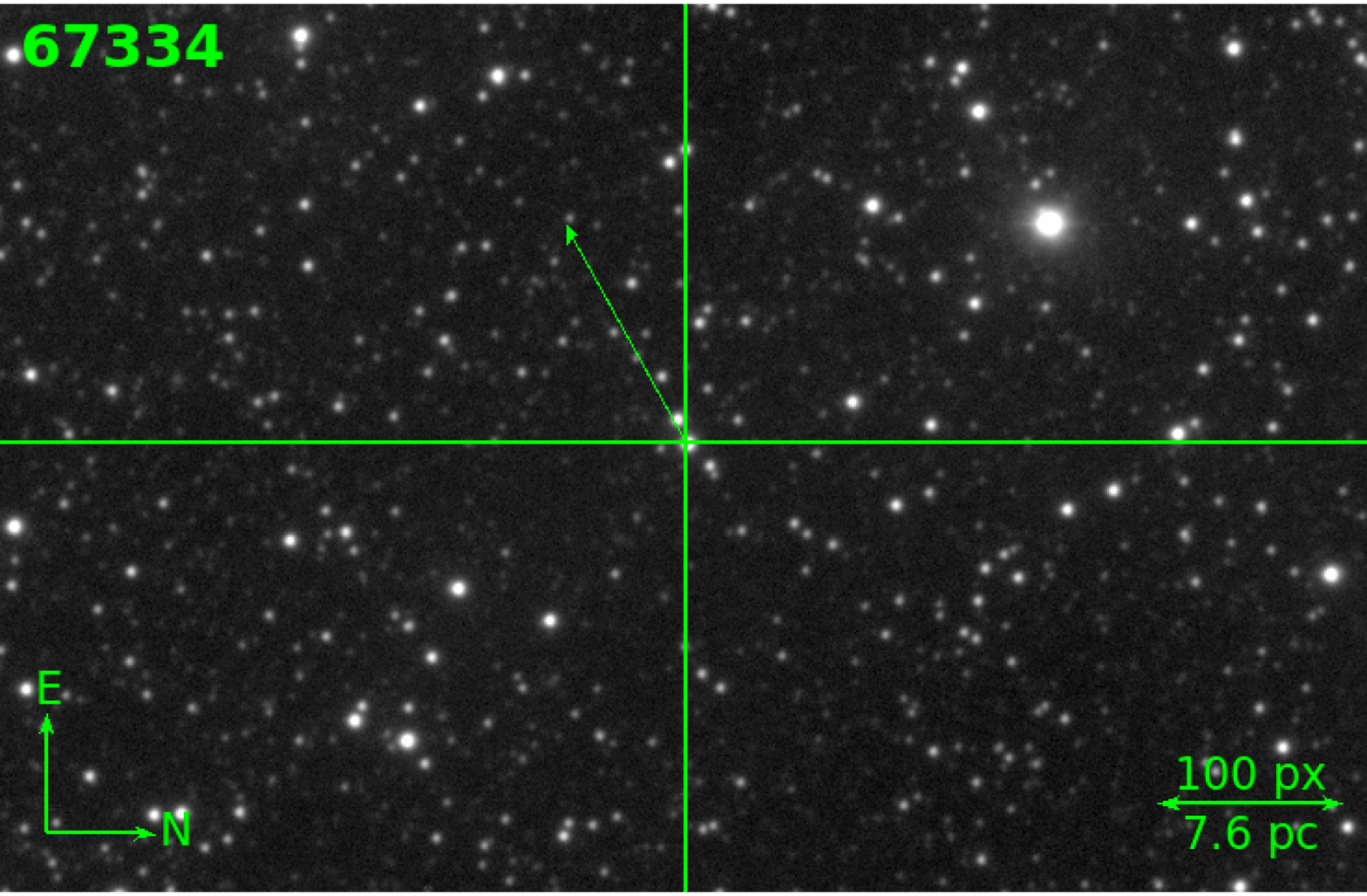}
}
\caption{The 5 {\it in situ} candidates that were classified as runaways in our own study. North is to the right and East is up. On the left side are the H$\alpha$ images and on the right are the $I$-band images. In the H$\alpha$ images, the target is highlighted by the circle. On the right, the green arrow shows the target's $v_{loc,\perp}$ (Table~\ref{table:insitu}).  Their kinematic data is shown in Table \ref{table:insitu}.}
\end{center}
\end{figure*}

\setcounter{figure}{11} 
\begin{figure*}
\begin{center}
\gridline{
\includegraphics[scale=0.3,angle=0]{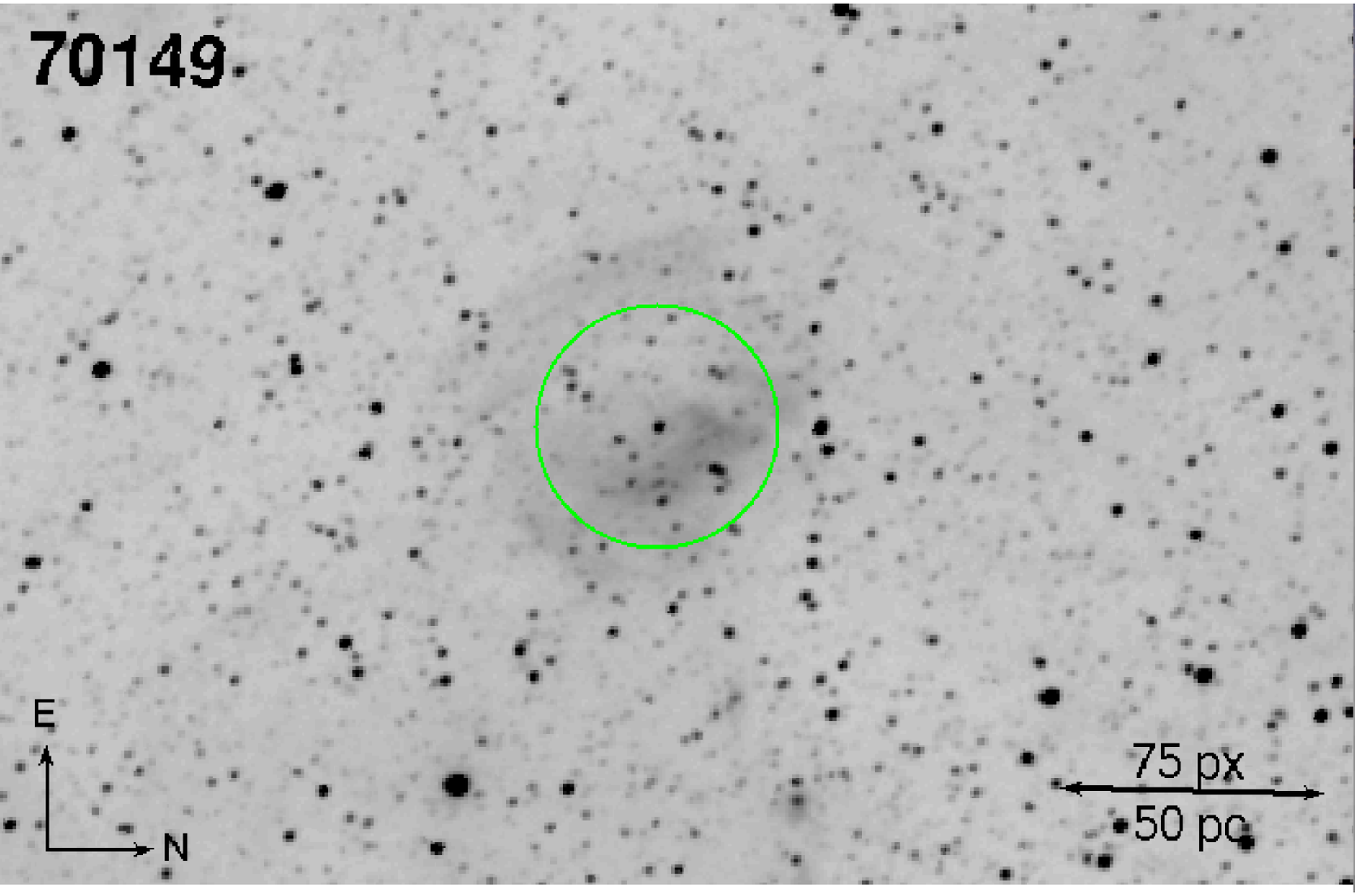}
\includegraphics[scale=0.3,angle=0]{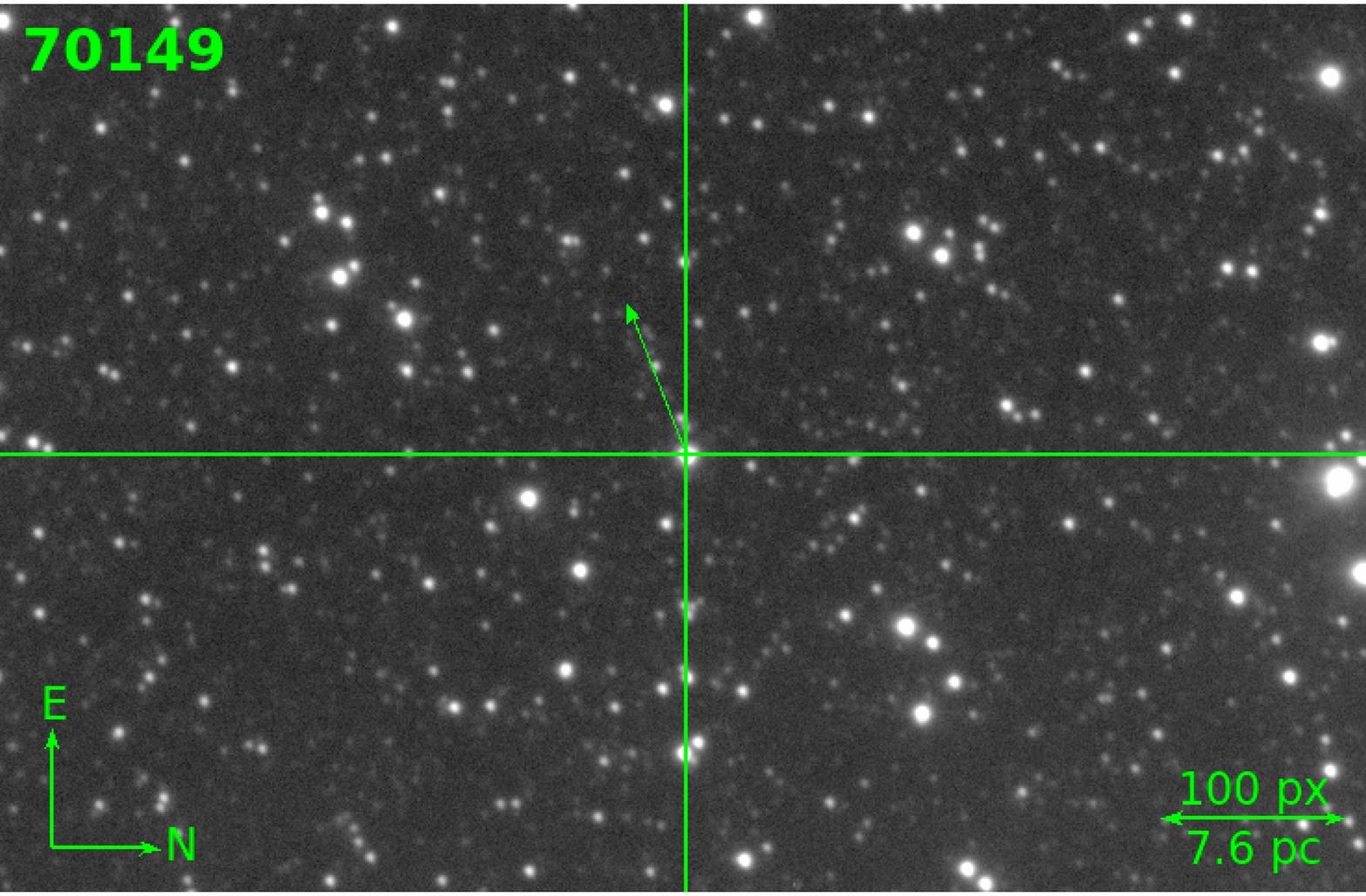}
}
\gridline{
\includegraphics[scale=0.3,angle=0]{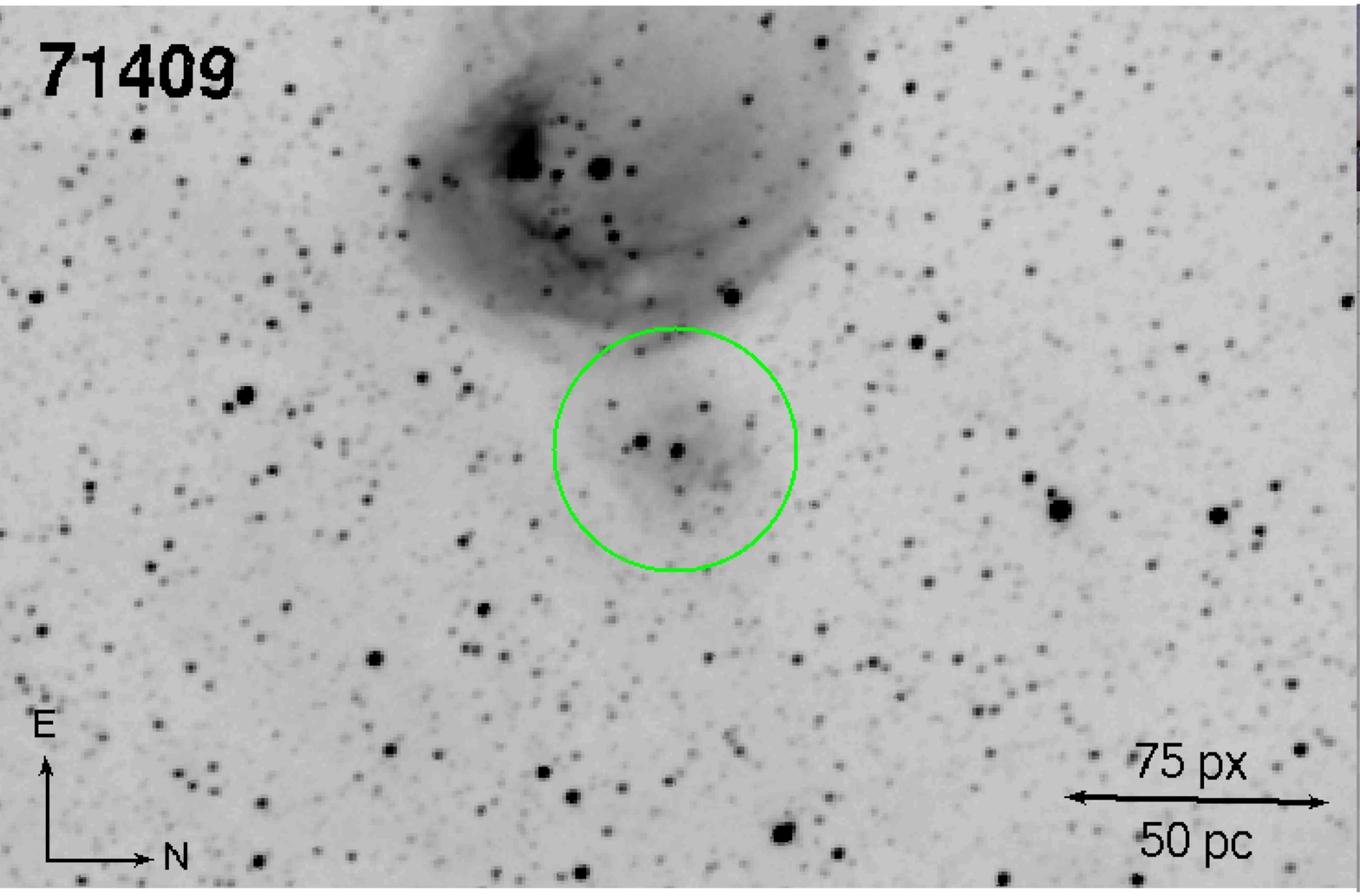}
\includegraphics[scale=0.3,angle=0]{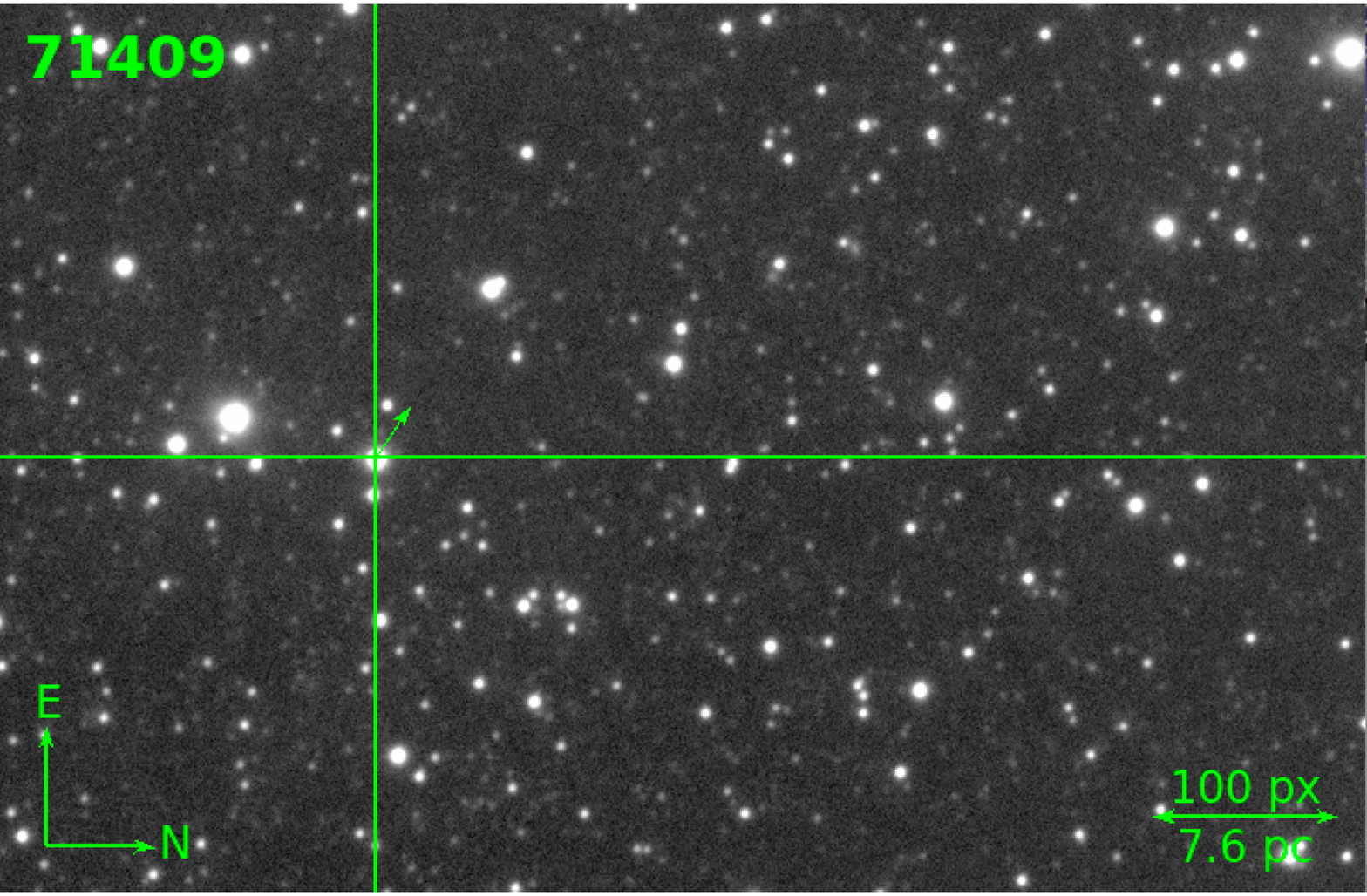}
}
\caption{- continued.
\label{fig:insitucand}}
\end{center}
\end{figure*}
The runaway frequency in this subsample is larger than what we find in Section \ref{subsec:runaways}. 
This is likely because the {\it in situ} candidate sample was selected to have no visual evidence of TIBs, 
and therefore these objects are more likely to be either runaways or candidates for isolated {\it in situ} star formation. Since we know the fraction of runaways in our sample is high \citep{Oey2018, DorigoJones2020}, 
this further enhances the likelihood that objects selected to appear isolated are runaways.  Indeed, the fact that only half of the objects are confirmed runaways implies that the rest remain candidates for highly isolated, {\it in situ} star formation.
Such objects would not be runaways or walkaways, nor would they show TIBs.  For example, the \hii\ regions of targets [M2002] SMC-66415 
and 69598 show ``elephant trunks''  pointing toward the targets as shown in Figure~1 from
\citep{Oey2013}, which are difficult to explain if the objects originated far away.

\section{Fraction of TIB Clusters in the SMC} \label{subsec:finalpercentages}

The non-runaway sample shows slightly positive skews in the NN stellar density distributions.
These are statistically significant, and, as argued above, consistent with the possible presence of TIB clusters.
If there are any real clusters in our sample, we can estimate their potential number by determining the excess induced by the positive skew of the observed distributions relative to those of the random data. We obtain the excess number of target fields having values beyond the midpoint between the medians of our observed data and the random dataset. Table \ref{table:percentages} summarizes our estimates for the percentage of TIB clusters identified in our full sample and subsamples. These estimates assume Poisson errors and do not include any systematic effects.

When estimating their frequencies, we obtain relatively large values in both runaway and non-runaway distributions, as shown in Table \ref{table:percentages}. 
For the runaways, we do not believe that this excess corresponds to true cluster detections,
since our targets were selected to be far from any OB association and runaways are unlikely to originate as TIBs, and instead believe that these are caused by systematic effects. These could, for example, be due to the background stars having positions that are not purely random.
As noted above, Figure~\ref{fig:NNrunaways} shows that the runaway density distribution matches those of the random fields.
Instead, the positive runaway detections likely originate from objects that 
have moved into the line of sight toward a density enhancement.

\startlongtable
\begin{deluxetable*}{l|ccccc}
\tablecaption{Estimated Percentage of TIB Clusters\tablenotemark{a}
\label{table:percentages}}
\tablehead{
\colhead{} & \colhead{Full Sample} & \colhead{Non-Runaways }& \colhead{Runaways }& \colhead{Full Data }& \colhead{Non-Runaways } \\
\colhead{} & \colhead{\%} & \colhead{\%}& \colhead{\%}& \colhead{Subtracted \tablenotemark{b} }& \colhead{Subtracted\tablenotemark{b} }
}
\startdata
 NN Average vs Random 1  & $15 \pm 2.9$ & $22 \pm 5.1$&$11 \pm 3.4$&$4 \pm 4.5$&$11 \pm $ 6.1\\
NN Average vs Random 2& $12 \pm 2.6$ & $21 \pm 5.0$ &$8.3 \pm 2.9$ &$3.7 \pm 3.9 $&$13 \pm $ 5.8 \\
 NN Average vs Random 3 & $12 \pm 2.6$ & $18 \pm 4.6$&$8.3 \pm 2.9$&$3.7 \pm 3.9 $&$9.7 \pm $ 5.4 \\
 NN Median vs Random 1  & $17 \pm 3.1$& $23 \pm$ 5.3 &$13 \pm 3.7$ &$4.0 \pm 4.8$&$10 \pm $ 6.5  \\ 
 NN Median vs Random 2& $15 \pm 2.9$ & $22 \pm 5.1$ &$12 \pm 3.5$ &$3.0 \pm 4.5 $&$10 \pm $ 6.2 \\
 NN Median vs Random 3 & $11\pm 2.4$ & $16 \pm 4.3$ &$10 \pm 3.2$ &$1.0 \pm 4.0 $&$6.0 \pm 5.4 $  
\enddata
\tablenotetext{a}{``NN Average'' or ``Median'' specifies the method of combining the NN stellar density calculations for $j=8 - 12$.
``Random 1'', ``Random 2'', and ``Random 3'' refer to the respective random field datasets.}
\tablenotetext{b}{Columns 5 and 6 show the estimated frequencies obtained by subtracting the runaway frequencies from those of the full sample, and non-runaway samples, respectively. 
}
\end{deluxetable*}

As shown earlier, half of the top 20 potential cluster candidates from NN were actually runaways (Table \ref{table:runawayID}), showing that it is possible for runaways to appear to be in TIB clusters due to happenstance. We take the percentage of positive detections for runaways to be the percentage of TIBs that arise by chance. To obtain an estimate for the TIB cluster frequency, we therefore subtract this frequency of false positives
from the raw estimates from the non-runaways and the full sample. The final resulting estimates are also shown in Table \ref{table:percentages}, which support the existence of a low fraction of TIBs in our sample.

In general, the final, corrected TIB frequency estimates in Table~\ref{table:percentages} for non-runaways are 2 -- 3 times that for the full sample.
This is consistent with the bulk of these estimates representing real clusters, since non-runaways are about half the sample (101 out of 210 targets). 

We show the top 10 cluster candidates identified by NN in Table \ref{table: top10_NN_NR}, and their images in Figure \ref{fig:top10NN}.
A few of these candidates visually appear to be in clusters. However, the rest of them have ambiguous status, again consistent with the scenario that the TIB cluster frequency is small, and of marginal statistical significance. 
Among the top 10 candidates identified by both NN algorithms, 4 are not included in the top 20 FOF candidates. Interestingly, three of these objects,
[M2002] SMC-81646, 
75984 and 69598 are in less dense fields (Figure \ref{fig:top10NN}).

We note in Figure~\ref{fig:top10NN} that some of our field OB stars have companions associated by NN
that appear brighter than our targets in $I$-band.  A few of these red and yellow stars may be evolved massive stars, implying that our target OB star may not always be strongly isolated from other high-mass stars, since our selection criteria are based only on separation from luminous blue stars \citep{Lamb2016}.
For example, in at least one case, [M2002] SMC-81646, the bright companion is a likely SMC red supergiant, based on its radial velocity \citep{Massey2002,Massey2003}. Another object, [M2002] SMC-58947, has a possible yellow supergiant companion \citep[$B - V = 0.65$;][]{Massey2002}, but this could also be a foreground G star at a distance of 380 pc, which places it 270 pc below the Galactic plane ($b = -44.7^{\circ}$). The other two cases in Figure~\ref{fig:top10NN}, [M2002] SMC-6908 and 46241, show candidate SMC AGB and RGB stars respectively \citep{Boyer2011}, which are therefore likely field stars in the line of sight. These stars are much less luminous in $B$ and $V$ than our targets.  But in general, we caution that occasional evolved supergiants may be present near other target stars in our sample.  

With the estimates from our non-runaway data in Table \ref{table:percentages}, we can set an upper limit on the frequency of TIB clusters in our sample.
For the non-runaways, the average excess over the random fields is $11\% \pm 3.3\%$, combining the values based on the average results for $j=8-12$, corrected for the false positive rate; while for the excess based on the median calculations, the average TIB cluster frequency is $8.7\% \pm 3.5\%$.  
Overall, we can use these estimates as an upper limit to the frequency of TIB clusters in the entire field sample which is on the order of $\sim4-5\%$ since non-runaways represent about half of the full sample.

Although Figure~\ref{fig:top10NN} and Table~\ref{table:percentages} show few clear examples of TIBs,
our results suggest
that our estimated $4-5$\% fraction of TIB clusters might be real. On the one hand, statistical tests indicate a lack of positive detections for TIB clusters. 
There are mixed results for NN, with positive results from Wilcoxon but at least one negative result from the Rosenbaum test for all three datasets. These results seem to indicate a lack of evidence for TIBs within our sample. 
But on the other hand, there is a contrast between our runaway and non-runaway populations that is consistent with expectations if TIB clusters are present. 
The positive skews lead to estimated TIB cluster fractions for non-runaways from all the algorithms that are roughly double the value for the full data set, consistent with TIB clusters being associated with non-runaways, as expected. We also see that the non-runaways show a statistically significant distinction from the runaway NN density distributions using the KS test, and $p$-values below 0.1 for the AD test. Furthermore, the stacked fields show clear, centrally concentrated densities only for the non-runaways.  
Thus, although the statistical results are quantitatively inconclusive, the evidence does support a TIB cluster frequency of up to 4 or 5\%.

\startlongtable
\begin{deluxetable*}{c|cccccccc}
\tablecaption{Top 10 Candidates from NN for Non-Runaway Targets\tablenotemark{a} \label{table: top10_NN_NR} }
\tablehead{
\colhead{Target}\tablenotemark{b} & \multicolumn{2}{l|}{NN Median}&\multicolumn{2}{l|}{NN Average}& \multicolumn{2}{l|}{FOF $N_*$} &
\multicolumn{2}{l|}{FOF $M$-value}   \\
\colhead{ } & \colhead{Value}&\colhead{Rank} &\colhead{Value}&\colhead{Rank}& \colhead{Value}&\colhead{Rank} &
\colhead{Value}&\colhead{Rank}
}
\startdata
47459&4.0 &$\bigstar$  &4.0&$\bigstar$ &19& $\bigstar$ &25 &$\bigstar$\\ 
6908 &2.9&$\bigstar$&2.6 &$\bigstar$ &11& $\bullet$ &17&$\bullet$ \\
81646& 2.2& $\bigstar$ &2.3&$\bigstar$&5&\nodata&2.5&\nodata \\
46241 &2.2&$\bigstar$ &2.2&$\bigstar$ &14&$\bigstar$& 16  &$\bullet$ \\
75984 &2.2 &$\bigstar$ &2.2&$\bigstar$&8&\nodata&5.2& \nodata\\
58947 &1.9 &$\bullet$ &1.8&$\bullet$ &16& $\bigstar$ &20 &$\bigstar$\\
25639 &1.8 &$\bullet$ &1.8 &$\bullet$ &11 &$\bullet$&16  &$\bullet$ \\ 
42260 &1.5 &$\bullet$ &1.6& $\bullet$ &3 &\nodata  &4.0 &\nodata \\
69598 &1.5& $\bullet$ & 1.6& $\bullet$ &12& $\bullet$ &8.2& \nodata \\
24213 &1.5 &$\bullet$&1.5&$\circ$ &6 & \nodata & 11.2&$\circ$ \\
13774 & 1.3&$\circ$ & 1.5&$\bullet$& 12& $\bullet$ &21&$\bigstar$ \\
\enddata
\tablenotetext{a}{Open circles, filled circles, and stars correspond to objects identified among the top 20, 10, and 5 TIB candidates, respectively. There are 11 candidates since the median and average results share the same top 10 with one exception.}
\tablenotetext{b}{ID from \citet{Massey2002}.}
\end{deluxetable*}

\begin{figure*}[h!]
\begin{center}
\gridline{
\includegraphics[scale=0.3,angle=0]{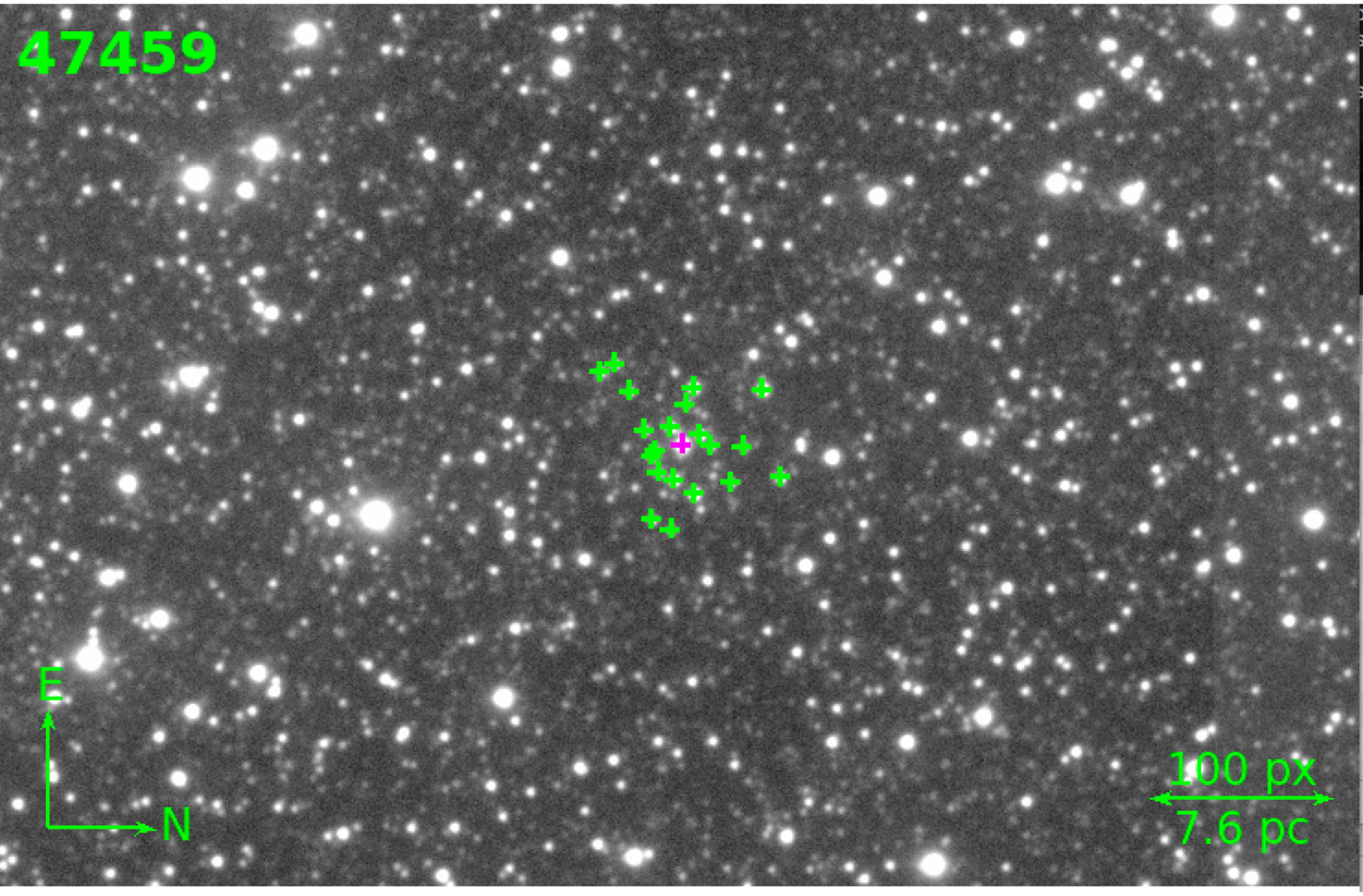}
\includegraphics[scale=0.3,angle=0]{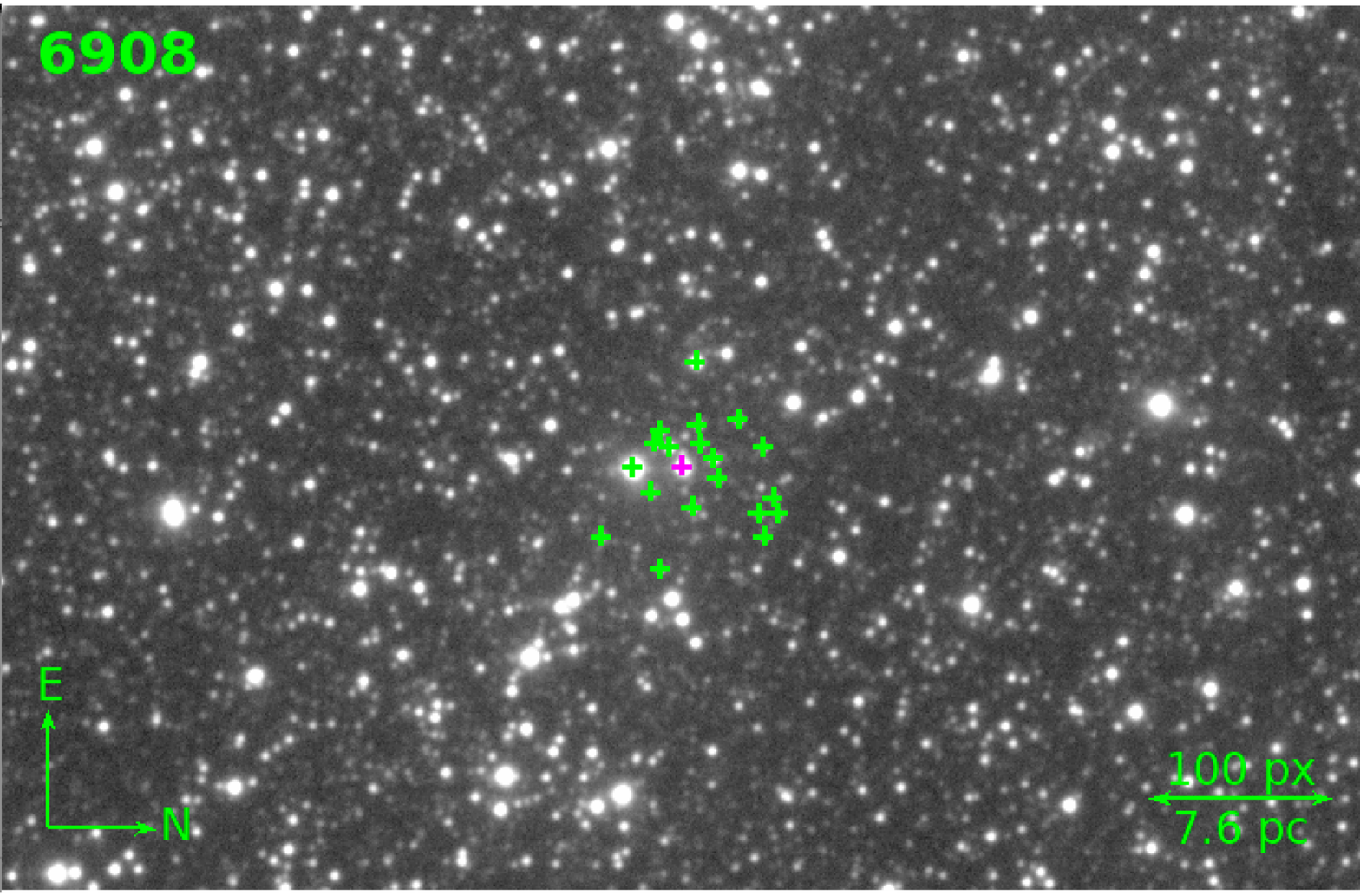}
}
\gridline{
\includegraphics[scale=0.3,angle=0]{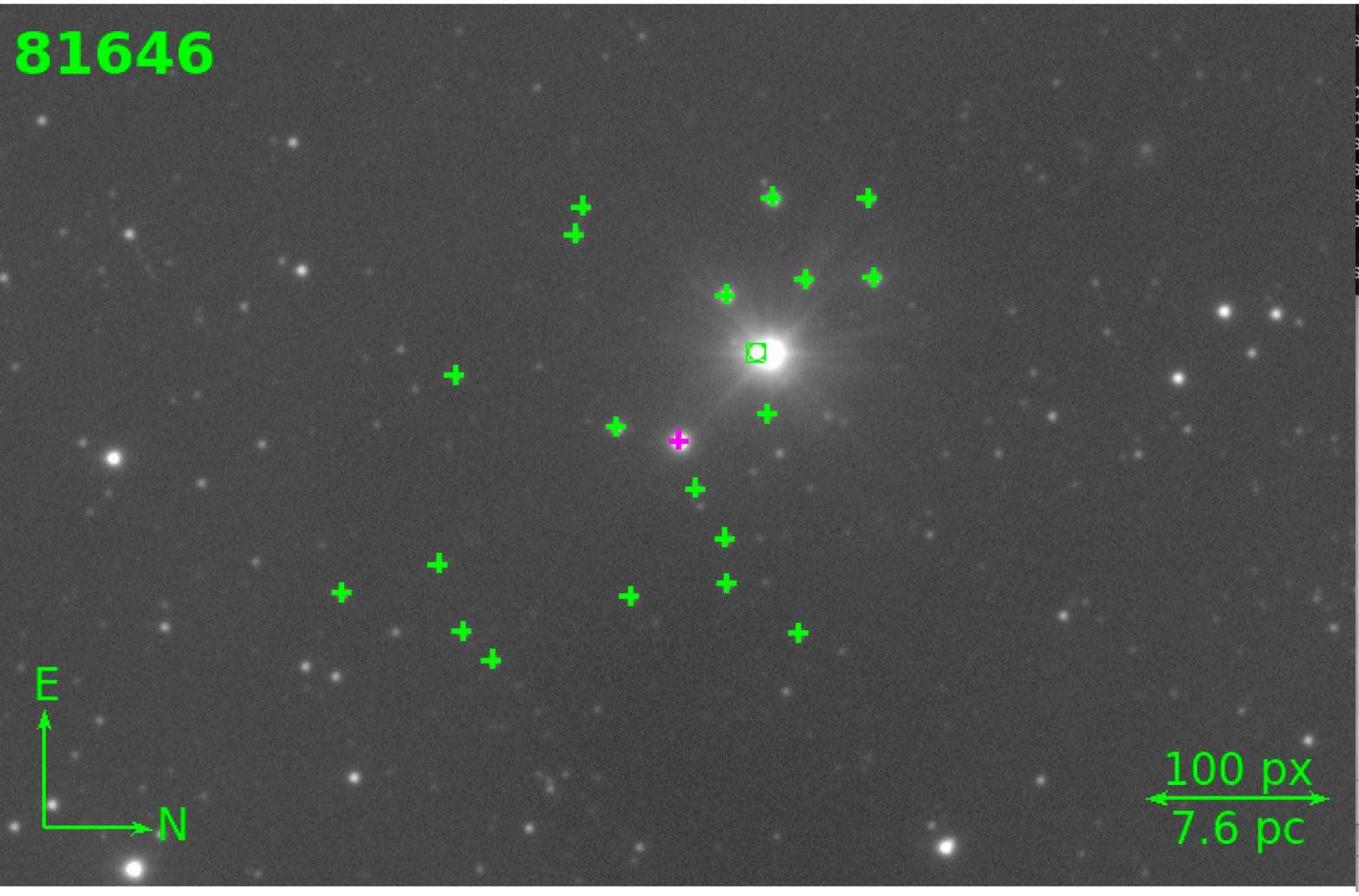}
\includegraphics[scale=0.3,angle=0]{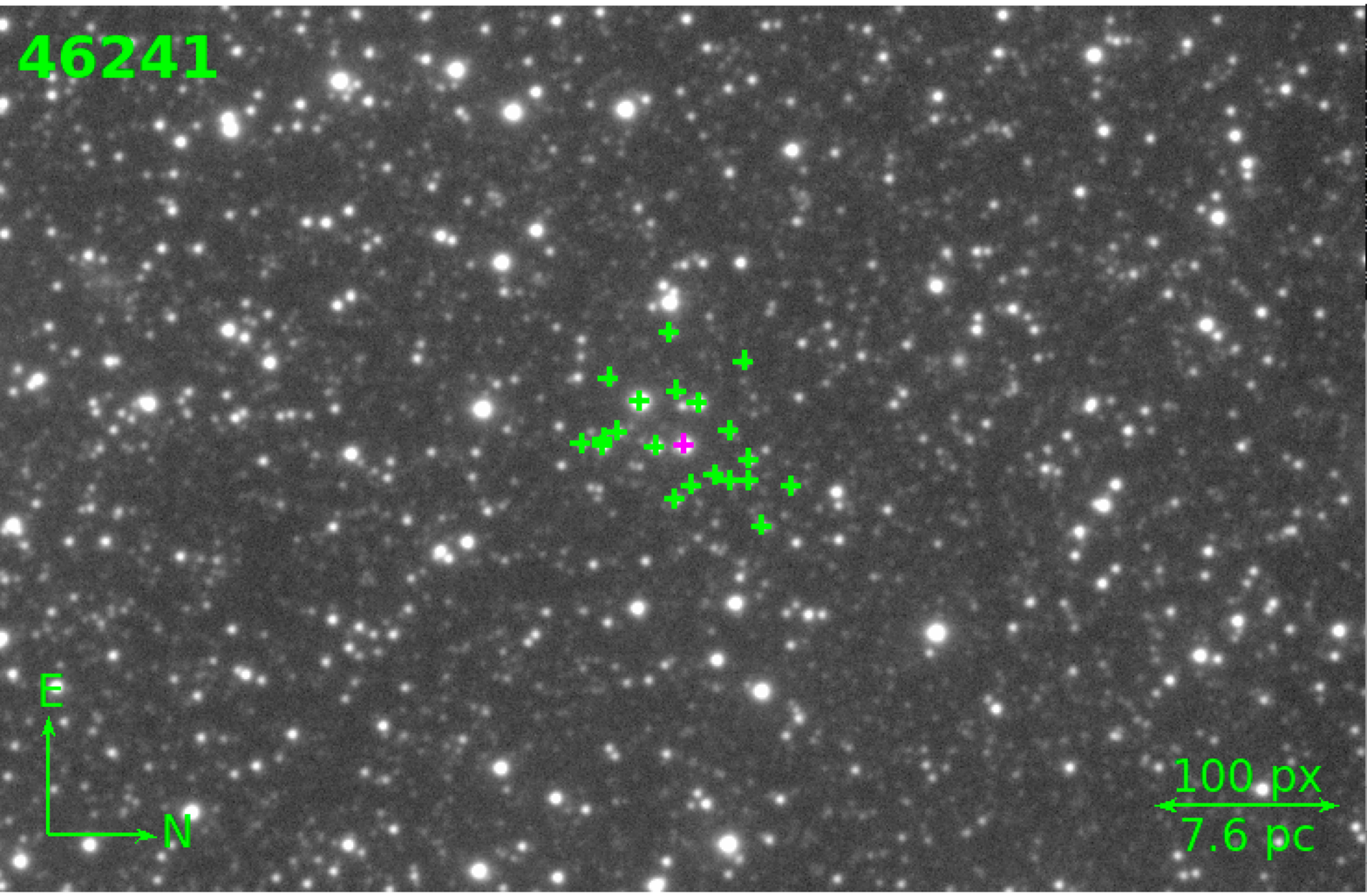}
}
\gridline{
\includegraphics[scale=0.3,angle=0]{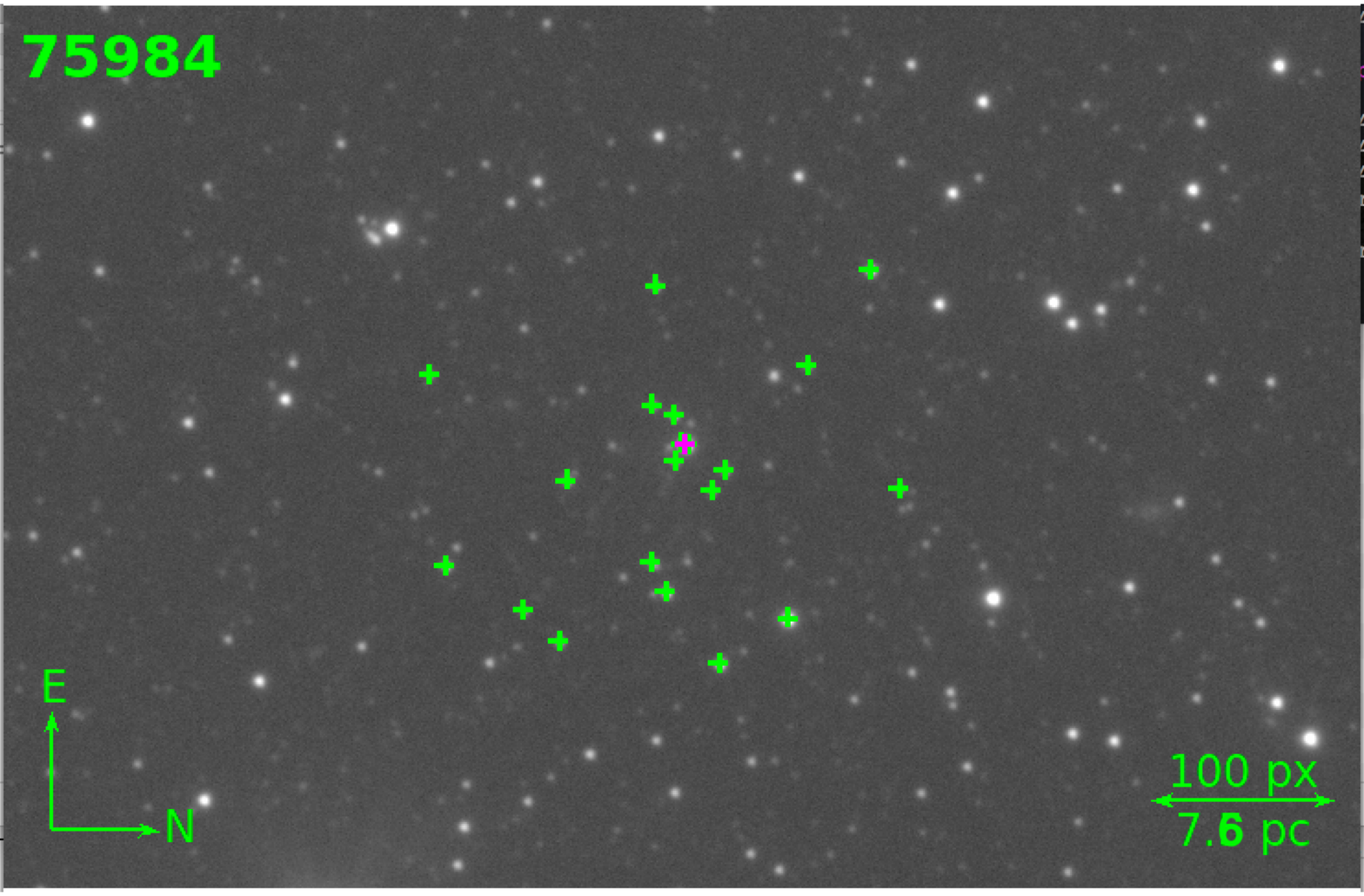}
\includegraphics[scale=0.3,angle=0]{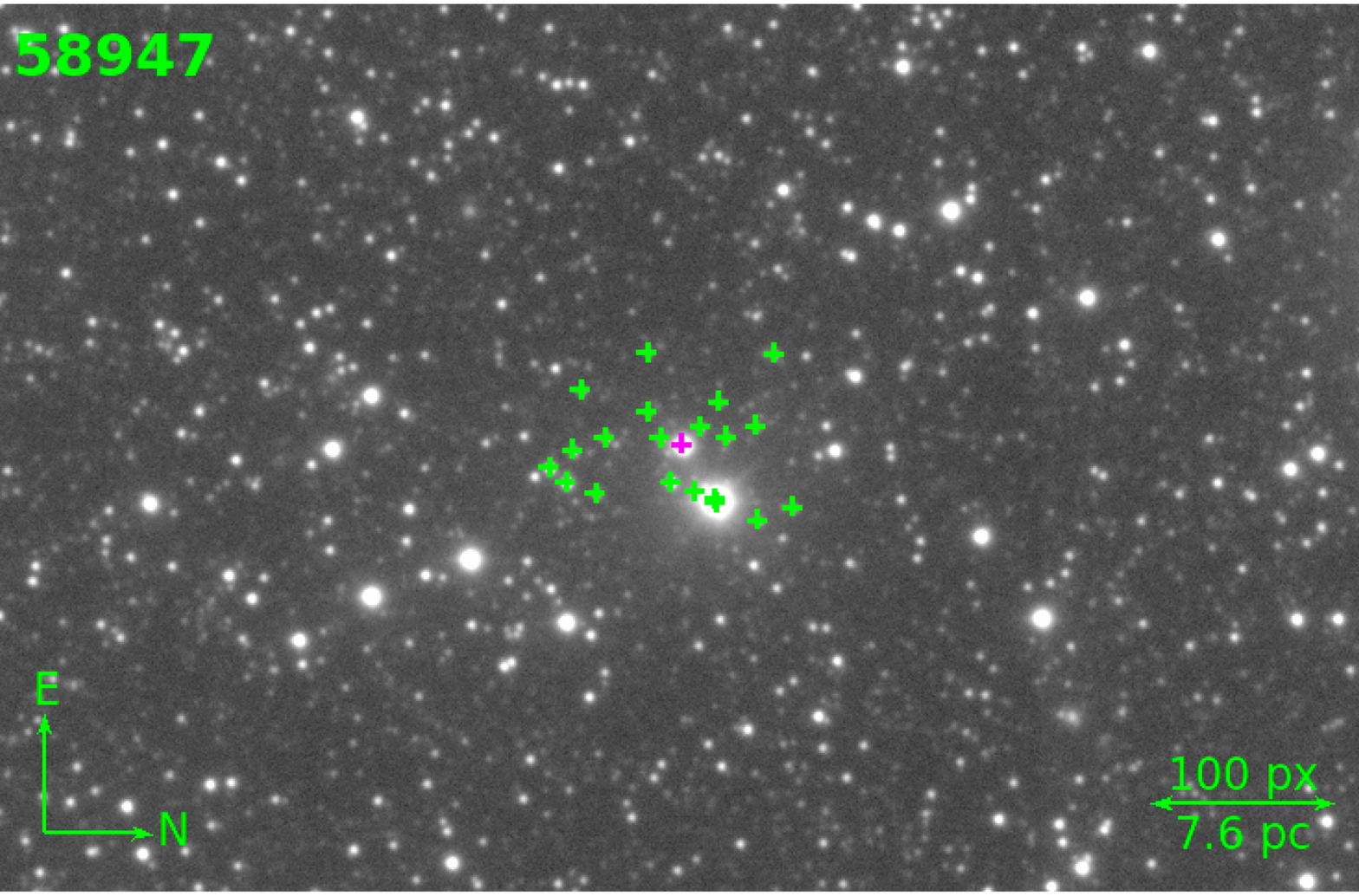}
}
\caption{The top 10 cluster candidates from NN, for the non-runaways in our study. There are 11 candidates shown since the median and average $j=8-12$ results share the same top 10 targets, except for one. North is to the right and East is up. 100 px corresponds to $26\arcsec$ in angular scale. Each target is showed in magenta with its 20 $j$th nearest neighbors in green. Their top-20 rank in each of our criteria are shown in Table \ref{table: top10_NN_NR} \label{fig:top10NN}.
}
\end{center}
\end{figure*}

\setcounter{figure}{12} 
\begin{figure*}
\begin{center}
\gridline{
\includegraphics[scale=0.3,angle=0]{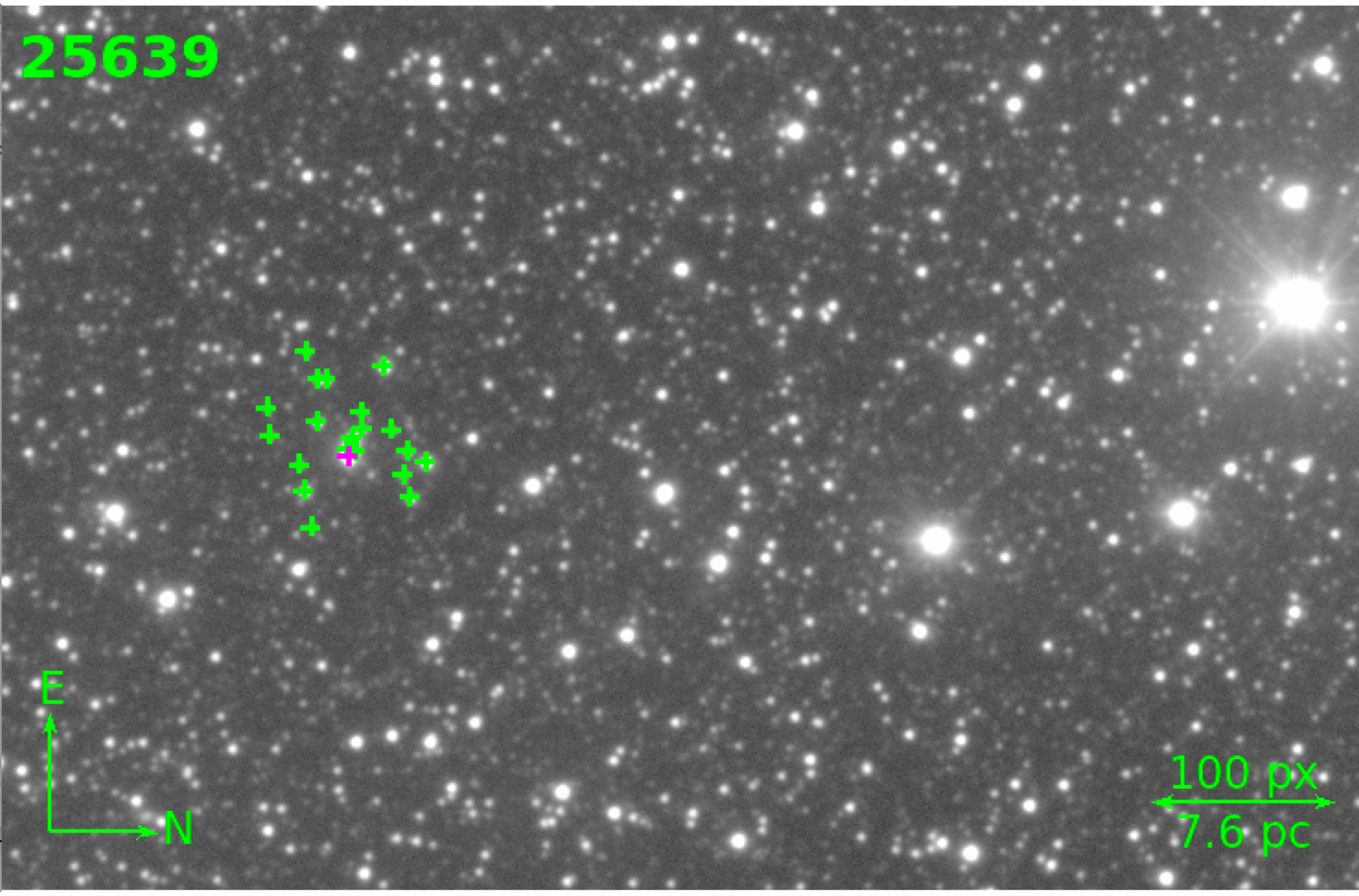}
\includegraphics[scale=0.3,angle=0]{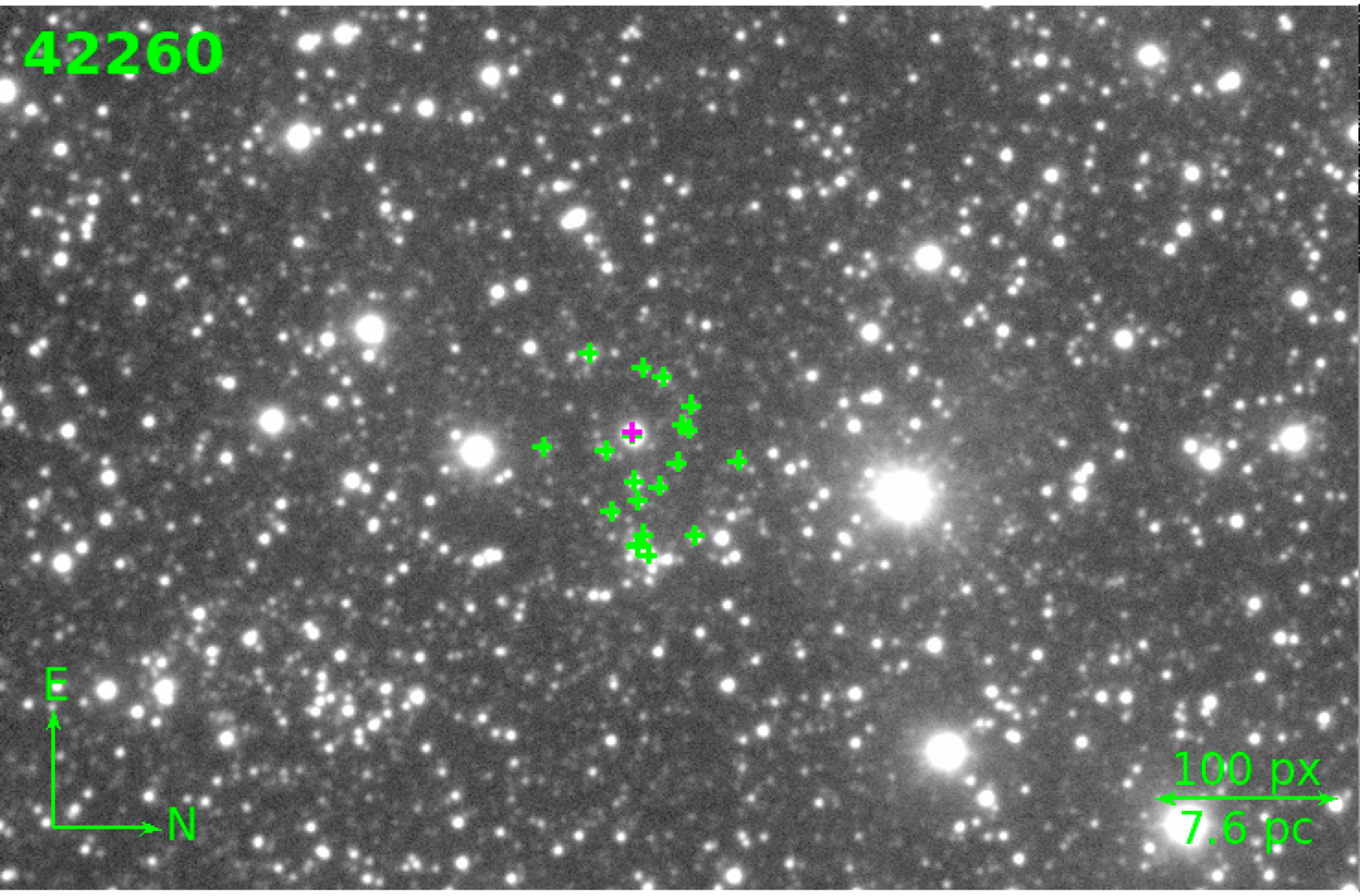}
}
\gridline{
\includegraphics[scale=0.3,angle=0]{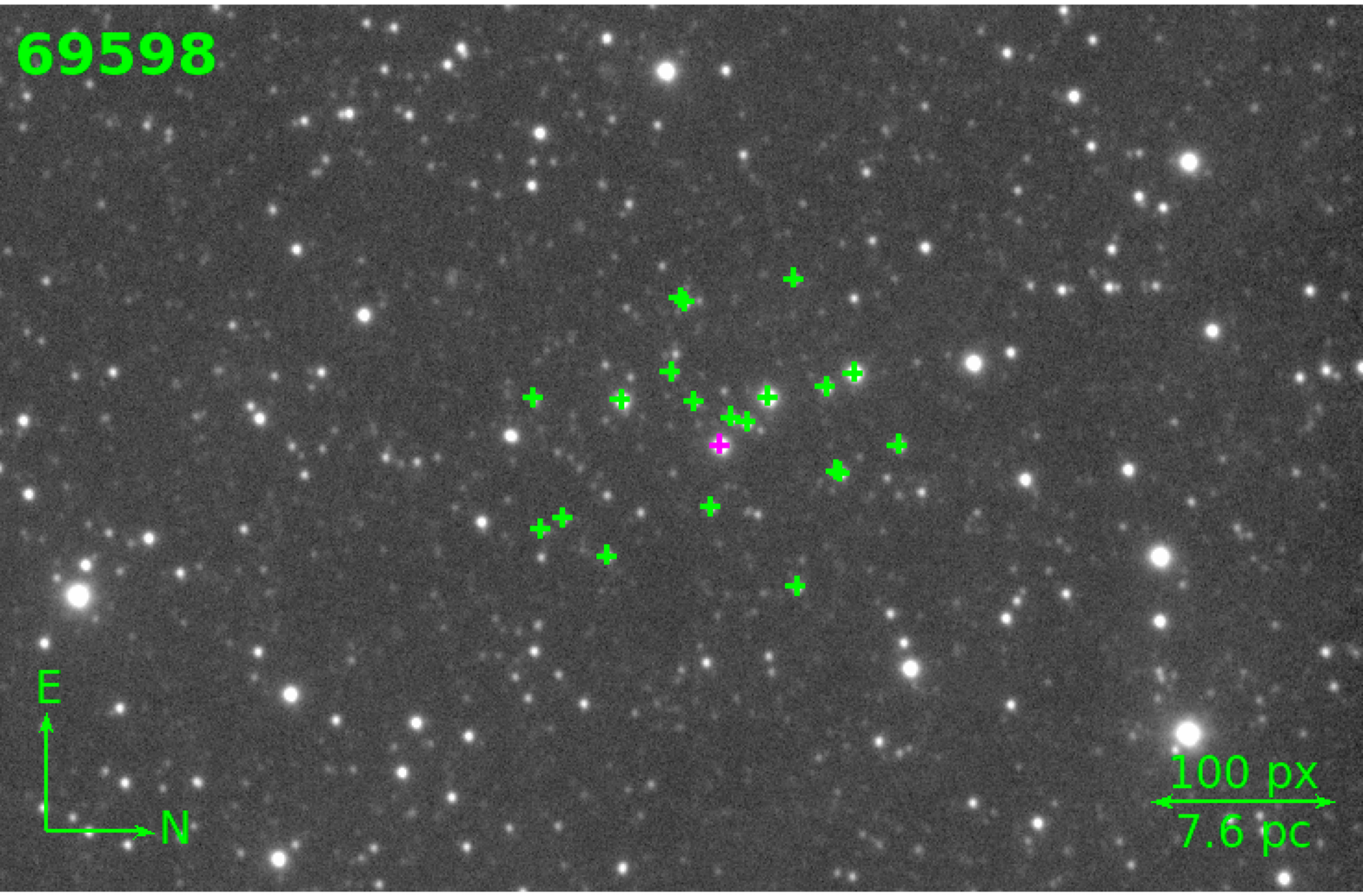}
\includegraphics[scale=0.3,angle=0]{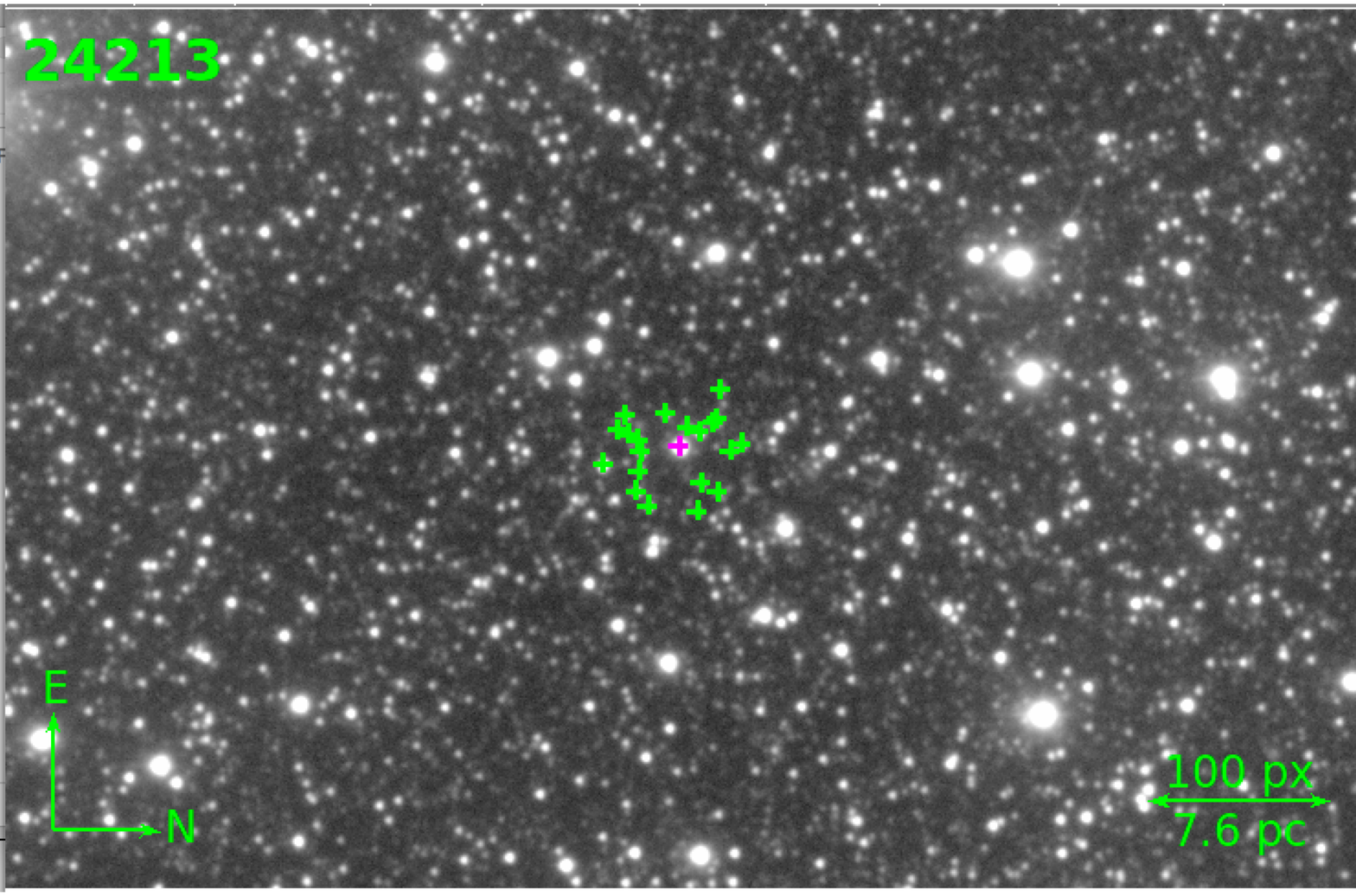}
}
\gridline{
\includegraphics[scale=0.3,angle=0]{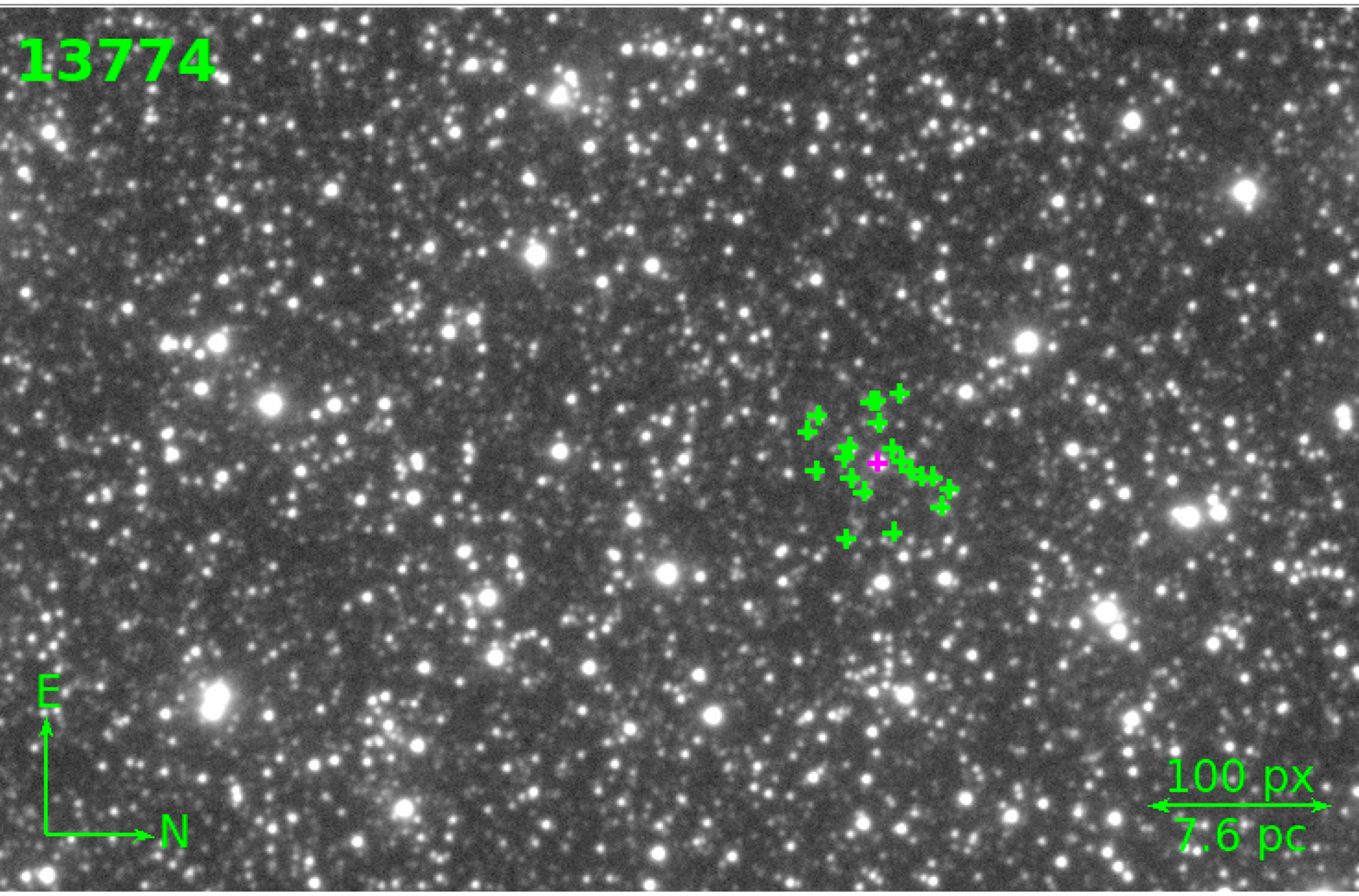}
}

\caption{- continued.}
\end{center}
\end{figure*}

\section{Discussion} \label{sec:implications}

Our results show that {\it in situ} star formation is rare at best, with at most $4-5\%$ of our target field OB stars being in small, TIB clusters. 
This result is consistent with the work of \citet{DorigoJones2020} who use stellar kinematics to determine that runaways and walkaways comprise the overwhelming majority of our sample.

Furthermore, the fact that 5 out of the 14 candidates for {\it in situ} field OB stars found by \citet{Oey2013} turn out to be runaways (Section~\ref{subsec:InSitu}) suggests that their criteria for identifying {\it in situ} field OB stars are surprisingly ineffective, therefore casting some doubt on the remaining 9 candidates in their sample. This result is consistent with \citet{Gvaramadze2012} who determined that many isolated {\it in situ} candidates are actually runaways. Moreover, for the other 5 of the {\it in situ} candidates that are in our sample, 3 are not among our top cluster candidates, suggesting that they are not TIB stars. This however does not rule out the possibility that they are actually rare cases of isolated field OB-star formation and that the 5 runaways may be a product of the selection bias within the sample.

There are two possible explanations for the apparent lack of TIB clusters:  either they could be evaporating on very short timescales, or they simply do not form. \citet{Oey2004} found that the cluster mass function is fully consistent with the existence of TIBs, which would represent the lowest-mass clusters containing single O stars. Indeed, TIBs could comprise up to $\sim$50\%  of field OB stars for the observed cluster MF \citep{Lamb2016}. 
Our results 
may be consistent with the smallest clusters undergoing infant mortality, and therefore causing their OB stars to appear isolated.
\citet{deGrijisGoodwin2008} show that a large fraction of clusters in the SMC evaporate on $3 - 10$ Myr timescales. Therefore, the presence of small, unbound associations with OB stars would not be unlikely, as also suggested by \citet{Ward2020}.

On the other hand, the smallest clusters that form OB stars may have masses larger 
than those probed by our sample selection criteria. \citet{lamb2010} estimated a lower limit of $\sim20\ M_{\odot}$ for the cluster MF, based on observations and Monte Carlo simulations that assumed power laws for both the cluster and stellar MF. If the cluster lower-mass limit has a larger value, then this could be an effect of the low metallicity environment in the SMC, which could inhibit the formation of the smallest clusters.

Alternatively, since we find that few, or none, of our targets correspond to small clusters stochastically forming OB stars, then this may support the $m_{\rm max}\propto M_{\rm cl}^{2/3}$ relation from \citet{Bonnell2004}.  In this scenario, the smallest clusters do exist, but never form OB stars. Stellar mergers have been proposed to explain observed exceptions \citep{OhKroupa2018}, perhaps including the estimated fractions in Section \ref{subsec:finalpercentages}.

Since the core collapse model for massive star formation does allow the occasional formation of OB stars as TIBs, the observed lack of TIB clusters may favor the competitive accretion model.
However, it may be that our non-runaway field OB sample is not large enough to distinguish between these models, and, as discussed in Section~\ref{subsec:finalpercentages}, relatively isolated star formation could still occur in very rare situations.  Our results remain consistent with the estimate of \citet{deWit2004},
who found that $4\% \pm 2\%$ of their sample cannot be traced to a formation to a cluster/OB association, suggesting that these could be either TIBs or candidates for isolated {\it in-situ} star formation. Additionally, our results are also consistent with OB stars forming in small, unbound associations, which would support the formation of massive stars by monolithic cloud collapse \citep{Ward2020}. The small fraction of TIB clusters in our observed dataset would occur if these associations disperse quickly, thereby leaving apparently isolated OB stars.
In any case, our new limits on the existence of TIBs set much more stringent constraints on the formation of massive stars in relative isolation.  

\section{Conclusion} \label{sec:conclusions}

In summary, we use two cluster finding algorithms, friends-of-friends and nearest neighbors, to determine whether our field OB stars are the ``tips of icebergs'' on tiny clusters based on stars having $I < 19.0$. Our 210 target stars are a subset of the statistically complete, RIOTS4 survey of field OB stars in the SMC \citep{Lamb2016}, that are also included in the $I$-band imaging from the OGLE-III survey \citep{Udalski2008}.  We compare our observed data to three realizations of random-field datasets for each field.
We also measure the stellar density as a function of radius from the targets in the stacked fields to search for a signal of TIB clusters in our sample. 

Our results show that there are very few TIBs in our sample, but that a small number likely do exist. 
Results for both cluster-finding algorithms show strong statistical similarities in the spatial distribution of our observed data and random-field datasets.  Indeed, the FOF algorithm, which we confirm to be less effective than NN \citep{Schmeja2011}, is unable to statistically identify a difference between runaway and non-runaway subsamples, highlighting the low occurrence of TIB clusters.

However, the NN algorithm and the stacked fields analysis do show significant differences between the runaways and non-runaways, suggesting the presence of a small number of TIB clusters. The 101 non-runaway stars show higher stellar-density environments,
consistent with the expectation that any TIB OB stars cannot be runaways.  The stacked fields also show an excess density relative to the random fields at radii $< 60$ px (4.6 pc).
In general, the estimated fraction of TIB clusters for non-runaway fields is 2 -- 3 times the estimated frequency in the full sample, which is again consistent with real clusters being present, since non-runaways make up roughly half of our full sample.
Overall, our results show that $\sim4-5\%$ of the field OB stars in the SMC are members of small clusters, and thus almost all are runaways and walkaways. 

The low detection rate of TIB clusters implies that either such clusters evaporate on very short timescales, or they form rarely or not at all.  This may imply that the cluster lower-mass limit is higher than that probed in our sample selection criteria.
If so, these results would be consistent with the $m_{\rm max}\propto M_{\rm cl}^{2/3}$ relation \citep{Bonnell2004}, which would support the competitive accretion model of massive star formation.  However, our sample may not be large enough to rule out the alternative, core collapse model.  

On the other hand, we note that our findings do support a frequency of $\sim4-5$\% for the presence of TIB clusters, and moreover, we cannot definitively rule out the possibility that some OB stars may form in highly isolated conditions, which would not be identified as TIBs.  Although our findings cast doubt on many such candidates identified in our earlier work \citep{Oey2013}, a few still remain as compelling possible candidates of isolated OB star formation.  However, it may be expected that these occur with even lower frequencies than TIB clusters.  Thus, our results set strong constraints on the formation of massive stars in relative isolation.

\acknowledgments

We are grateful to Johnny Dorigo Jones and Matt Dallas for providing data related to the RIOTS4 survey. We thank our anonymous referee for insightful comments and suggestions. This research was supported by the National Science Foundation, grant AST-1514838 to M.S.O., and by the University of Michigan.  M.S.O. also gratefully acknowledges hospitality from the University of Arizona that enhanced this work.

\vspace{5mm}
\facilities{Magellan:Baade, Magellan:Clay, OGLE}

\bibliography{Reference}

\end{document}